\documentclass[structabstract]{aa}
\usepackage{natbib}
\usepackage{graphicx}
\usepackage{natbib}
\bibpunct{(}{)}{;}{a}{}{,}
\usepackage{float}
\usepackage{amsmath,amssymb}
\usepackage{aalongtable}
\usepackage[varg]{txfonts}

\def\msun{M$_{\odot}$}

\def\HD~elta{H${\delta}$}

\def\smash{{\sc smash+}}
\def\naco{{\it NACO}}
\def\pionier{{\it PIONIER}}
\def\sam{{\it SAM}}
\def\gaia{{\it Gaia}}

\defcitealias{sana2014}{Paper~I}

\begin{document}

\title{Southern massive stars at high angular resolution: Physical separations and mass ratios}
\author{F. Tramper \inst{\ref{cab},\ref{leuven}}\corrauth{ftramper@cab.inta-csic.es} 
\and H. Sana \inst{\ref{leuven},\ref{lgi}}\email{hugues.sana@kuleuven.be}
\and A. de Koter \inst{\ref{api},\ref{leuven}}\email{a.dekoter@uva.nl}
\and T. Pauwels\email{tinne.pauwels@kuleuven.be}
\inst{\ref{leuven}} 
                }
\institute{Centro de Astrobiolog{\'i}a (CAB), CSIC-INTA, Carretera de Ajalvir km 4, E-28850 Torrej{\'o}n de Ardoz, Madrid, Spain\label{cab}
\and Institute for Astronomy, Universiteit Leuven, Celestijnenlaan 200 D, B-3001, Leuven, Belgium\label{leuven}
\and
Leuven Gravity Institute, KU Leuven, Celestijnenlaan 200D box 2415, 3001 Leuven, Belgium\label{lgi} 
\and 
Anton Pannekoek Institute for Astronomy, University of Amsterdam, Science Park 904, PO Box 94249, 1090 GE Amsterdam, The Netherlands\label{api} 
}

\date{Received 1 October 2025 / 
Accepted 2 June 2026}

\abstract
{A key property of massive stars is their high degree of multiplicity, which can impact their evolution and end-of-life products.  The Southern Massive Stars at High angular resolution (\smash) survey uses interferometric and high-angular-resolution techniques to  detect companions at intermediate separations, from about 1 mas to 8\arcsec, a domain that so far has remained largely unexplored.}
{For this study, we converted the angular  separations and magnitude contrasts into physical units, i.e.,\ projected  physical separations and mass ratios. We also derived the sensitivity of the survey for various physical and orbital parameters and we investigated the orbital separation and mass ratio distributions.}
{We developed a spectral-type and luminosity class -- $H$-band luminosity -- mass  calibration based on existing grids of physical parameters for O stars, and we used it to obtain the photometric distance to each system, correcting for all known companions within the 2MASS point-spread function. We also derived the individual masses of the primaries and of each detected companion. }
{The probability of detecting companions is very uniform within the sensitivity limits of the \smash\ survey, which can be considered almost complete for binaries with $1< \log (a/AU)<4$ and $q=M_2/M_1>0.2$. The projected separations follow a uniform distribution in log separation. The  mass ratios follow a power-law distribution $f_q \propto q^\kappa$ with $\kappa$ values that seem to decrease toward larger separations. For resolved companions within 100~AU, we find $\kappa_{<100} =-0.6^{+0.9}_{-0.7}$ which is compatible with both the power-law distributions derived for spectroscopic binaries ($\kappa\approx0$) and with what was proposed in an earlier study ($\kappa =-1.4\pm0.4$) based on about half the \smash\ sample in the same range. Beyond $\approx$1000~AU, we observe a clear lack of (near-)equal-mass companions, with an upper mass-ratio limit declining toward larger separations.}
{}


\titlerunning{\smash: Physical separations and mass ratios}

\maketitle
\nolinenumbers

\section{Introduction}\label{sec:intro}
The evolution of single massive stars is regulated by their mass, rotation, and mass loss \citep[e.g.,][]{langer2012,ekstroem2012}.  Most massive stars are also part of multiple systems with at least one nearby companion \citep[e.g.,][]{mason2009, sana2011}. Because of this high multiplicity fraction, interactions through mass transfer or mergers are thought to fundamentally modify the evolution of high-mass stars \citep{sana2012}. When this interaction happens, and what its outcome is, primarily depends on the initial properties of a given system. Therefore, a good understanding of the overall multiplicity properties of massive star populations is fundamental to predict their evolutionary life cycle and final outcome. 

In the last decade, the characterization of the multiplicity properties of massive stars has mostly been performed through  spectroscopy \citep[e.g.,][]{banyard2022,dunstall2015, kiminki2018,kobulnicky2014,MA2019,villasenor2021,villasenor2025} and high-angular-resolution imaging techniques \citep{MA2018, rainot2022, pauwels2024}. If limited to these approaches, the view of massive star multiplicity would be biased toward tight ($<$ 1 AU) and wide ($> 10^3$ AU) systems \citep{mason1998}. For nearby stars, interferometry offers a way to probe the intermediate separation range. However, the higher contrast ($\Delta m > 2$ mag) and closest angular separation ($d < 75$~milli-arcsec) regimes were challenging to probe with the first generations of optical interferometers \citep{sanaevans2011}.

However, progress in the robustness and sensitivity of long-baseline interferometry in the 2000s offered new possibilities.  The Southern MAssive Stars at High angular resolution {\sc (smash+)}  survey \citep[][henceforth \citetalias{sana2014}]{sana2014} combined observations performed with the sparse aperture masking (\sam) mode \citep{lacour2011} of \naco\ at the Very Large Telescope (VLT) and with  the four-beam combiner \pionier\ \citep{lebouquin2011} at the VLT Interferometer (VLTI). \naco\ probes separations in the range 30-250~mas, while \pionier\ opens the 1-45 mas window. Additionally, \naco\ provides adaptive-optics- (AO-) corrected imaging at large working angles, which allows the detection of binaries at separations $> 300$ mas.  {\sc smash+} targeted all nearby O stars in the southern hemisphere ($\delta < 0\degr$) with H-band magnitudes brighter than $m_H = 7.5$. In total, 117 O-type stars were observed with \pionier, and 162 O-type stars with \naco/\sam. Of these objects, 105 were observed with both instruments.
The \smash\ sample is summarized in Table~\ref{t:sample}.
\begin{table}[t!]
    \centering
    \caption{Overview of the sample.}
    \label{t:sample}
    \begin{tabular}{r c c c}
    \hline
    Lum.  & Pionier & NACO & Pionier \\ 
    Class & only & only & + NACO \\
    \hline
    V-IV   & 1 & 35 & 36  \\ 
    III-II & 3 & 11 & 34 \\
    I      & 8 &  9 & 33  \\
    No LC  & 0 &  2 &  2 \\
    \hline
    All    & 12 & 57& 105 \\
    \hline
    \end{tabular}
\end{table}

\begin{figure}
        \center
        \resizebox{1\hsize}{!}{\includegraphics{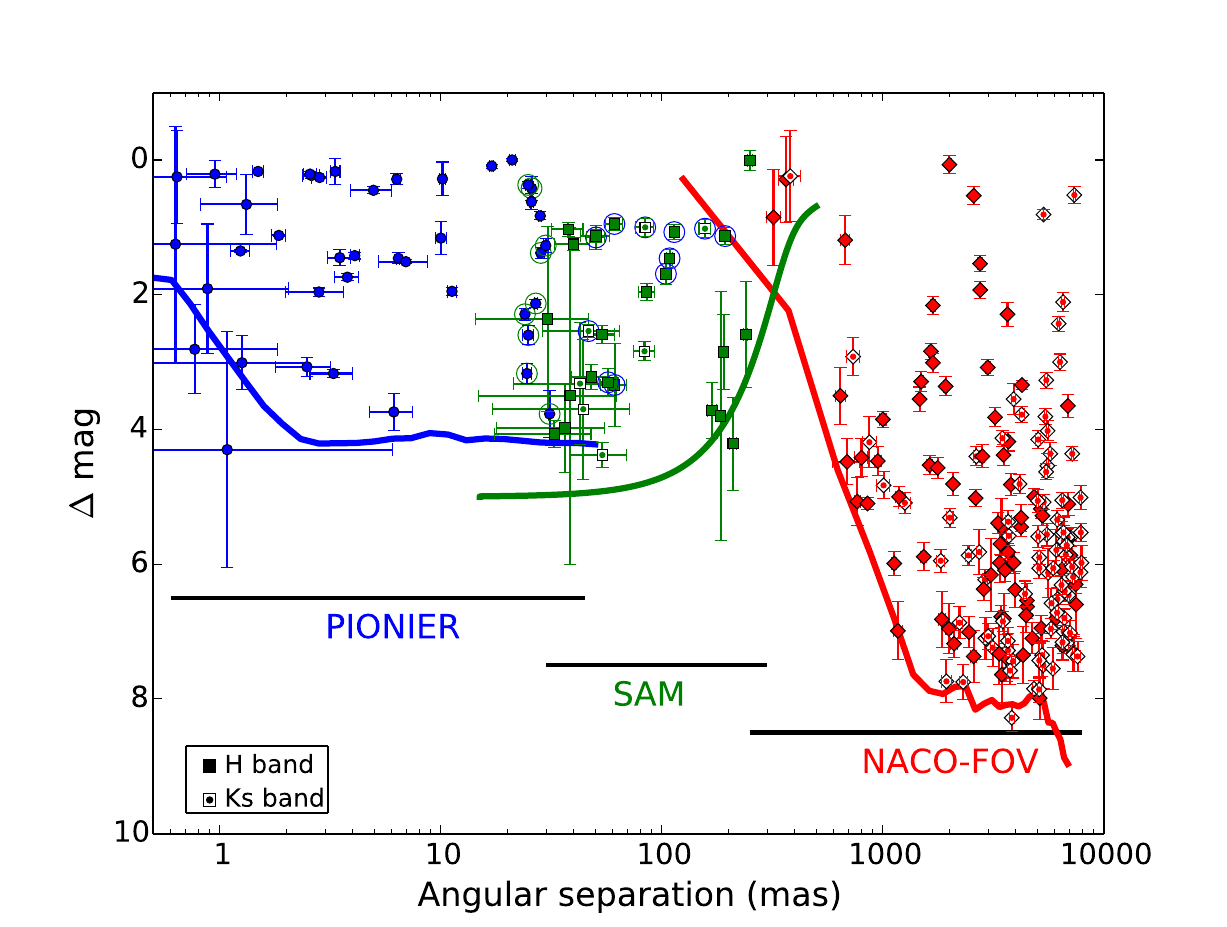}}
        \caption{Magnitude difference versus angular separation for the pairs detected in the \smash\ survey. Circled objects are detected by both \sam\ and \pionier. The solid lines indicate the median H-band sensitivity of the various instruments. Figure reproduced from \cite{sana2014}.}
        \label{fig:smash}
\end{figure}

The resolved pairs found by the {\sc smash+} survey are shown in Fig.~\ref{fig:smash}, which also shows the separation ranges and median sensitivities of the  instruments. The survey revealed a large number of companions, with a fraction of 0.53 stars having at least one resolved companion within 200 mas. Including known spectroscopic or eclipsing companions, the multiplicity fraction reaches 0.91 at separations smaller than 8\arcsec\ and the average number of companions is $2.2\pm0.3$. For luminosity class (LC) V, the fraction of bound companions reaches 100\%\ at 30~mas.  The {\sc smash+} survey therefore demonstrated that massive stars form almost exclusively in multiple systems and that the majority of massive stars are in triples or higher-order multiples.

A decade later, the \smash\ survey has remained a reference for the separation range beyond the spectroscopic regime, providing a  database to constrain the distributions of wide binaries and higher-order systems \citep[e.g.,][]{moe2017,perets2025}. For this study, we converted the observed angular separations and magnitude differences of individual companions detected in the \smash\ survey to projected physical separations and mass ratios, respectively, to derive their distributions and investigate possible correlations between physical and orbital properties. The methods used are described in Sect.~\ref{sec:methods}. Section~\ref{sec:bias} investigates the sensitivity domain of the \smash\ survey while  Sect.~\ref{sec:results} presents and discusses orbital parameter distributions. We summarize our findings in Sect.~\ref{sec:conclusions}.

\section{Methods}\label{sec:methods}

In this section we describe the methods used to convert the observed angular separations and magnitude differences to physical units, i.e., projected separation expressed in astronomical units (AU) and mass ratios. The distance determination needed to convert the angular separations to physical separations is presented in Sect.~\ref{sec:distances}, followed by the mass determinations based on the magnitudes in Sect.~\ref{sec:mass}.

\subsection{Distance determination}\label{sec:distances}

To convert the observed projected separations of the companions on the sky to a projected physical separation, the distances to the targets are needed. While \gaia\ distance estimates are valuable for many targets, for bright stars and for binary systems they can be unreliable, especially when dealing with binary systems in the separation range covered here. Specifically, the astrometric measurements of the photocenter can be impacted by the scanning direction of \gaia\ with respect to the instantaneous orientation of the binary system on the sky. The \gaia\ measurements for bright, hot stars might further be subject to a number of systematics \citep{als3}. Some of the stars in our sample are also simply not in Gaia DR3, or have extreme uncertainties. To provide a homogeneous distance determination of the whole sample of stars, we developed an alternative strategy based on an absolute magnitude calibration as a function of spectral type (SpT) and luminosity class (LC). We then used the stars that do have reliable {\it Gaia} distances to assess the accuracy of our calibration.

\begin{figure}
        \resizebox{\hsize}{!}{\includegraphics{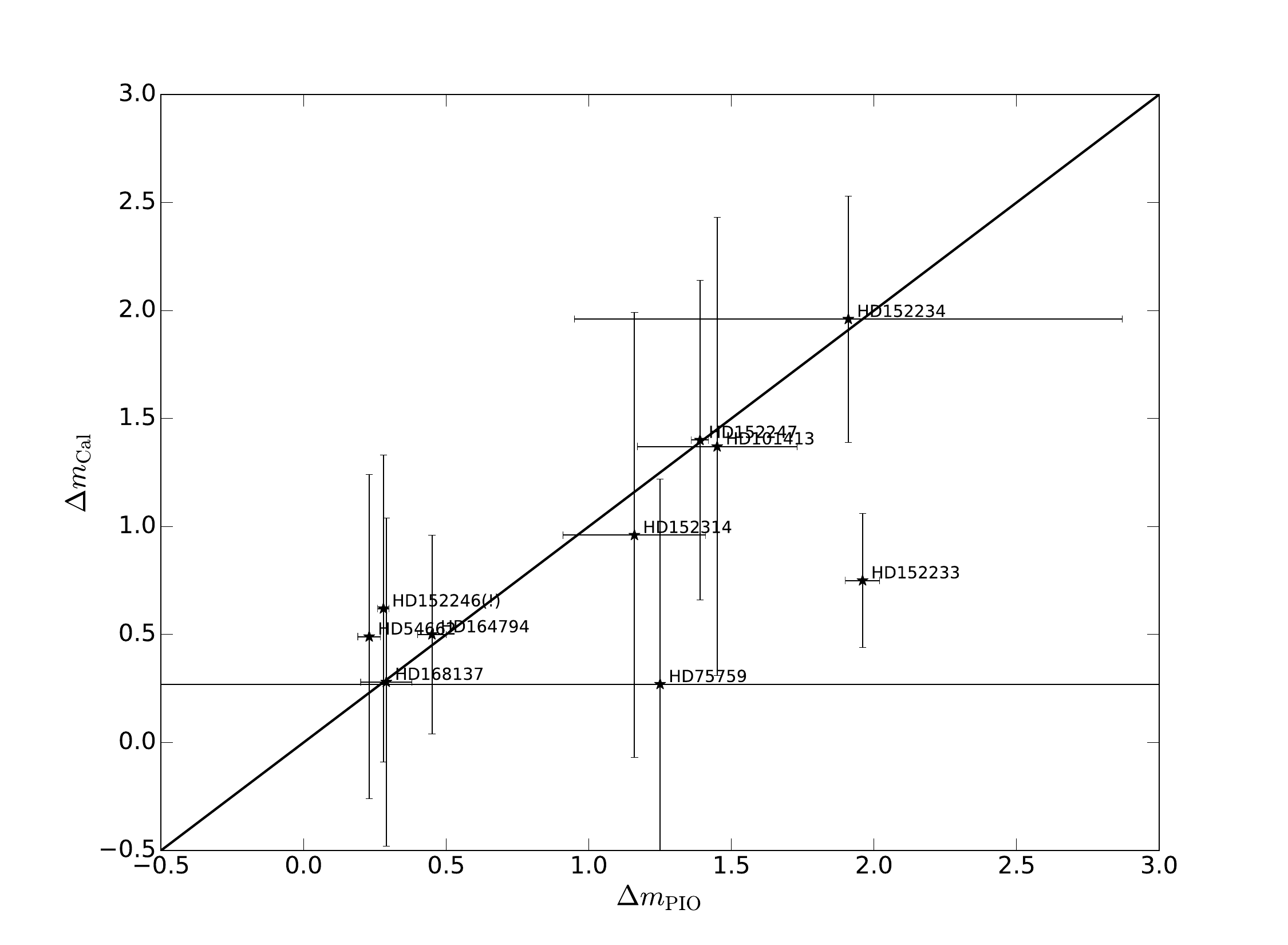}}
        \caption{Magnitude difference of SB2 systems resolved by  \pionier\ versus those obtained from the absolute magnitude calibration.}
        \label{fig:PIOvsCal}
\end{figure}

\subsubsection{Spectrophotometric distances}
Our calibration was derived from the absolute magnitudes of \citet[][their table~2]{martins2006}. As magnitudes are only given for LCs V, III, and I, we first derived magnitudes for luminosity classes IV and II by linearly interpolating between the closest luminosity classes. We then derived a relation for the absolute magnitude as a function of spectral type for each of the luminosity classes, so that we were able to derive absolute magnitudes for those spectral types not listed in \citet[][e.g., O2, O9.7]{martins2006}. The procedure and resulting relations are described in detail in Appendix~\ref{sec:absmag}. The relations were derived for the K band, but they can also be applied for the H band using $(H-K)_0 = -0.10$, which is valid for all spectral subtypes and luminosity classes in the O-star regime \citep{martins2006}. We adopted an uncertainty of half a spectral subtype and half a luminosity class to estimate the uncertainty on the absolute magnitude.

To test the above calibration, we applied it to both the primary and secondary in double-lined spectroscopic binary (SB2) systems that have been resolved by \pionier, and for which the spectral type and luminosity class of the secondary is known. The resulting predicted magnitude difference, which only depends on the spectral classification of both components and is independent of extinction and apparent magnitudes, can then be compared to that observed by \pionier, as shown in Fig.~\ref{fig:PIOvsCal}. In general, we find good agreement between the predicted and observed magnitude difference.

Before we can use the derived absolute magnitudes of the primaries to determine the distance, the apparent magnitudes need to be corrected for the contribution from nearby companions. The photometry listed in Table~1 of \citetalias{sana2014} originates from the 2MASS survey, which has an aperture of 4\arcsec. We thus corrected these magnitudes for the observed magnitude difference of all resolved companions within 4\arcsec\ (see Appendix~\ref{sec:magcorrect}).

\begin{figure}
        \resizebox{\hsize}{!}{\includegraphics{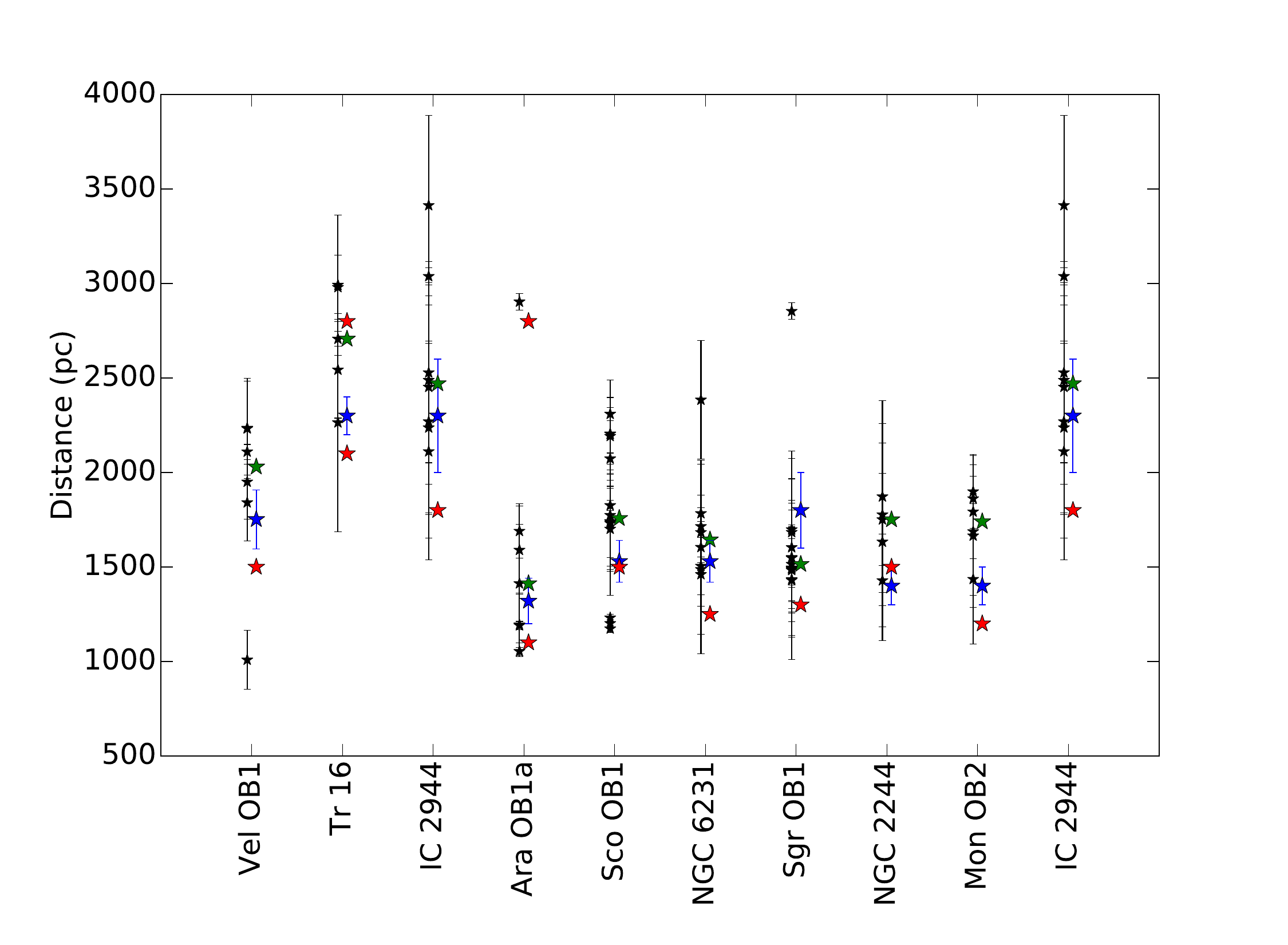}}
        \caption{Derived distances for clusters with at least four members in the {\sc smash+} sample. Black stars indicate the distances for the individual stars and green stars the median. Blue stars indicate literature values for the cluster distances, and red stars the distances based on Tycho-2 \citep{kharchenko2005, melnik2009}.}
        \label{fig:clusterdist}
\end{figure}
Apart from the resolved companions, the observed magnitudes may also contain a contribution from unresolved companions, i.e.,\ the inner spectroscopic binaries. To correct the observed magnitude, we further used the calibration given in Appendix~\ref{sec:absmag} to estimate the magnitude difference. Therefore, we could only do this for systems for which the spectral types of both components are known, i.e.,\ double-lined spectroscopic binaries (SB2). This method cannot be applied to systems for which the companion spectral type could not be determined, mostly single-lined spectroscopic binaries (SB1), unless the spectroscopic binary is so wide that the secondary was resolved by \pionier. This typically occurs for systems with orbital periods of a few months to a couple of years, and we could identify four cases in the \smash\ database. For the remaining SB1 systems, however, the non-detection of the companion in the spectrum likely implies that it is faint compared to the primary star, and thus will only have a small contribution to the combined magnitude, and we thus neglected their contribution. These systems are marked in Table~\ref{tab:distance} for reference.

As a last step, we corrected the magnitudes for extinction using the available VJHK-band photometry and the \citet{fitzpatrick1999} extinction curve (see Appendix~\ref{sec:extinction}). We then calculated the distance to the system using the absolute H-band magnitude and the apparent H-band magnitude of the primary star corrected for extinction and the contribution from companions. The resulting distances are given in Table~\ref{tab:distance}.

\subsubsection{Accuracy  of the distance determination}
To assess the accuracy of the derived distances, we compared the median distances of stars in clusters to literature values of the distance to those clusters. We applied this to each cluster that has at least four members in the {\sc smash+} sample. This comparison is shown in Fig.~\ref{fig:clusterdist}. We find that on average the derived distances are slightly larger than the literature values, with a mean offset of +186 pc for the 10 clusters. 

\begin{figure}
        \resizebox{\hsize}{!}{\includegraphics{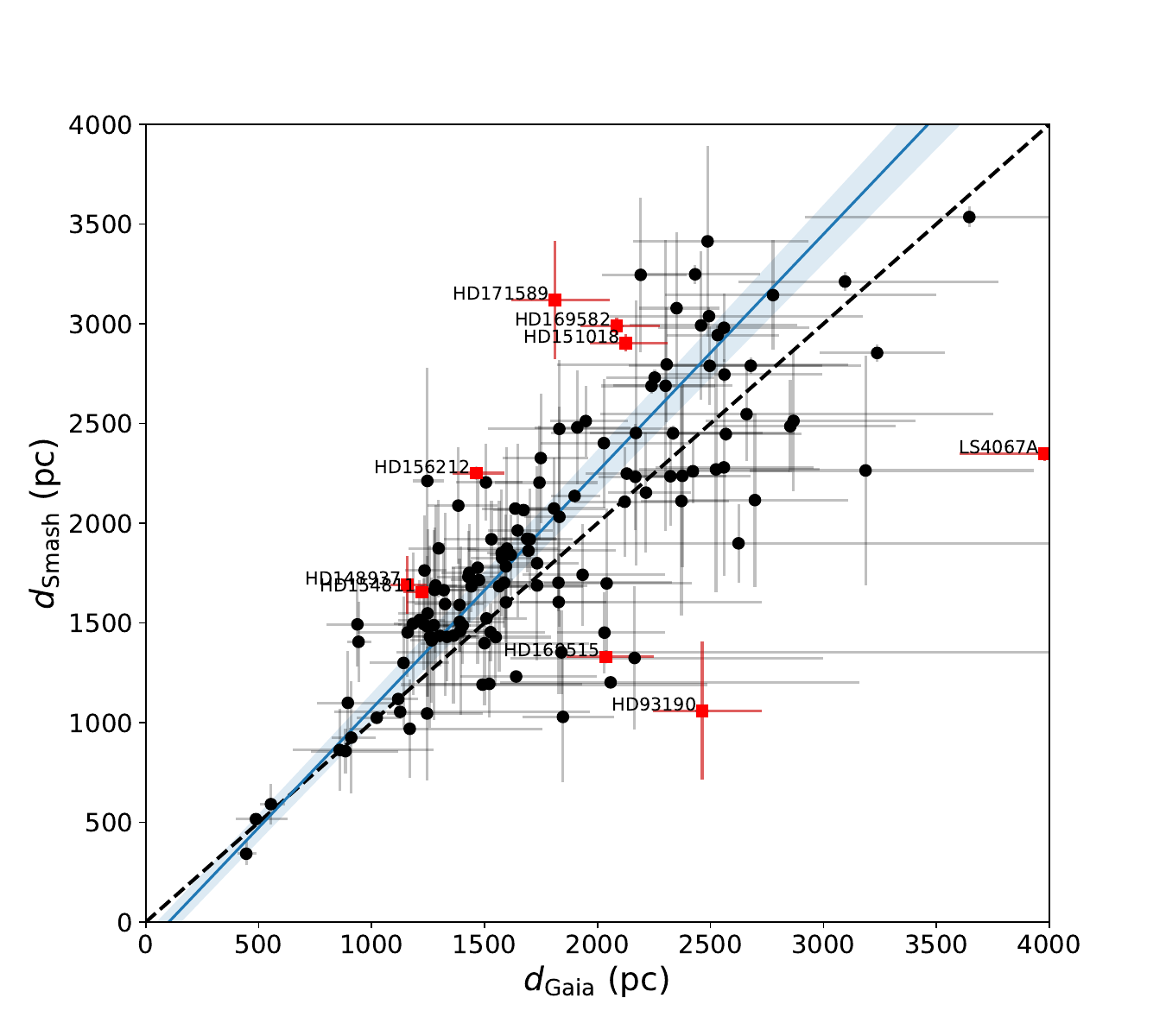}}
        \caption{Comparison of the spectro-photometric distances derived in this work to the \gaia\ distances from \citet{als3}. The dashed line gives the 1:1 relation. The red symbols indicate the locations of the outliers discussed in the text while the blue line and associated shaded area are the systematic deviations presented in Eq.~\ref{eq:gaia}.
    }
        \label{fig:gaia_compar}
\end{figure}
In Fig.~\ref{fig:gaia_compar}, we also compare our individual distance with recently available distance estimates from the Alma Luminous Star III catalog \cite[][ALS~III]{als3}. The latter relies on  \gaia\ but applies a series of corrections appropriate for bright stars. We rejected stars with $RUWE>1.4$ in this comparison as this is generally indicative of a poorer astrometric solution, as well as a small number of stars with {\it ALSIII-Gaia} distances over 4~kpc and large distance errors. While the dispersion is significant, most of it results from typical error sizes of $\sim$200~pc. 

We observe a small but systematic offset of  $\langle d_\mathrm{H} - d_\mathrm{Gaia}\rangle = 156 \pm 22$~pc. Applying a 3$\sigma$ clipping to remove outliers (see below), this difference lowers to  $111 \pm 25$~pc. While statistically significant, this offset is smaller than the mean uncertainties from either method. 
A closer look at the 9 outliers with  $>3\sigma$ deviations between the two distance estimates provides insight into the physical cause of this disagreement.

Three objects that have an {\it ALSIII-Gaia} distance significantly larger than the spectro-photometric distance are all Oe stars
(HD169515, O9.7Ibep;
HD93190, O9.7:V:(n)e;
LS4067A, O4.5Kfpe). These objects are too faint compared to the Martins' calibrations that we adopted, which probably results from a lower surface brightness due to their rapid rotation or to partial extinction by their disks. 

Six stars have an {\it ALSIII-Gaia} distance smaller than their spectro-photometric distance, i.e., they are brighter than expected from the calibration. Four of these six stars are  supergiants (HD151018 O9Ib,
HD154811, OC9.7Ib;
HD156212, B0Iab;
HD169582, O6Iaf) and one is a bright giant (HD171589 O7.5II(f)).
The supergiants are known to display significant spread in their luminosities at a given SpT as objects of different masses evolve almost horizontally while nearing the end of the main sequence, so that supergiants of a given spectral subtype can have different luminosities.

\begin{figure}
        \resizebox{\hsize}{!}{\includegraphics{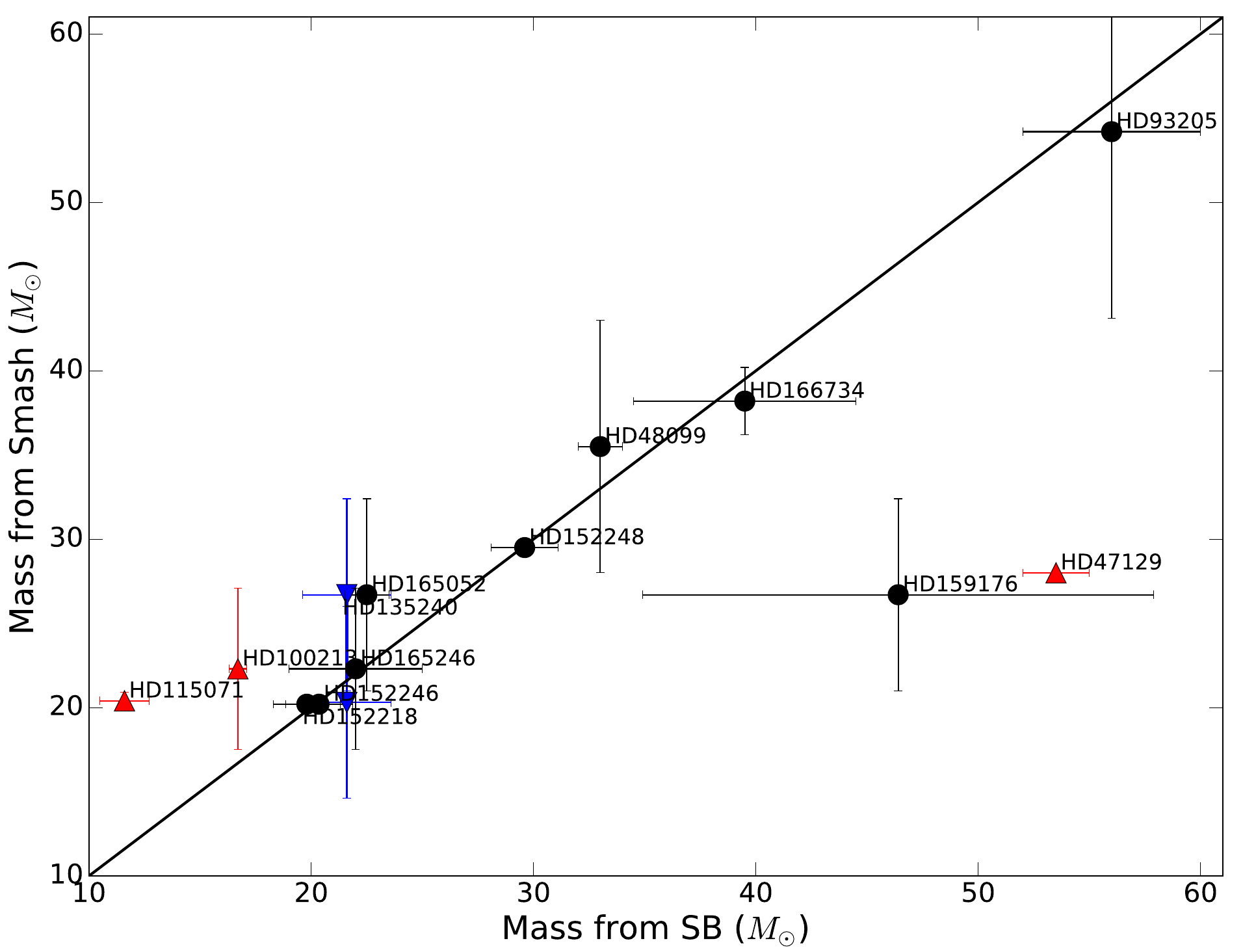}}
        \caption{Dynamical masses versus the masses derived in this work. Systems that are in a (post-)interaction state (see text) are indicated by red triangles. HD~135240 has an ambiguous luminosity class (III-V), the blue triangles indicate masses derived for LC III and V (20.3 and 26.7~\msun, respectively). The dynamical mass of HD~159176 relies on the analysis of ellipsoidal variation and is quite uncertain.}
        \label{fig:masscheck}
\end{figure}

The last system (HD148937, O6f?p) is a magnetic system where the magnetic star is believed to be a merger product \citep{frost2024}. In addition, it has been suggested that some blue supergiants could also be the result of binary interaction, as, for example, suggested by the strong decrease in nearby companions for O9.7 supergiants in the VFTS survey \citep{sana2013}. 

Most of these outliers seem to have physical reasons to deviate from the calibrations. As an interesting corollary, the spectro-photometric calibration that we provide could be used in conjunction with independent distance estimates such as those from \gaia\ to single out candidate objects that have a different evolutionary history for further investigations.

In summary, both the cluster distances and the recent individual {\it ALSIII-Gaia} distances suggest that the spectro-photometric distances that we derived are slightly overestimated. For a mean distance of 1915~pc of the stars in the full sample, the systematic error corresponds to less than 10\%\ (clusters) and by about 5\%\ ({\it ALSIII-Gaia}). 
This suggests that the Martins calibrations are actually slightly too bright, which is an interesting finding per se. The systematics could be calibrated out by correcting the spectro-photometric distance estimate by the slope of the relation
\begin{equation}
    d_\mathrm{H} = -122.3\pm64.7 + (1.19\pm0.04)d_{Gaia}.
\label{eq:gaia}
\end{equation}
However, the difference is small enough that it does not impact the empirical distributions of logarithmic (projected) separations ($\log Sep$) as the systematics only amount to $\Delta \log Sep\approx+0.02$. As a consequence we proceed without applying the correction and use the spectro-photometric distances $d_\mathrm{H}$ to convert the angular separations $\rho$ to projected physical separation as
\begin{equation}
    \left(\frac{Sep}{\mathrm{AU}}\right)=\left(\frac{\rho}{\mathrm{mas}}\right) \left(\frac{d_H}{\mathrm{kpc}}\right).\label{eq:sep}
\end{equation}
The resulting projected separations are given in Table~\ref{tab:distance}.

\begin{table}
\centering
\caption{Known absolute primary masses of SB2s. Known (post-)interaction systems are marked with an asterisk.}\label{tab:SBmass}
\begin{tabular}{l c c c}
\hline\hline \\[-8pt]
ID                      &       $M_{\textrm{lit}}$      &       $M_{\textrm{smash}}$            & Reference\\
                        &       ($M_{\odot}$)           &       $(M_{\odot})$   \\
\hline \\[-8pt]
HD~47129$^*$            &       $52-55$                 &       28.0                                            & \cite{linder2008} \\
HD~48099                &       $33.0\pm1.0$            &       35.5                                            & \cite{mahy2010} \\
HD~93205                &       $52-60$                 &       54.2                                            & \cite{morrell2001} \\
HD~100213$^*$   &       $16.7\pm0.4$            &       22.3                                            & \cite{penny2008} \\
HD~115071$^*$   &       $11.6\pm1.1$            &       20.4                                            & \cite{penny2002} \\
HD~135240       &       $21.6\pm2.0$            &       20.3 -- 26.7                            & \cite{penny2001} \\
HD~152218       &       $19.8\pm1.5$            &       20.2                                            & \cite{rauw2016} \\
HD~152246       &       $20.35\pm1.5$           &       20.2                                            & \cite{nasseri2014} \\
HD~152248       &       $29.6\pm1.5$            &       29.5                                            & \cite{sana2001} \\
HD~165052       &       $22.5\pm1.0$            &       26.7                                            & \cite{ferrero2013} \\
HD~159176$^a$   &       $46.4^{+14.3}_{-9.5}$ & 26.7                                            & \cite{penny2016} \\
HD~165246       &       $19-25$                 &       22.3                                            & \cite{johnston2017} \\
HD~166734       &       $39.5^{+5.4}_{-4.4}$            &       38.2                                    & \cite{mahy2017}\\
\hline
\end{tabular}\\
\flushleft
{\sc notes:} a. The dynamical mass of HD~159176 relies on the analysis of ellipsoidal variations and is quite uncertain.
\end{table}

\subsection{Mass determination}\label{sec:mass}
In this section we provide an overview of the method to determine the masses of the primaries and of their companions. The details of the derivation, and the resulting calibrations, are given in Appendices~\ref{sec:bolcor} and \ref{sec:masslum}.

\subsubsection{Primary masses}\label{sec:massprim}

For the primary stars, we used the available spectral type and luminosity class to estimate the mass from the \citet{martins2005} and \cite{martins2006} calibrations. First,  H-band bolometric corrections were used to derive calibrations as a function of spectral type and luminosity class. Similar to the absolute magnitude calibration, we first derived H-band bolometric corrections for luminosity classes IV and II which are not in \citet[][see Fig.~\ref{fig:BC_LC}]{martins2006}. Then we derived relations for the bolometric correction for each of the luminosity classes as a function of spectral type (Fig.~\ref{fig:BC_SpT}). These relations were then used to convert the absolute H-band magnitude to the bolometric magnitude, and hence the absolute luminosities.

The masses of the primary stars can now be derived from the mass-luminosity relation. However, the exponent of the mass-luminosity relation changes both with luminosity, and with luminosity class (see Appendix~\ref{sec:masslum}).  For luminosity classes III and I, we used the \cite{martins2005} calibration to derive the exponent of the mass-luminosity relation as a function of luminosity. For luminosity class V, we used the \cite{brott2011} models evaluated at $\log{g} = 3.92$, which is the value of the surface gravity $g$ for all dwarf spectral subtypes in \cite{martins2005}. This allowed us to use the dwarf calibration at low luminosities, which could then also be applied to the companion stars. The latter are indeed expected to be predominantly dwarfs (see Sect.~\ref{sec:masscomp}). For luminosity classes II and IV, we again interpolated between LC I--III and III--V, respectively. As above, we adopted half a luminosity class and half a spectral subtype to estimate uncertainties on the obtained masses. The resulting masses are given in Table~\ref{tab:distance}.

To assess the accuracy of the above approach, we compared the derived primary masses to literature values of SB systems with dynamical mass measurements 
(i.e.,\ for systems with known inclination; see Table~\ref{tab:SBmass}). This comparison is shown in Fig.~\ref{fig:masscheck}. In general, our masses are in good agreement with the literature values. However, three stars (HD~115071, HD~100213, and HD~47129) deviate by more than 2$\sigma$. Interestingly, each of these stars is known to be either currently interacting or in a post-interaction state. HD~~115071 is a semi-detached, over-luminous system, where mass-transfer has previously taken place \citep{penny2002}. HD~~100213 is a contact binary currently experiencing Roche-Lobe overflow \citep{penny2008}. Lastly, HD~~47129 (Plaskett's star) is in a post-interaction state, and is a nitrogen-rich, evolved star \citep{linder2008} with a partially stripped donor and a magnetic accretor. That the relation that we derived from main-sequence calibrations does not work well for these interacting binaries is unsurprising, and we conclude that our calibration works well for pre-interaction systems, which is likely the case for the majority of the systems in our sample.

\begin{figure}[t!]
        \resizebox{\hsize}{!}{\includegraphics{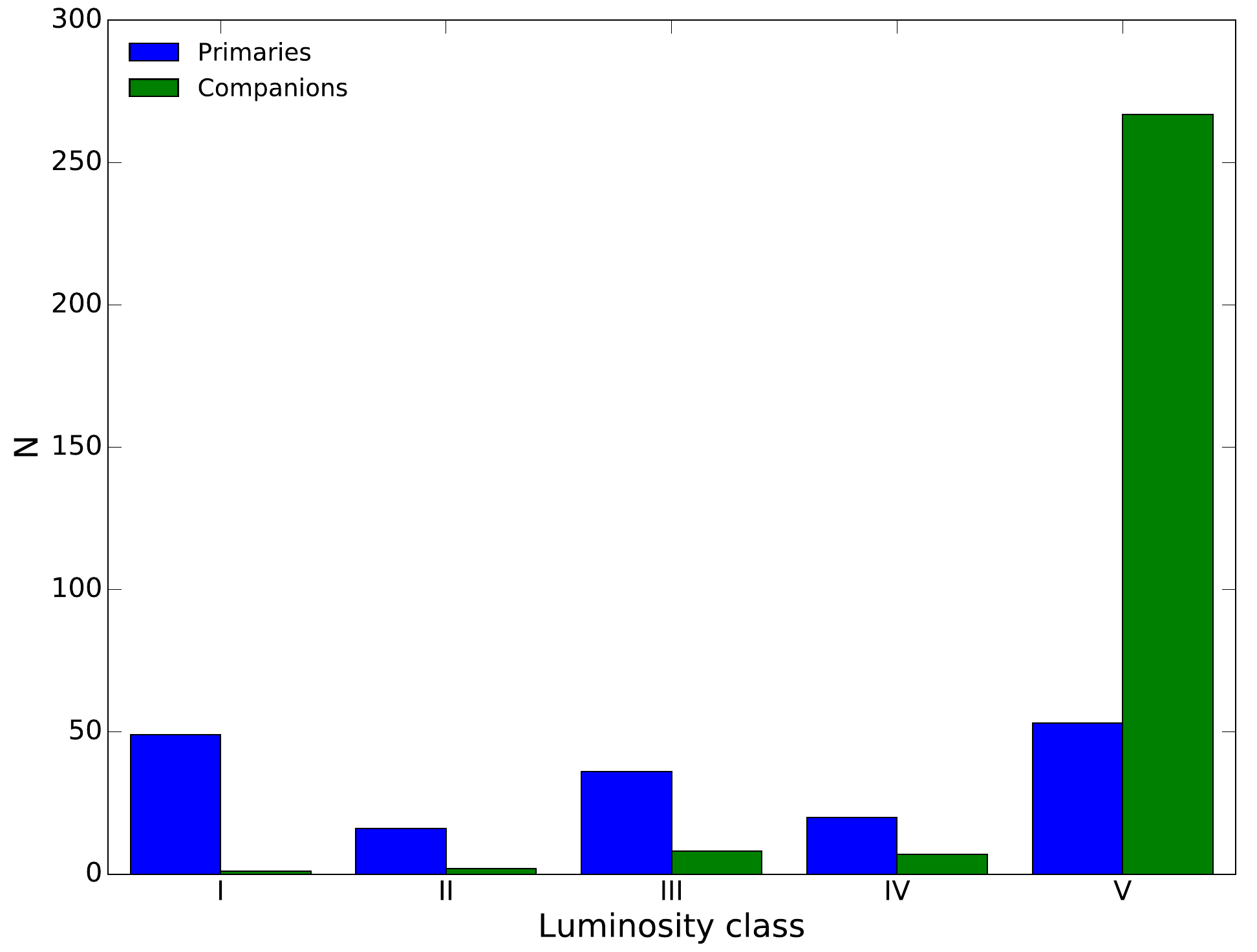}}
        \caption{Histogram of the luminosity classes of the primaries and companions.}
        \label{fig:compLC}
\end{figure}

\subsubsection{Companion masses}\label{sec:masscomp}

As the luminosity classes of the companion stars are unknown, the above method cannot be directly applied to derive their masses. However, those companions that have a large magnitude difference with the primary should be considerably less massive, and therefore still be dwarfs regardless of the luminosity class of the primary. This is the case for the vast majority of our sample, and masses for these stars can directly be obtained from the calibration based on the \cite{brott2011} models derived above. 

However, for the brighter companions, the assumption of a dwarf luminosity may not be correct, in particular for systems with more evolved primaries. For example, for a system with an O9~I primary and a fainter companion with $\Delta m = 1.0$ mag, the assumption of a dwarf luminosity class for the companion would result in a mass ratio greater than unity. If this were true, the more massive companion should have evolved faster than the primary, and thus should have a more evolved luminosity class than the primary. Therefore, the assumption of a dwarf luminosity class does not hold. In these cases, we adopted the least evolved (i.e.,\ the most massive for the given magnitude) luminosity class that gives a mass ratio smaller than unity, and adopted the resulting mass as an upper limit.

Figure~\ref{fig:compLC} shows the distribution of adopted luminosity classes of the companions compared to the known luminosity classes of the primaries. The majority of companions have luminosity class V. Note that the distribution of primary luminosity classes is biased toward more evolved (visually brighter) stages as a result of the magnitude-limited approach of the survey \citepalias[see][]{sana2014}, and hence is not expected to follow a canonical  mass function.

\section{SMASH+ detection limits}\label{sec:bias}
The instrumental detection curves in the observational plane ($\Delta H$ versus $\rho$; e.g., Fig.~\ref{fig:smash}) are well constrained and quite homogeneous throughout the survey \citepalias[][]{sana2014}. However, each object has been observed only once, so that some binaries may have eluded detection due to unfavorable geometry or because they are (temporarily) within the inner working angle or beyond the outer angle limits of the  instruments. In this section, we first investigate the relations between the  instantaneous projected separation ($Sep$), the instantaneous true separation ($r$) and the semi-major axis ($a$) of the relative orbit.  We then investigate the detection probability as a function of the orbital properties, accounting for the observational strategy of the survey.

\begin{figure}
    \centering
    \includegraphics[width=\linewidth]{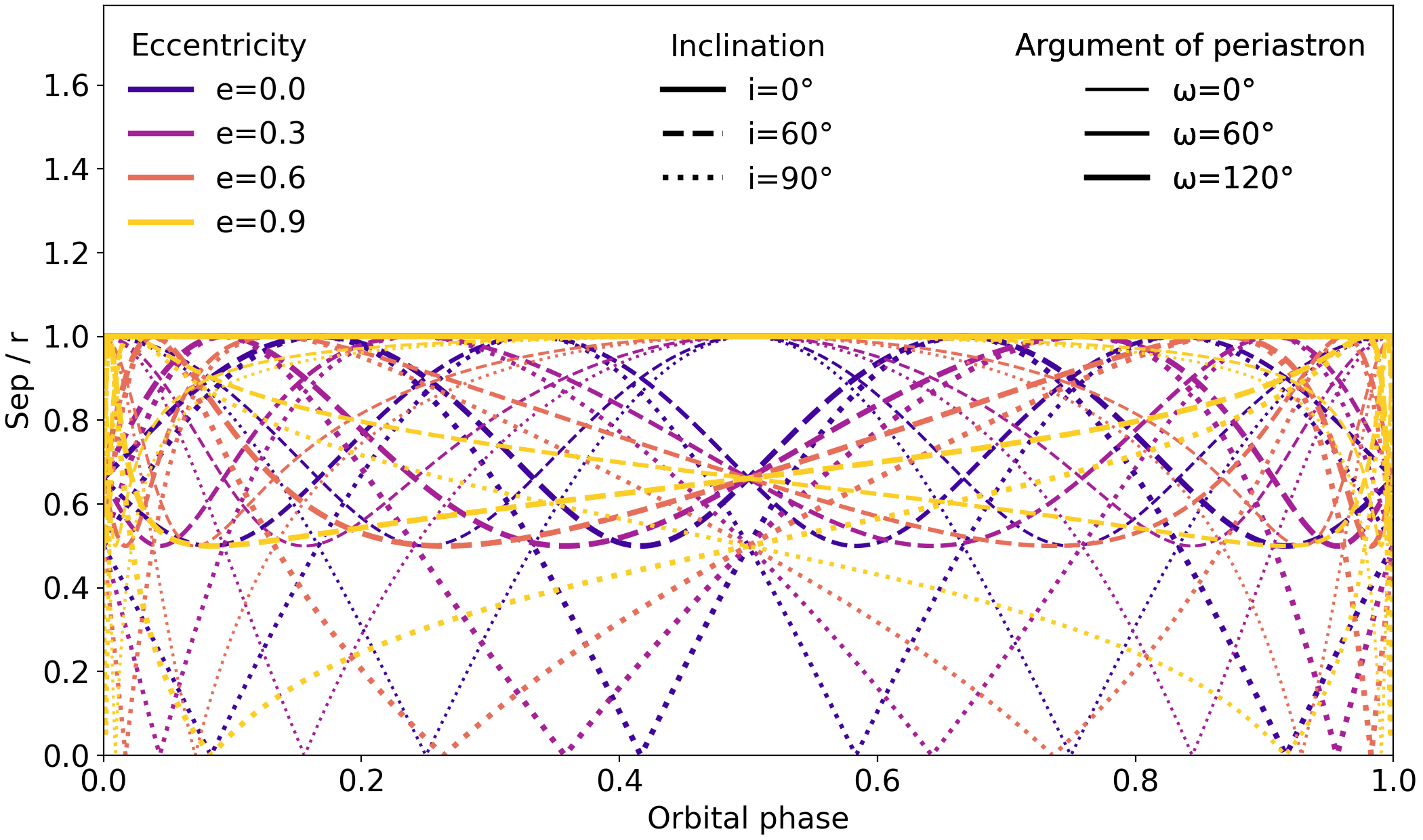}
    \includegraphics[width=\linewidth]{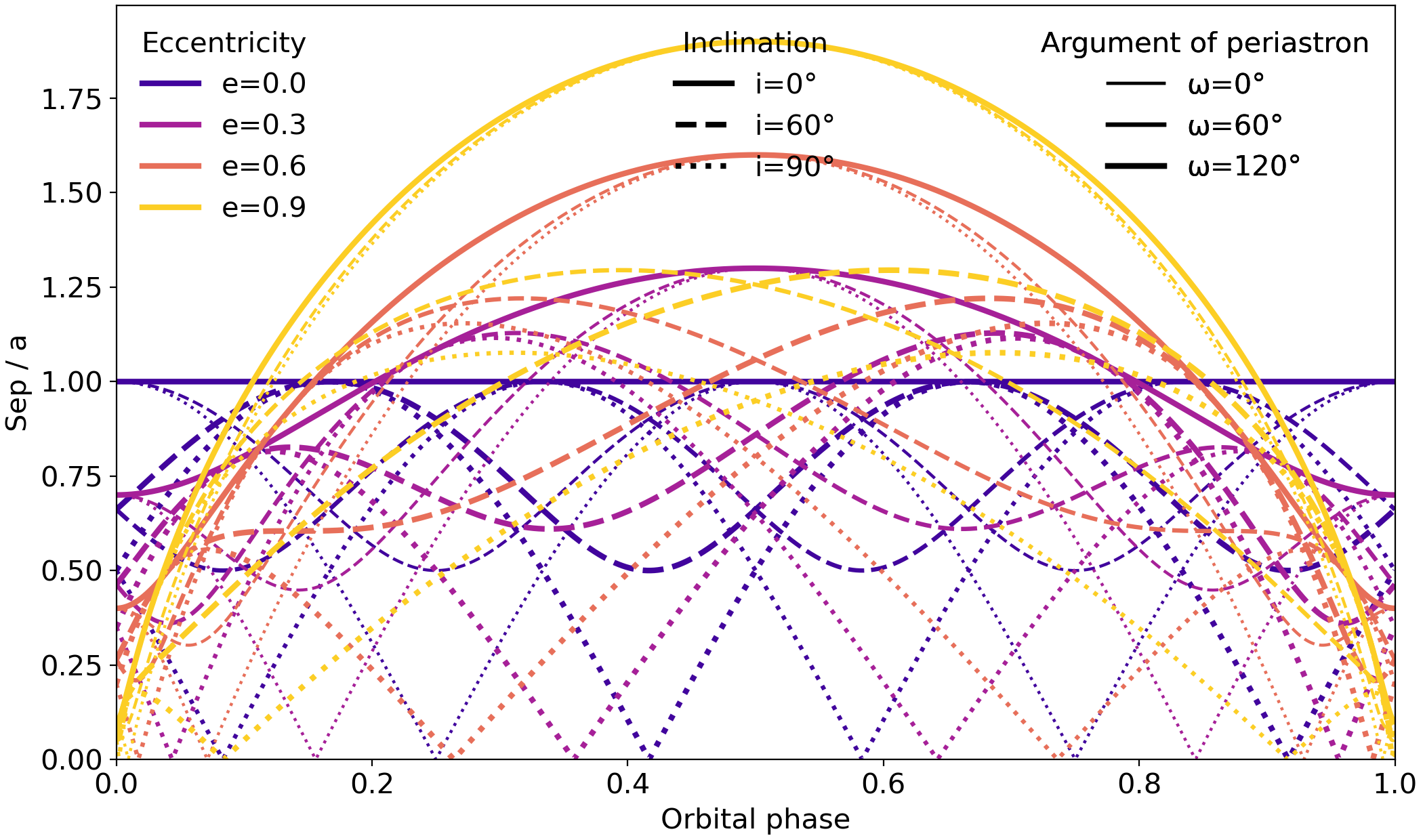}
    \caption{Ratio of the projected instantaneous separation $Sep$  to  the  instantaneous separation $r$ (top panel) and the semi-major axis $a$ (bottom panel) as a function of the orbital phase for various orbital geometries (see legend).}
    \label{fig:sepa}
\end{figure}

\begin{figure}
    \centering
    \includegraphics[width=0.96\linewidth]{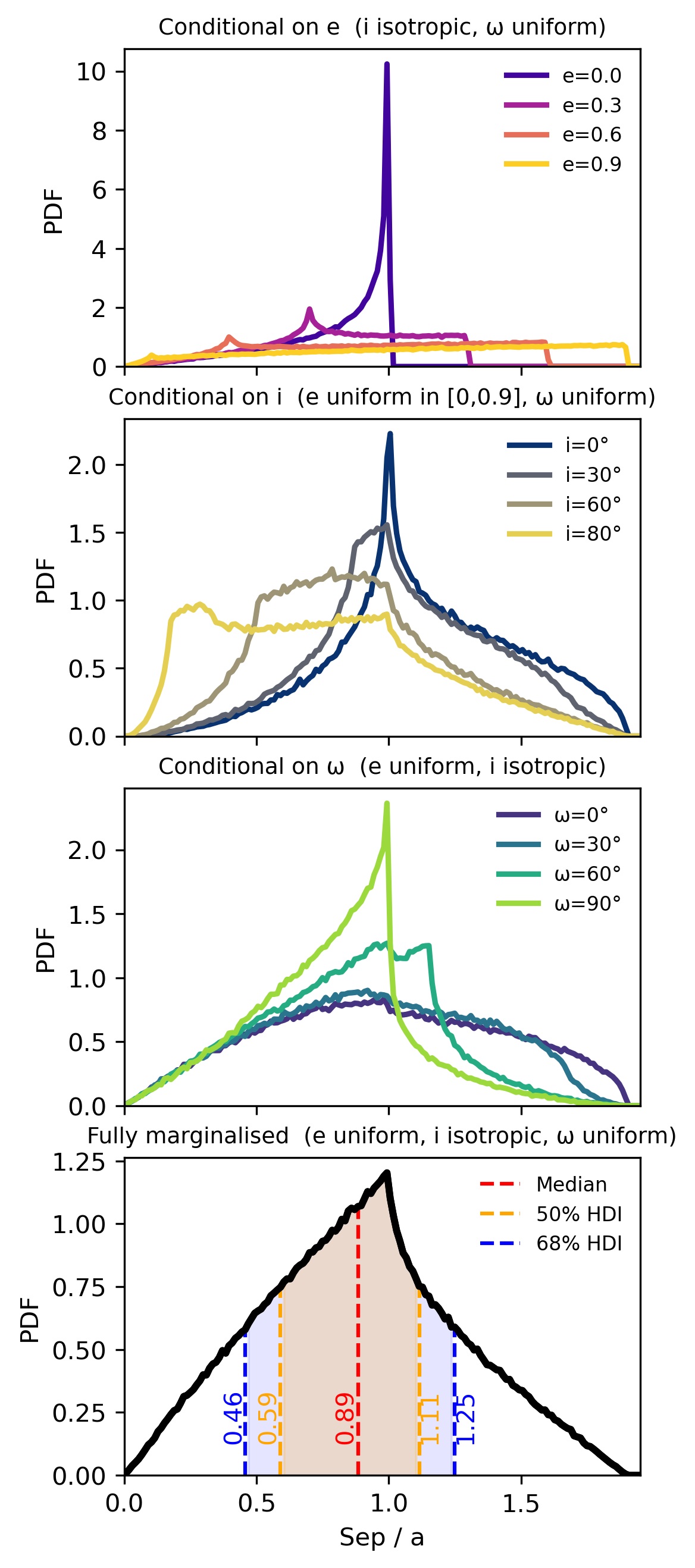}
    \caption{Probability density functions (PDFs) of $Sep/a$ for various orbital configurations. The bottom panel gives the PDF marginalized over the 3D orientation and the uniform eccentricity distribution between $e=0.0$ and $0.9$. The bottom panel also provides the boundaries of the 50 and 68\%\ high density intervals (HDIs).}
    \label{fig:sepa2}
\end{figure}

\subsection{Projection effects}
The \smash\ observations are snapshot measurements that provide the instantaneous  angular separations $\rho(t)$. The latter are easily related to the projected physical separation $Sep(t)$ through Eq.~\ref{eq:sep}. The relation between the instantaneous separation and the semi-major axis of the orbit is well known:
\begin{equation}
    \frac{r(t)}{a}=\frac{(1-e^2)}{1+e\cos\nu(t)},\label{eq:r_o_a}
\end{equation}
where $e$ is the eccentricity and $\nu$, the true anomaly. The ratio between the projected separation and the true separation is given by
\begin{equation}
    \frac{Sep}{r}(t)=\left(1-\sin^2 i \sin^2 (\omega+\nu(t)) \right)^{1/2},\label{eq:sep_o_r}
\end{equation}
where $i$ is the orbital inclination, and $\omega$ the argument of periastron passage. The relation to the semi-major axis of the relative orbit $a$ is \citep[e.g.,][]{SanaVrancken2026}
\begin{equation}
    \frac{Sep(t)}{a}=\frac{1-e^2}{1+e \cos \nu(t)} \left(1-\sin^2 i \sin^2\left(\omega+\nu(t)\right) \right)^{1/2}. \label{eq:sep_o_a}
\end{equation}
This results in a rather complex dependence of the measured projected separation on the orbital properties $e, i$, and $\omega$. Figure~\ref{fig:sepa} illustrates these various effects and shows the nontrivial mapping of a single snapshot measurement to the physical properties of the orbits. 

Fortunately, the sample is large enough that we can hope to marginalize over the random 3D orientation of the orbits, defined by the set of Kepler's angles $(i,\omega,\Omega)$, and over the distribution of eccentricities. The argument of the ascending node $\Omega$ is critical for the orientation of the projected orbit on the plane of the sky but has no impact on the distribution of projected separations so that we ignore it.  Figure~\ref{fig:sepa2} provides the probability density functions (PDFs) of $Sep/a$ marginalized over the various elements defining the orbital configurations ($i, \omega, e$).

\begin{figure*}
        \resizebox{.33\textwidth}{!}{\includegraphics{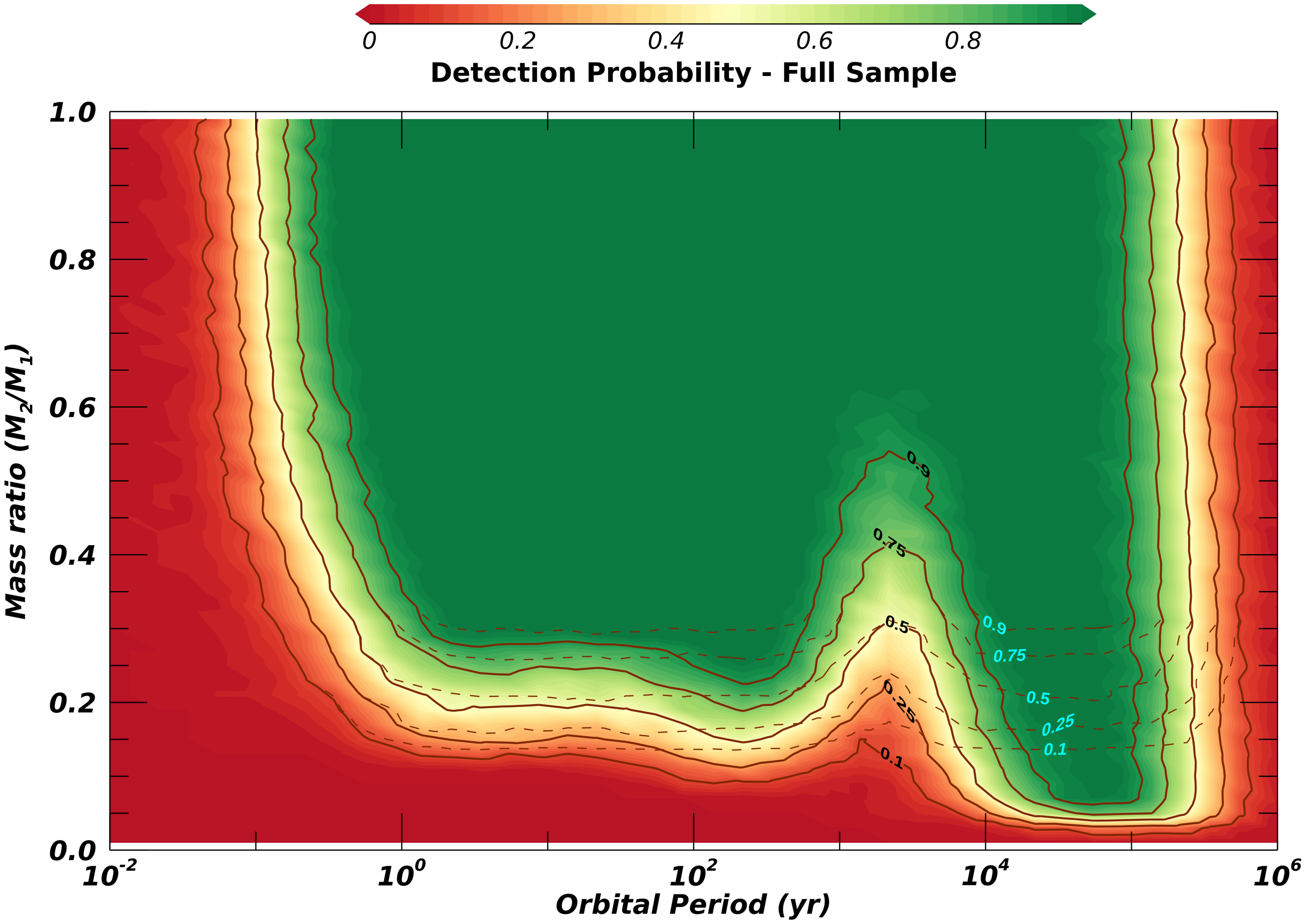}}
        \resizebox{.33\textwidth}{!}{\includegraphics{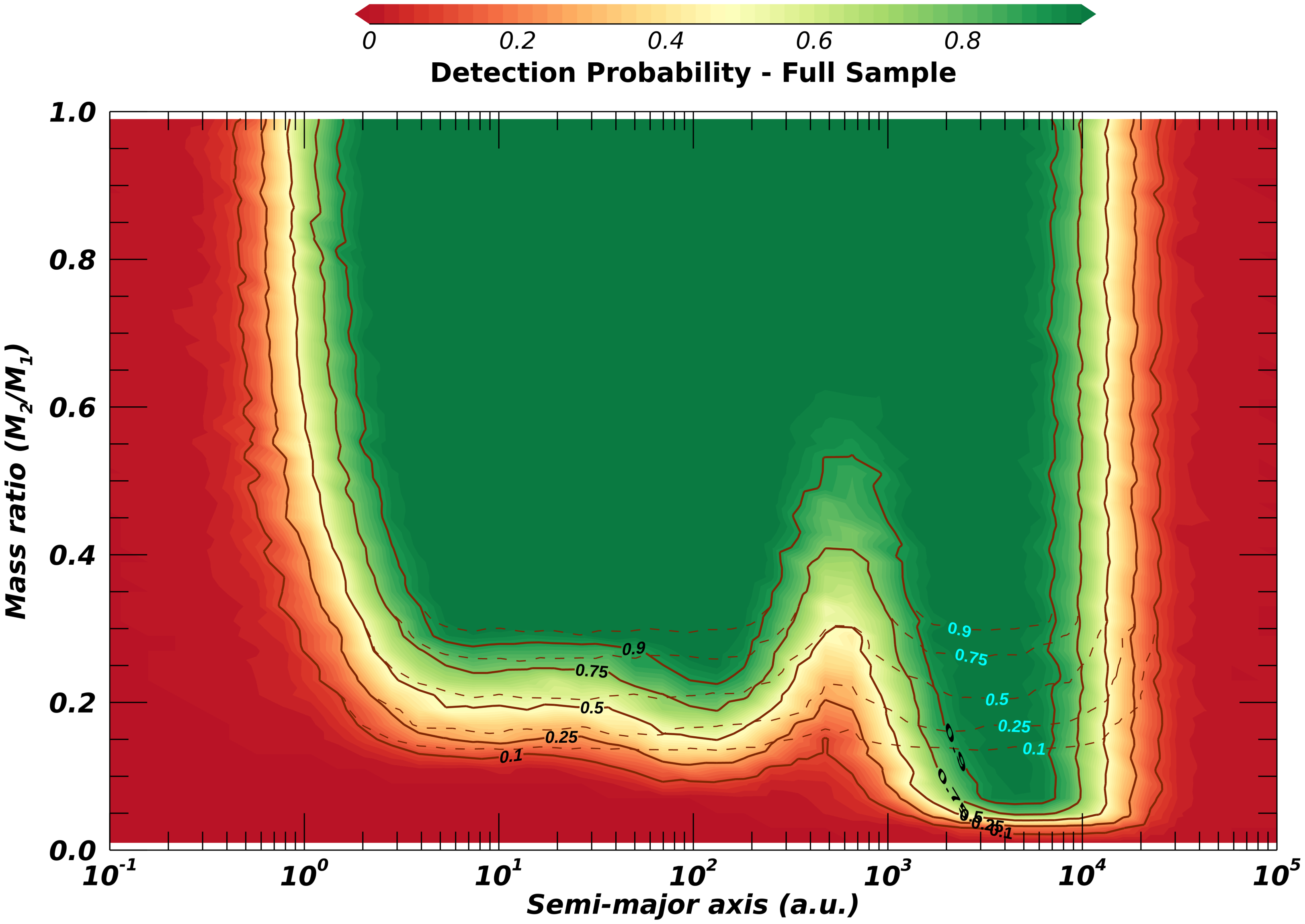}}
        \resizebox{.33\textwidth}{!}{\includegraphics{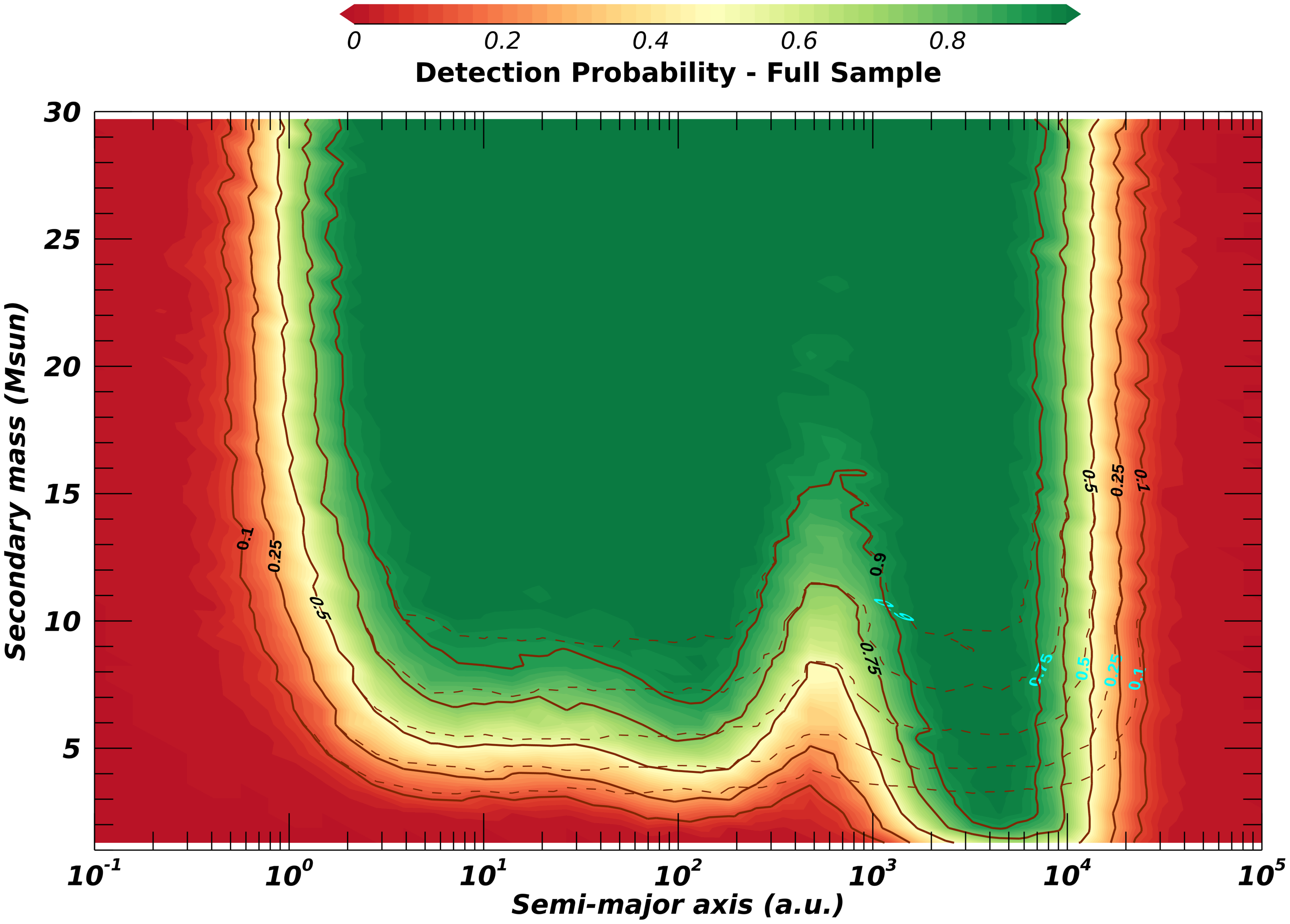}}
        \resizebox{.33\textwidth}{!}{\includegraphics{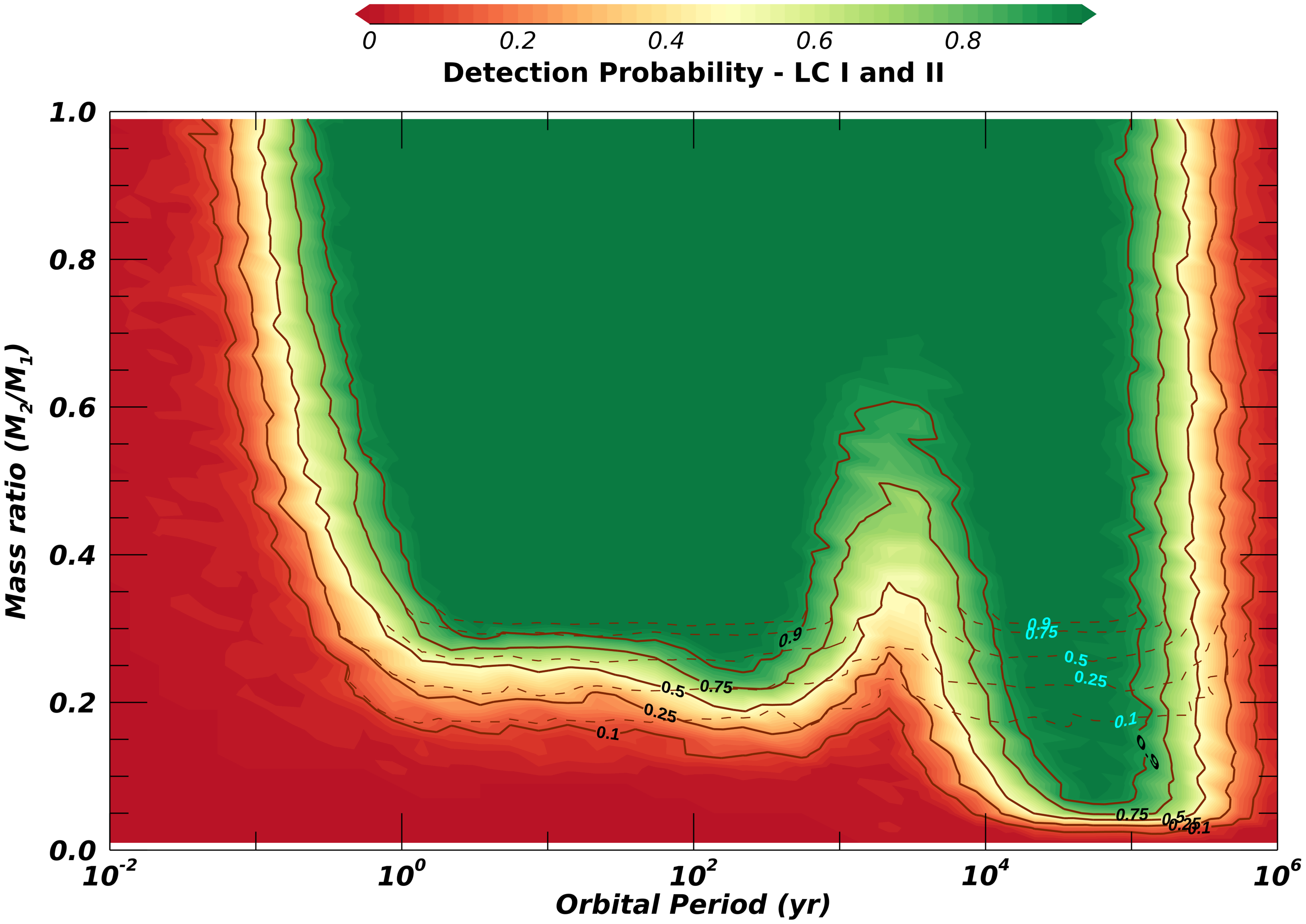}}
        \resizebox{.33\textwidth}{!}{\includegraphics{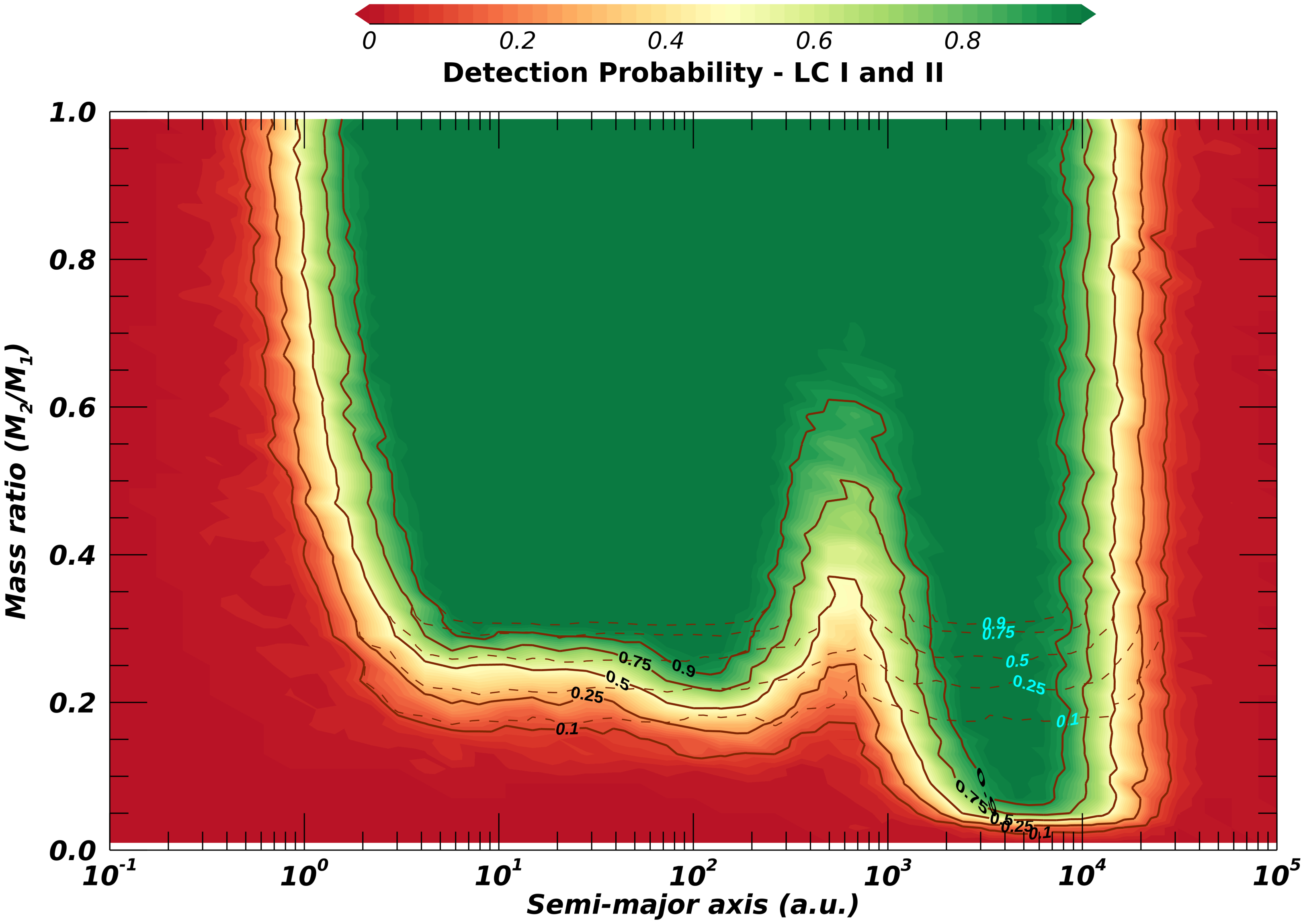}}
        \resizebox{.33\textwidth}{!}{\includegraphics{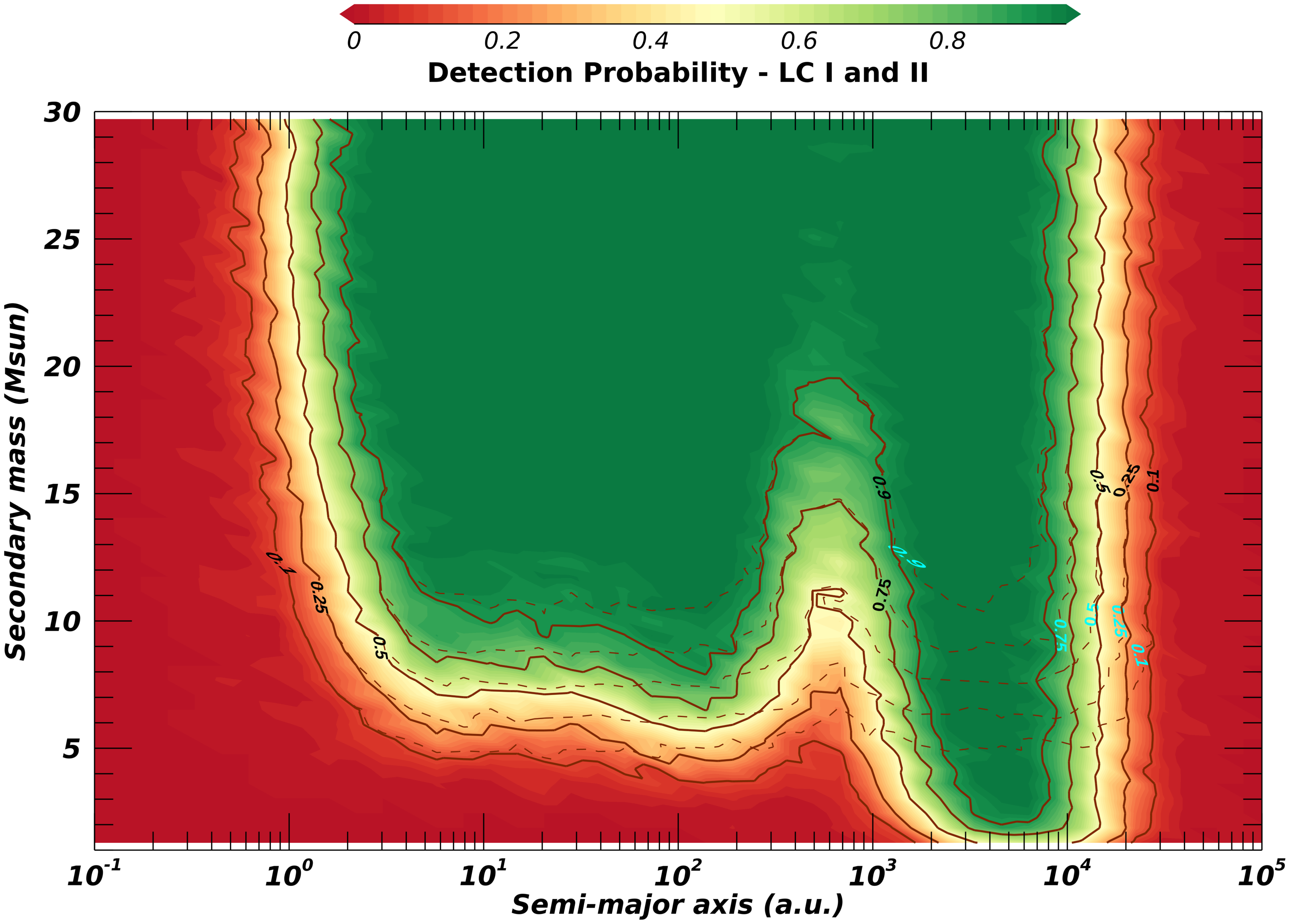}}
        \resizebox{.33\textwidth}{!}{\includegraphics{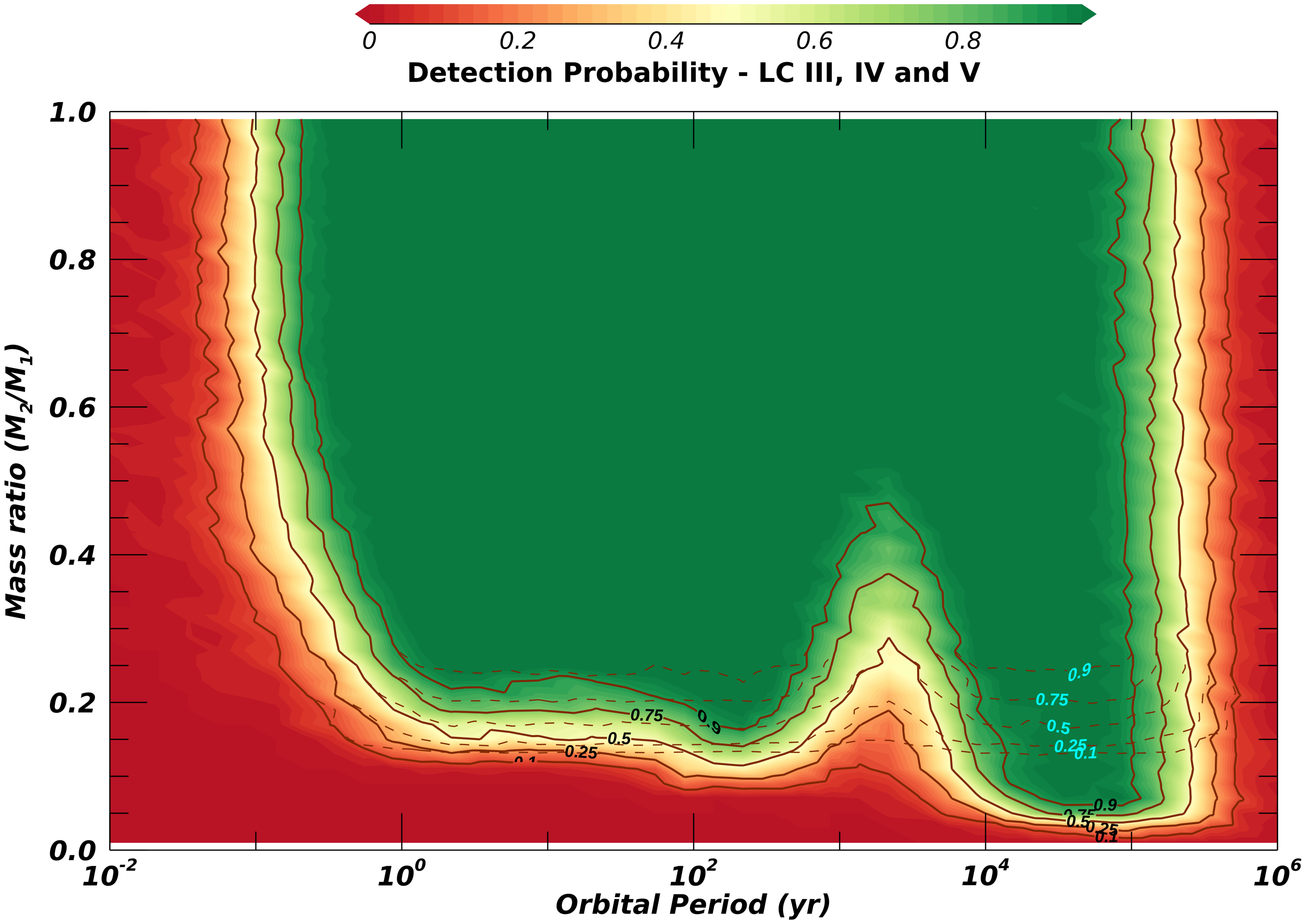}}
        \resizebox{.33\textwidth}{!}{\includegraphics{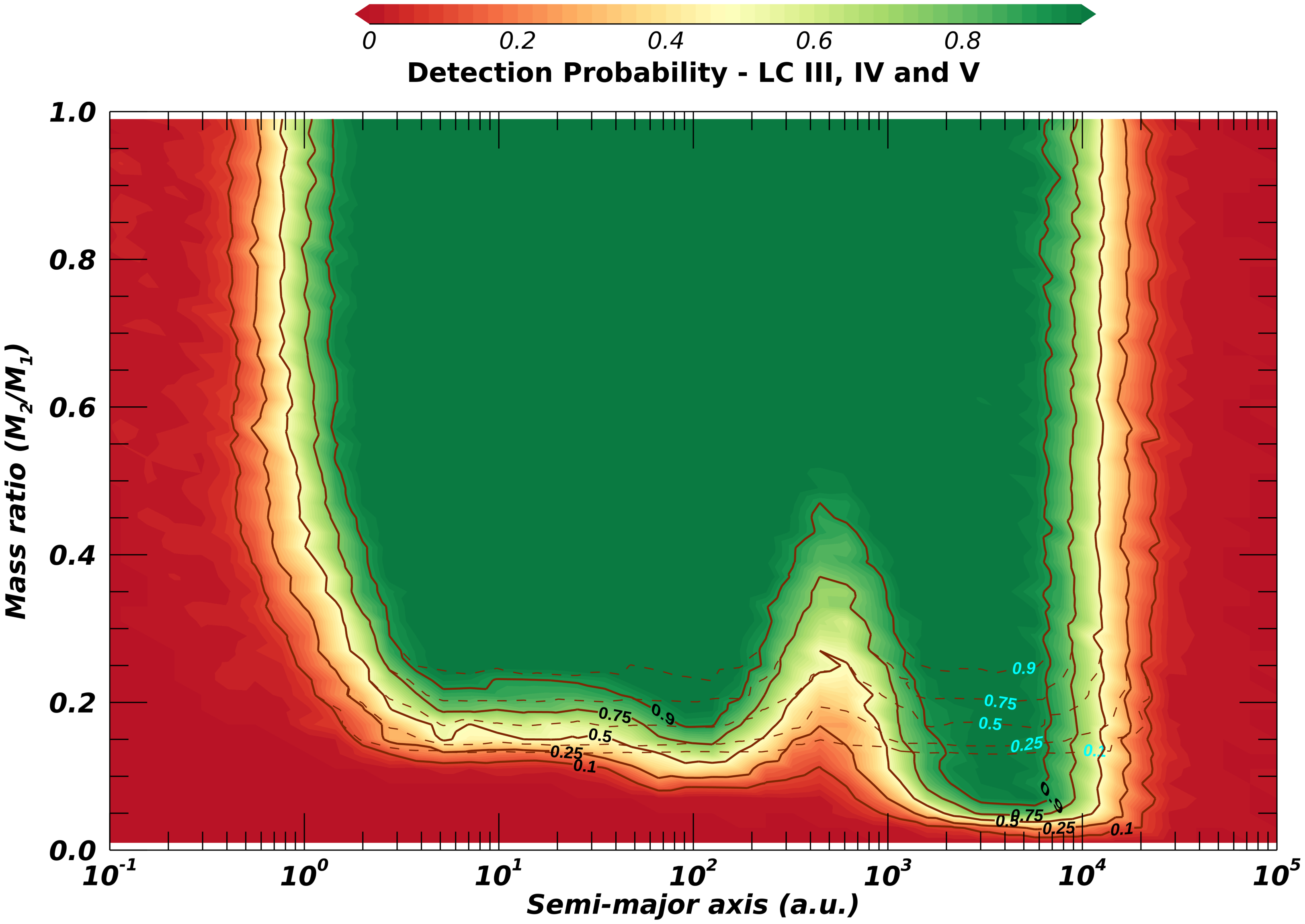}}
        \resizebox{.33\textwidth}{!}{\includegraphics{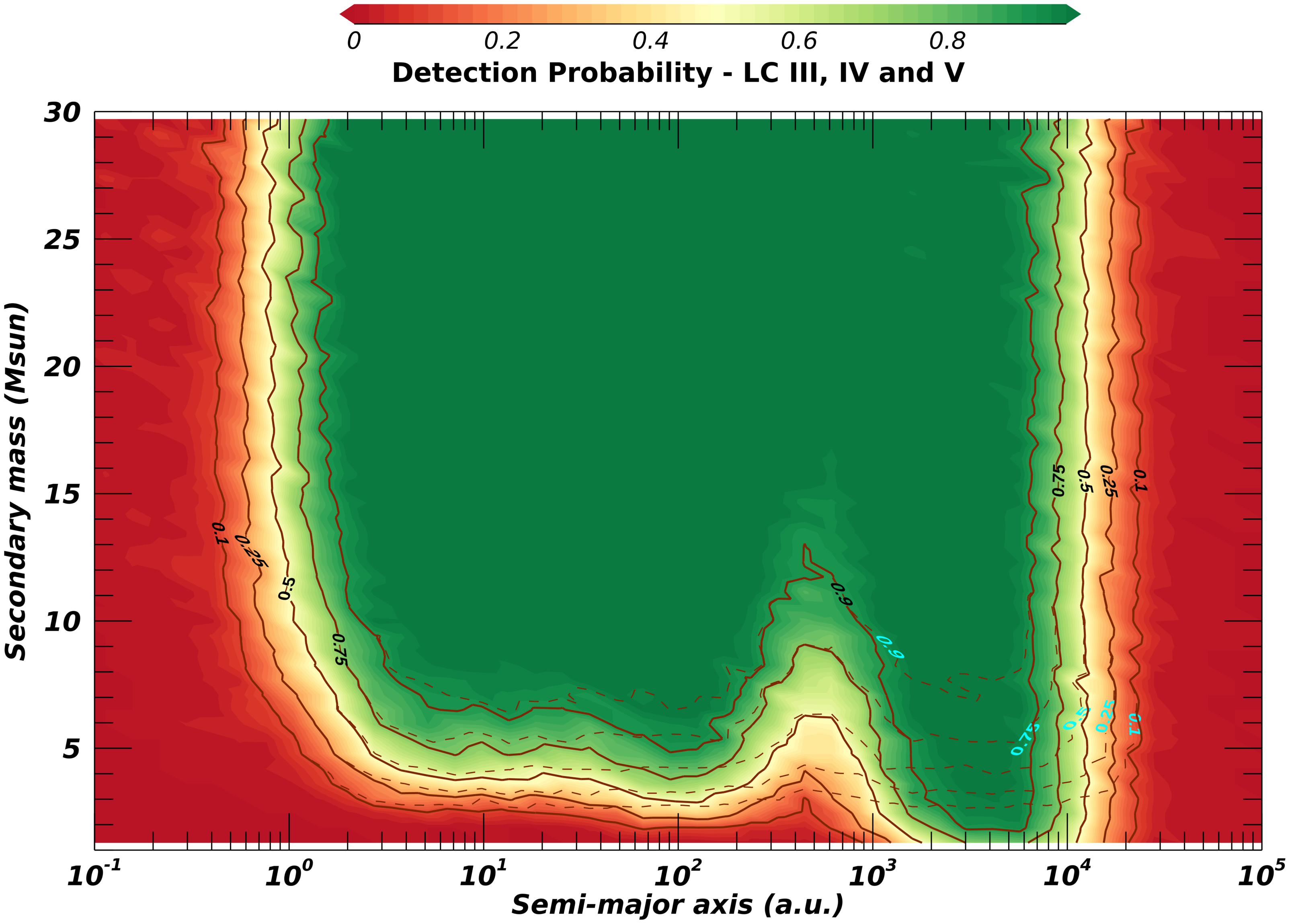}}
        \caption{Binary detection probabilities of the \smash\ survey projected on the mass ratio versus the orbital period (left), versus the semi-major axis (middle), and the companion mass versus the semi-major axis (right) planes. The colored background and solid equi-probability curves are based on the sample of stars that have been observed both by \pionier\ and by \naco. The dashed equi-probability curves (and cyan labels) show the detectability when restricting the sample to $\Delta H \leq4$ (i.e., the "cleaned" sample). From top to bottom: the full sample, luminosity classes I and II, and luminosity classes III to V.  }
        \label{fig:bias}
\end{figure*}
It illustrates the critical impact of eccentricities, which widen the distribution, while large, more probable inclinations make the distribution asymmetric. Assuming a random orientation of the orbit in 3D space ($f_{\cos i} \propto \mathcal{U}(-1,1)$; $f_{\omega} \propto \mathcal{U}(0,\pi)$), and a uniform distribution of the eccentricities  ($f_{e} \propto \mathcal{U}(0,0.9)$), the bottom panel provides the overall PDF marginalized over all orbital configurations and summarizes the overall likelihood that the measured $Sep$ is representative of the size of the orbit given by $a$. This PDF is the one that we can use to assess the overall impact of the geometry. Fortunately that distribution has a mode at $Sep/a=1$, and a median $Sep/a \approx 0.9$ . Thus, on average, the measured projected separation $Sep$ is a reasonable estimator of the semi-major axis $a$ of the relative orbit, albeit with a broad dispersion (50\%-HDI=[0.6:1.1]). As discussed in Sect.~\ref{sec:results:separations}, we
aim to constrain the distribution of semi-major axis in $\log a$.  As for the discussion of distance estimates in Sect.~\ref{sec:distances}, and given that  $\log a \propto \log Sep$ (Eq.~\ref{eq:sep_o_a}), the small systematic error introduced by using $Sep$ as a biased estimator of $a$ remains small compared to the signal. Incidentally, it even almost counterbalanced the systematics on the distance quantified by Eq.~\ref{eq:gaia} and we proceed without further corrections.

\subsection{Detection maps}

We used the Monte Carlo detection probability approach of \citet{sana2013}, adapted to relative astrometry methods as in \citet{frost2025}. Unlike \citet{sana2013} and \citet{frost2025} however, we did not draw the masses of the primary stars from a given initial mass function but we adopted the distance, primary mass, and $H$-band absolute magnitude of the stars in the \smash\ survey as given in Table~\ref{tab:distance}. Indeed the \smash\ primary masses do not follow a canonical mass function because of the overrepresentation of supergiants in the sample (see Sect.~\ref{sec:mass}).
We simulated 10\,000 mock \smash\ samples, adopting a uniform eccentricity distribution between $e=0.0$ and 0.9, and 
  random orbital orientations in 3D space. At each draw, the central stars were paired with a companion with a mass-ratio $q=M_2/M_1$ drawn uniformly between 0.01 and 1, and we computed the magnitude contrast and projected separation of a random observational epoch. These mock observables were then compared to the survey sensitivity curves presented in  Fig.~\ref{fig:smash} to decide whether the system would be detected as a binary or not. Throughout this process, we assumed that companions are all dwarfs, which is valid for the majority of the sample.

Figure~\ref{fig:bias} shows the resulting detection probability maps.  The sharp boundaries at orbital periods of about 1.5 months to 10$^5$~yr (equiv.\ semi-major axes of $\sim$1 to 10$^4$~AU) result from the  inner and outer working angles of the survey (approx.\ 0.001 and 8\arcsec).
Within these boundaries, the overall detection probability nears 100\%. Similarly, the detection probability is  uniformly excellent above the horizontal cut-off defined by the contrast limits of $\Delta H \approx 4$, 5, and 8 in the interferometric, coronagraphic, and adaptive-optics imaging regimes, respectively. These limits translate to mass-ratio limits of $q_\mathrm{lim}\approx 0.2$ and 0.005, respectively, with  $q_\mathrm{lim}$ values slightly dependent on the luminosity class (Fig.~\ref{fig:BC_H}).

The contrast limits of the instruments that we used depend on magnitude difference and not the absolute brightness of the companion. As a consequence, the limiting companion mass that we are able to detect depends on the brightness of the central star. For this reason, we also separated the \smash\ sample according to the luminosity class into roughly two samples of similar size. For each subsample, we recomputed the detection maps. The last column of Fig.~\ref{fig:bias} reveals that interferometric observations can detect a 4~\msun\ companion near an LC~V-III star but only a 7~\msun\ companion near a LC~II-I star. In the AO regime, \smash\ is sensitive to solar-mass companions near LC~I-II stars and to subsolar mass companions near LC~III-V stars. 

Overall, our simulation results show that, within the sensitivity limits of the survey, the binary detection probabilities are very uniform. The one exception is the structure around  $P\approx10^{3.5}$~yr ($a\approx 500$~AU) that corresponds to the transition from \naco/\sam\ to \naco\ AO-imaging (Fig.~\ref{fig:smash}). Given these results, the observed distributions obtained in the next sections can be considered to be directly representative of the true distributions, and do not require bias-corrections, as long as one remains in the high-detection-probability regions outlined in Fig.~\ref{fig:bias}.

\section{Results and discussion}\label{sec:results}

By applying the methods presented in the previous section, the projected physical separations of all companions, and the masses of the primaries and of all their companions were determined. In this section we discuss these results.

\subsection{Companion separations}\label{sec:results:separations}

\begin{figure}
        \resizebox{\hsize}{!}{\includegraphics{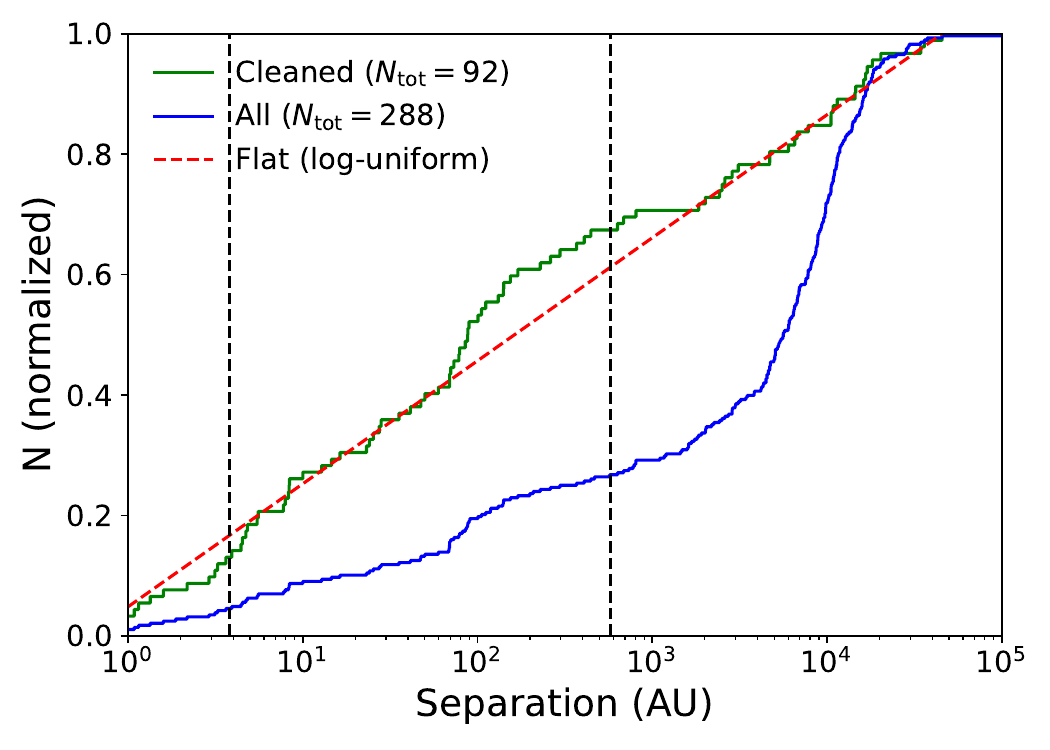}}
        \caption{Cumulative distribution of the projected separations of all the detected companions within 8\arcsec\, (blue), and of the detected companions with $\Delta H \leq 4$ mag, of systems that were observed by both \pionier\ and \sam\ (green). Dashed lines indicate the drop in detector efficiency below $\sim$2 mas and at $\sim$300 mas at the average sample distance of 1915 pc.}
        \label{fig:sepdistcorr}
\end{figure}

  The distribution of  projected separations of all found companions within 8\arcsec\, in the sample  (Fig.~\ref{fig:sepdistcorr}) shows two separation ranges where the number of detected companions increases more sharply, at $Sep\approx 100$~AU and $\approx5000$~AU. The first increase coincides with the switch between \pionier\ and \naco/\sam\ at the average distance of the sample, and can hence be explained by the higher number of stars that have been observed with the latter instrument. The second increase occurs at the switch between the \naco/\sam\ and \naco/FOV modes and can be explained by the much deeper detection contrast  of the latter mode (see Figs.~\ref{fig:smash} and \ref{fig:bias}). Many of the large number of faint objects at large separations are likely to be chance line-of-sight objects instead of true companions (see \citetalias{sana2014}, Fig. 8).

To provide distributions that are less affected by selection biases, we defined a { "cleaned"} sample, which contains the 105 systems (297 companions, 288 within 8\arcsec) that have been observed by both instruments and for which we considered only detected companions brighter than a contrast limit of 4 magnitudes, resulting in a sample of 92 companions. Aside from two small separation ranges (below 2~mas ($\sim$6 AU) and around 300 mas ($\sim$650 AU)), the detection probability is indeed uniform within the adopted boundaries (Fig.~\ref{fig:bias}). Furthermore, all companions in this sample are almost certainly physically bound instead of chance line-of-sight objects (see \citetalias{sana2014}, Sect. 4.1).

Overall, the distribution of projected separations in the cleaned sample seems to follow an O\"epik law\footnote{An O\"epik law corresponds to a uniform distribution in $\log Sep$ \citep{oepik}, or alternatively in $\log P_\mathrm{orb}$.}. There might be a small but significant  deviation at $\sim$80 to 100~AU where we note a sharper increase in the number of detected companions. Either this is  a remaining artifact of the transition between the \pionier\ and \sam\ instruments, or it may be intrinsic. A larger sample would be desirable to confirm the presence of such a bump in the cumulative distribution function.

\begin{figure*}
        \resizebox{\hsize}{!}{\includegraphics{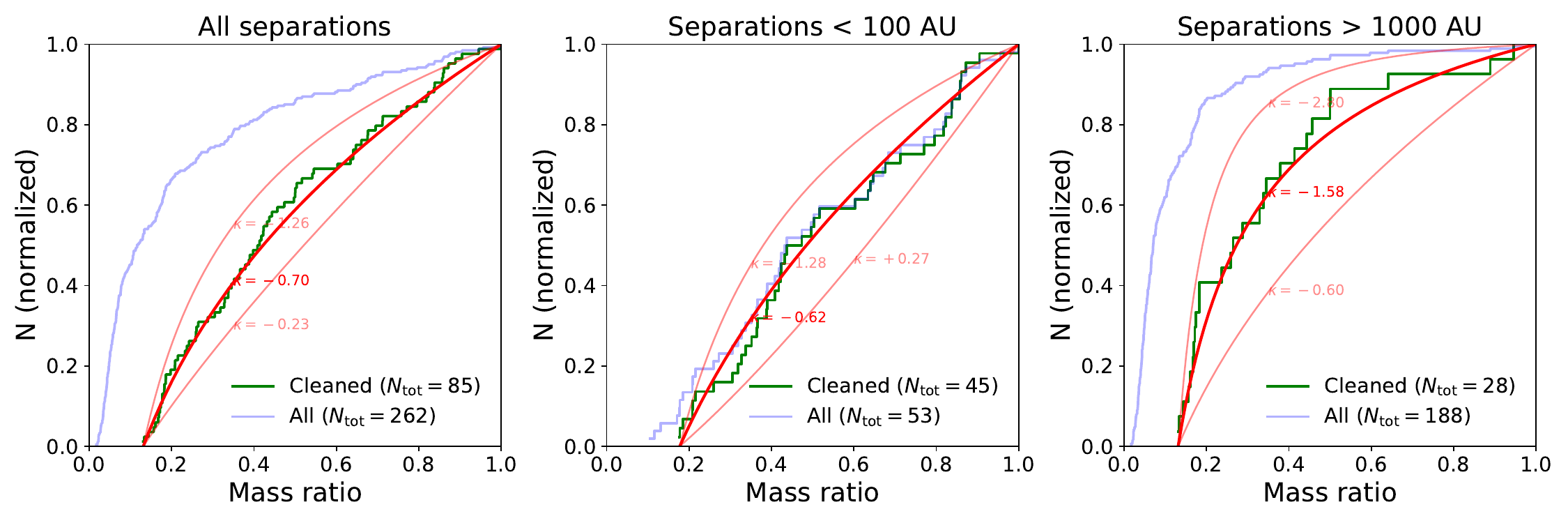}}
        \caption{Left: Cumulative distribution of the mass ratios of all the detected companions within 8\arcsec\ (blue), and of the detected companions with $\Delta H \leq 4$ mag, of systems that were observed by both \pionier\ and \sam\ (green). The red line shows a fit to the green distributions with its 95\% confidence intervals. Middle: Same as on the left, but for separations $<$ 100 AU. Right: Same as on the left, but for separations $>$ 1000 AU.}
        \label{fig:massdistcorr}
\end{figure*}

\subsection{Masses}\label{sec:results:masses}

 Figure~\ref{fig:massdistcorr} presents the cumulative distributions of the mass ratios, excluding 26 (7) companions in the full (cleaned) sample for which only upper limits on the mass ratios are derived.
 The overall distribution of mass ratios is skewed toward low-mass ratios, with almost 70\%\ of the detected companions having $q<0.2$ (Fig.~\ref{fig:massdistcorr}, left). This abundance of low-mass ratio companions results from the large sensitivity of the NACO-AO observations at larger separations (Fig.~\ref{fig:massdistcorr}, right). These companions disappear entirely when limiting ourselves to the inner 100~AU separation range as such low-mass-ratio companions are below the \smash\ detection limit.

 Limiting ourselves to the cleaned sample, we used the Kuiper probability $p_\mathrm{K}$ as a goodness-of-fit metric to fit a power-law distribution $f_q \propto q^\kappa$ to the mass-ratio distribution. Outer envelopes are taken as the $\kappa$ values for which $p_\mathrm{K}=0.1$. We obtained $\kappa \approx -0.7^{+0.5}_{-0.6}$, $\kappa_{<100} \approx -0.6^{+0.9}_{-0.7}$, and $\kappa_{>1000}=-1.6^{+1.0}_{-1.2}$ for the cleaned sample over the full separation, for $Sep<100$~AU and $Sep>1000$~AU, respectively. 
 This quantifies that companions at larger separations have smaller mass ratios than if randomly drawn from a uniform distribution \citep[see also discussion in][]{moe2017}. One has to be careful to directly compare the best-fit exponent $\kappa$ derived for $Sep<100$ AU to the other two values as the lower boundary of the fitted range varies and it is known that the best-fit exponent value depends on the adopted range \citep{almeida2017}. However, the differences here are large enough that we can conclude that the distribution is significantly more skewed at large separations compared to lower separations. To give an order of magnitude, the distributions in the inner ($Sep<100$~AU) and outer ($Sep>1000$~AU) ranges have  medians at $q\approx 0.5$ and 0.3, respectively. 
 
 Compared to mass-ratio distributions found for the tighter spectroscopic binaries ($P\lesssim3000$~d, $a \lesssim 10$~AU) in the Milky Way \citep[][ $\kappa=-0.1\pm0.6$]{sana2012} and Large Magellanic Cloud \citep[$\kappa=+0.2\pm0.2$][]{shenar2022}, the index of the power-law distribution seems to decrease toward larger separations, a point already made by \citet{moe2017}. However, the $\kappa$ value that we derived for the inner range $\kappa_{<100}=-0.6^{+0.9}_{-0.7}$ is much flatter than the value of $\kappa_{3-90}=-1.4\pm0.4$  derived by \citet{moe2017} using a subsample of 21 \smash\ targets. The reasons for such differences are unclear and may lie in different sample selection functions and survey sensitivity estimates combined with  small sample statistics. One can argue that both values are within 2$\sigma$ from one another, but the $\kappa_{<100}$ value that we derived would then also be compatible with the flat-mass ratio distribution of the tighter spectroscopic binaries. Determining the most appropriate mass-ratio distribution for intermediate-period binaries would be desirable as it impacts the outcome of case-B mass transfer, with potentially important differences for predictions of population synthesis computations.

\begin{figure}
        \resizebox{\hsize}{!}{\includegraphics{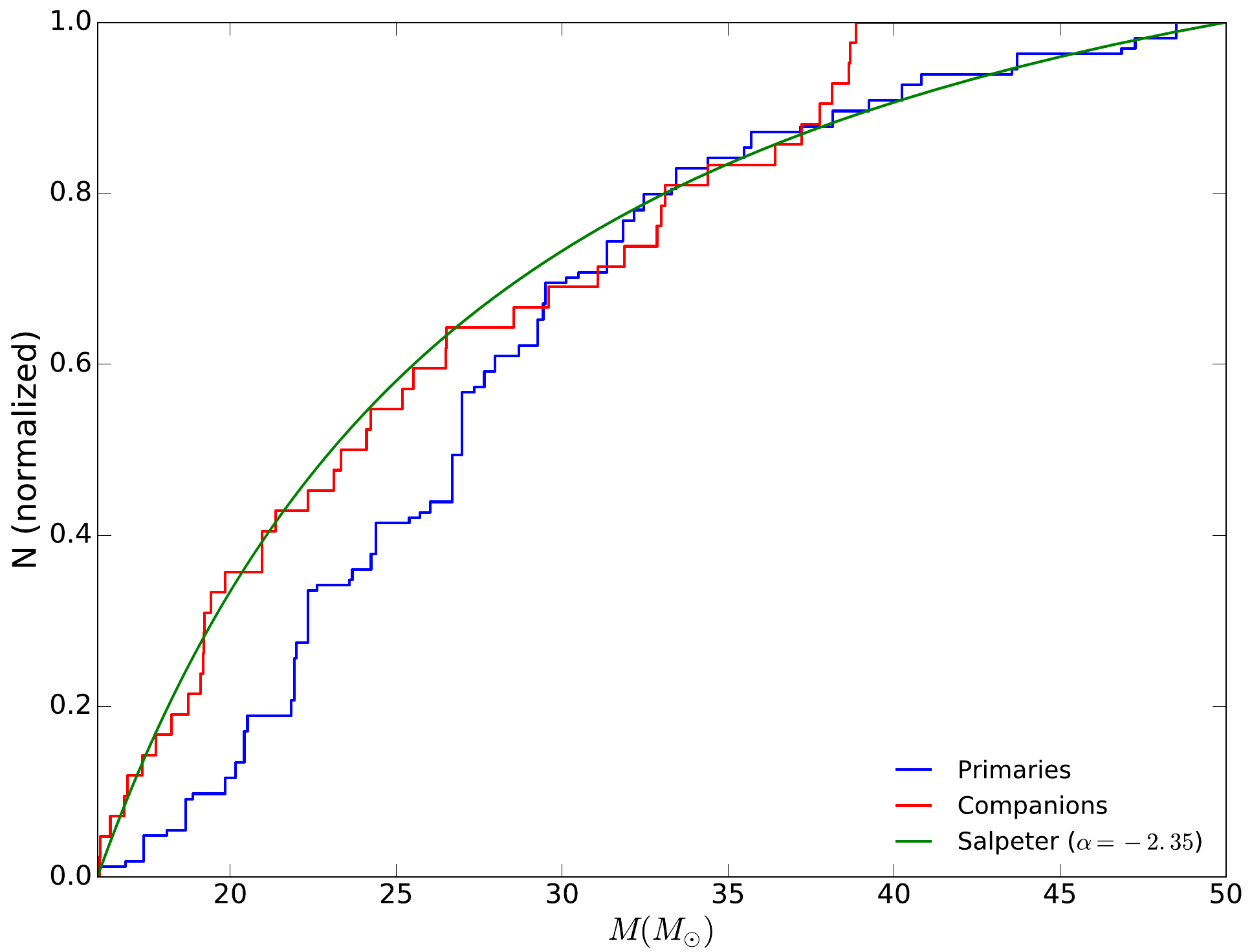}}
        \caption{Cumulative distributions of the masses of all primaries (blue) and companions (red) with masses $16 M_{\odot} \leq M \leq 50 M_{\odot}$. Also shown is the Salpeter mass function in this mass range (green).}
        \label{fig:massIMF}
\end{figure}

Finally, we compare the distribution of observed masses of the primaries and the observed companions in Fig.~\ref{fig:massIMF}. Also shown is a Salpeter initial mass function (IMF). As the primaries are all O-type stars, we only do this for masses greater than $16\, M_{\odot}$. This reveals a lack of primary stars with low masses ($M \leq 30 \,M_{\odot}$) compared to the observed companions. This is a direct result of the magnitude-limited approach of the survey, causing an overabundance of (more massive) supergiants compared to dwarfs as these can be seen up to larger distances. Interestingly, the companions seem to follow the Salpeter IMF across most of the observed mass range, in agreement with the findings of \citet{GRAVITY2018} from a sample of 22 OB stars in the Orion Nebula.

\begin{figure}
        \resizebox{\hsize}{!}{\includegraphics{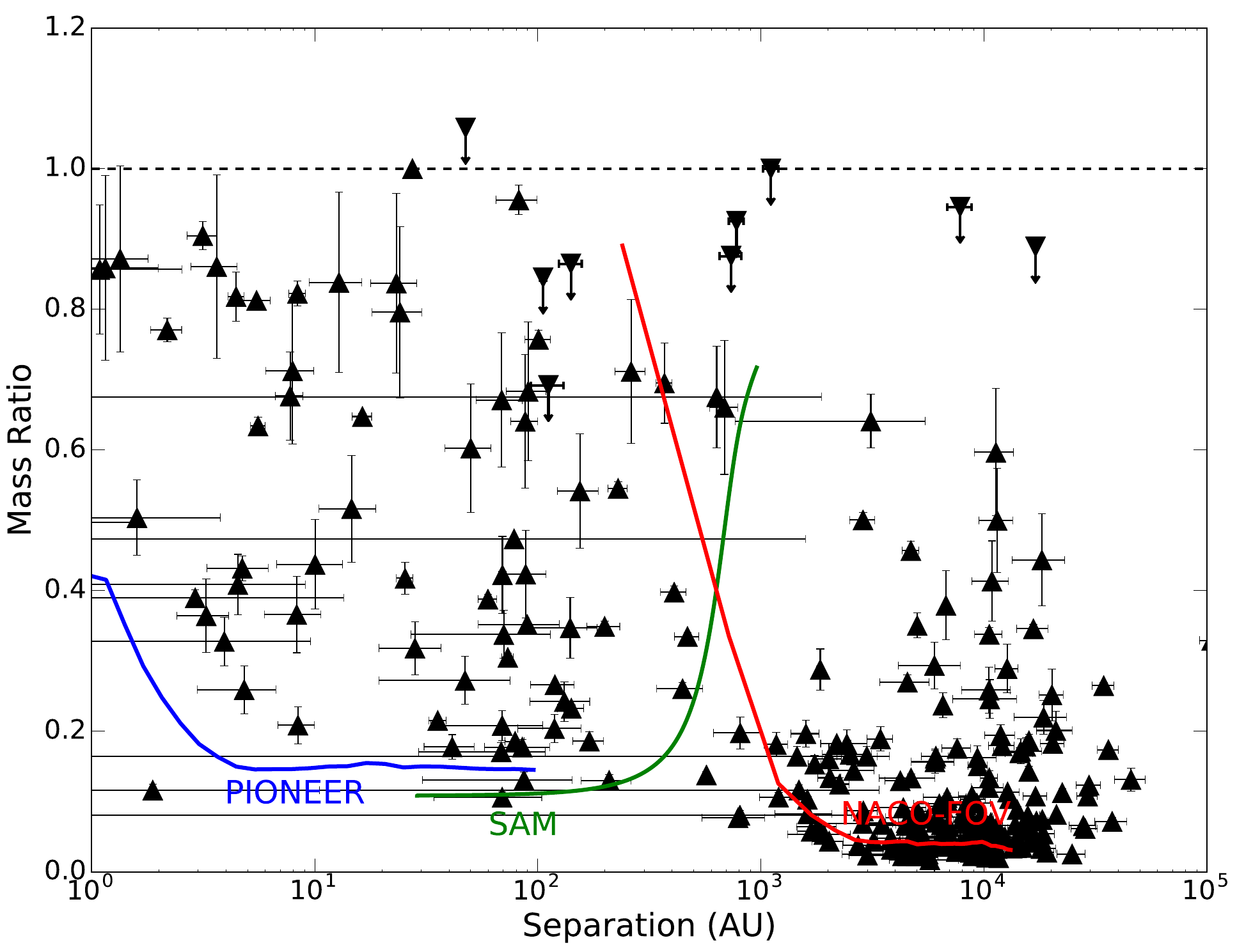}}
        \caption{Projected separation versus the mass ratio for all the detected companions within 8\arcsec. Also indicated are the median detector sensitivities in the H band for an average sample distance of 1915 pc and assuming a dwarf primary.}
        \label{fig:sepmass}
\end{figure}

\begin{figure}
        \resizebox{\hsize}{!}{\includegraphics{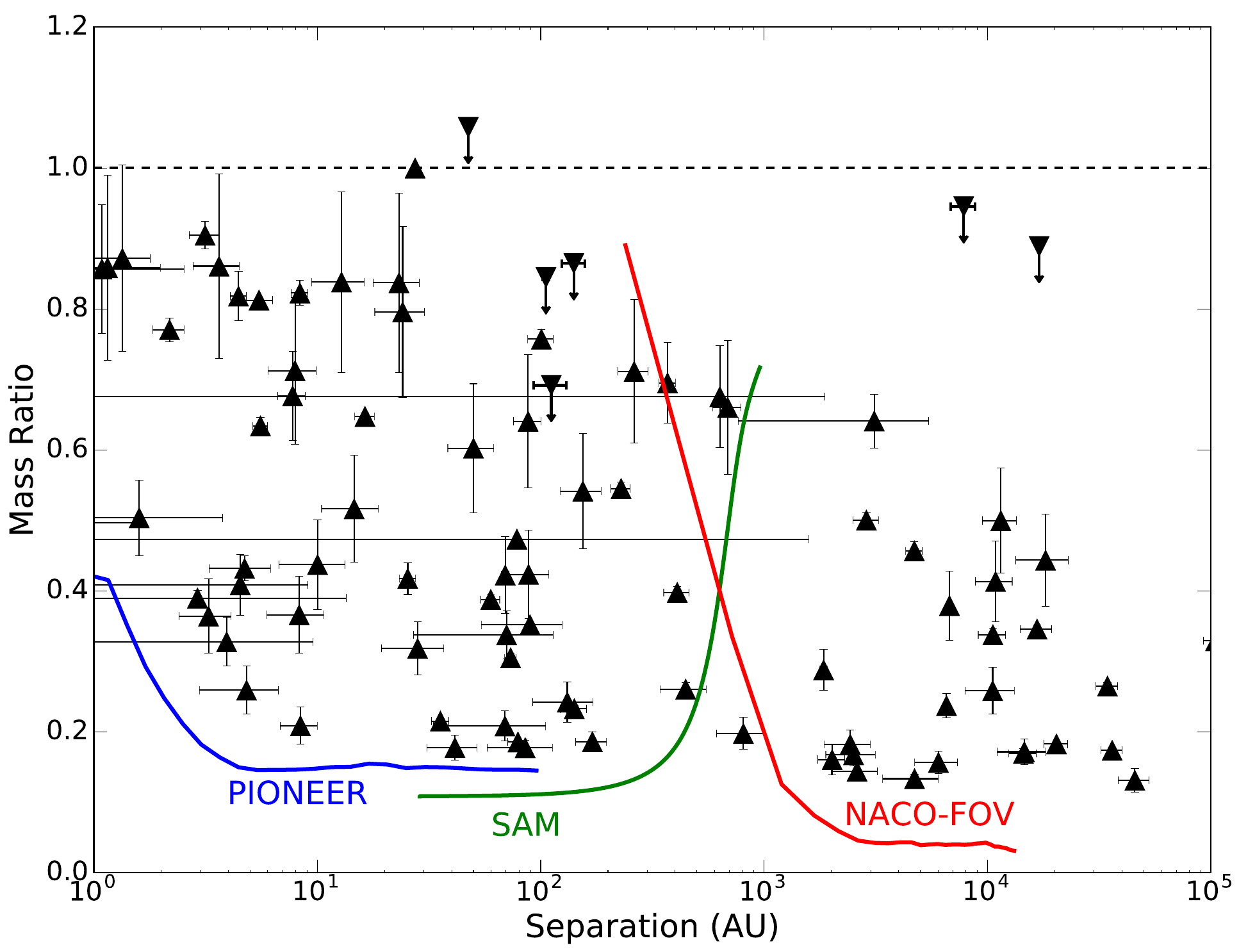}}
        \caption{Same as Fig.~\ref{fig:sepmass} but for detected companions with $\Delta H \leq 4$ mag, for systems that were observed by both \pionier\ and \naco.}
        \label{fig:sepmasscorr}
\end{figure}

\subsection{Separations versus mass ratios}

Combining the results of Sects.~\ref{sec:results:separations} and \ref{sec:results:masses} we can now show the results of Fig.~\ref{fig:smash} in physical units, i.e.,\ projected separation versus mass ratio. This is done in Figs.~\ref{fig:sepmass} and \ref{fig:sepmasscorr}, where  the latter figure focuses on the cleaned sample limited to $\Delta H\leq4$. While we note the absence of twin binaries ($q>0.95$), the most striking result is probably the lack of high-mass-ratio companions at projected separations greater than 100 AU, i.e.,\ in the top right area of the figure. At these separations, no companions with mass ratios greater than 0.8 are detected except for those with upper limits on their mass. This mass-ratio limit decreases toward larger separations, with no companions with mass ratios greater than $~\sim0.7$ detected above 1000~AU, and no companions with mass ratios greater than $\sim0.5$ detected above $10^4$ AU. 
We tentatively derived an upper envelope delimiting the avoidance zone in the upper-right corner from the populated area below the curve:
\begin{equation}
    q_\mathrm{crit}\approx 1-0.2 \left(\log_{10}\frac{Sep}{AU}-2\right)\hspace*{0.5cm}\mathrm{for}\ Sep > 100\ AU. 
\end{equation}
It is beyond the scope of this paper to interpret these trends. Ultimately, the physical reasons for these trends are rooted in a complex set of factors. It is typically assumed, however, that components of binaries wider than a few hundred AU form nearly independently of one another \citep[e.g.,][]{WhiteGhez2001, moe2017, ElBadryRix2019a, ElBadryRix2019b}. Components that assembled at closer separations probably form in a highly correlated way.  A decline in the maximum mass ratio $q_{\rm crit}$ for massive stars is also reported by \citet[][Fig.~4]{moe2017}, though the onset of the decline occurs at a much shorter separation (at about an orbital period of $10^{1.5}$~d).

\section{Summary and conclusions}\label{sec:conclusions}
In this paper we converted the \smash\ observational measurements of \citetalias{sana2014}  into physical units, i.e.,\ observed angular separations on the sky to projected physical separations, and observed $H$-band magnitude differences to mass ratios. To do this, we derived calibration relations based on physical properties of O-type stars presented in \cite{martins2005} and \cite{martins2006}, as well as the evolutionary models of \citet{brott2011} for lower-mass stars. 

The accuracy of the resulting distances was assessed by comparing them to known cluster distances, and to up-to-date {\it ALSIII-Gaia} distances. Overall, we find good agreement, with the spectrophotometric distances slightly overestimating the Gaia distances by just 5\%, except for Oe stars, for a small subset of supergiants, and for binary evolution products. For the derived masses, primary stars with known dynamical masses were used to assess the accuracy of the relations used. We find excellent agreement, except for known (post-)interaction systems. Although this was not the original aim, our calibrations can be used with independent mass and/or distance estimates to identify peculiar stars whose physical parameters are potentially affected by binary evolution.

The derived projected separations follow an O\"epik law, i.e.,\ a flat logarithmic distribution. The distribution of mass ratios follows a power-law distribution $f_q \propto q^\kappa$, with $\kappa$  changing with the separation range considered. For companions within 100~AU, the value that we derived ($\kappa \approx -0.6^{+0.9}_{-0.7}$) lies between the uniform  mass-ratio distributions derived for  spectroscopic binaries, and the steeper $\kappa = -1.4\pm0.4$ value reported by \cite{moe2017} based on the analysis of a subsample of the \smash\ data. The value that we derived is compatible within errors with both regimes and a larger data set is probably needed to reach a firm conclusion. 
The distribution of companion masses seems to be well reproduced by a Salpeter initial mass function. Finally, we find an absence of massive companions at separations larger than 100 AU. 

\section*{Data availability}
Table~\ref{tab:distance} is only available in electronic form at the CDS via anonymous ftp to cdsarc.u-strasbg.fr (130.79.128.5) or via http://cdsweb.u-strasbg.fr/cgi-bin/qcat?J/A+A/.

\begin{acknowledgements}
This project has received funding from the European Research Council under European Union's Horizon 2020 research programme (MULTIPLES, No 772225), and from the KU Leuven Research Council (grant METH/24/012: SOUL and C16/17/007: MAESTRO). 
FT  acknowledges support by grant PID2022-137779OB-C41, funded by the Spanish Ministry of Science, Innovation and Universities/State Agency of Research MICIU/AEI/10.13039/501100011033. This research has made use of the SIMBAD database, operated at CDS, Strasbourg, France. This work made use of Astropy:\footnote{http://www.astropy.org} a community-developed core Python package and an ecosystem of tools and resources for astronomy \citep{astropy:2013, astropy:2018, astropy:2022}.
\end{acknowledgements}

\bibliographystyle{aa}
\bibliography{smash_distance}

@INPROCEEDINGS{SanaVrancken2026,
       author = {{Sana}, Hugues and {Vrancken}, Jasmine},
        title = "{Observing binaries}",
    booktitle = {Encyclopedia of Astrophysics},
         year = 2026,
       volume = {2},
        month = jan,
        pages = {106-118},
          doi = {10.1016/B978-0-443-21439-4.00146-2},
       adsurl = {https://ui.adsabs.harvard.edu/abs/2026enap....2..106S}
}

@ARTICLE{GRAVITY2018,
       author = {{GRAVITY Collaboration} and {Karl}, Martina and {Pfuhl}, Oliver and {Eisenhauer}, Frank and {Genzel}, Reinhard and {Grellmann}, Rebekka and {Habibi}, Maryam and {Abuter}, Roberto and {Accardo}, Matteo and {Amorim}, Ant{\'o}nio and {Anugu}, Narsireddy and {{\'A}vila}, Gerardo and {Benisty}, Myriam and {Berger}, Jean-Philippe and {Blind}, Nicolas and {Bonnet}, Henri and {Bourget}, Pierre and {Brandner}, Wolfgang and {Brast}, Roland and {Buron}, Alexander and {Caratti O Garatti}, Alessio and {Chapron}, Fr{\'e}d{\'e}ric and {Cl{\'e}net}, Yann and {Collin}, Claude and {Coud{\'e} Du Foresto}, Vincent and {de Wit}, Willem-Jan and {de Zeeuw}, Tim and {Deen}, Casey and {Delplancke-Str{\"o}bele}, Fran{\c{c}}oise and {Dembet}, Roderick and {Derie}, Fr{\'e}d{\'e}ric and {Dexter}, Jason and {Duvert}, Gilles and {Ebert}, Monica and {Eckart}, Andreas and {Esselborn}, Michael and {F{\'e}dou}, Pierre and {Finger}, Gert and {Garcia}, Paulo and {Garcia Dabo}, Cesar Enrique and {Garcia Lopez}, Rebeca and {Gao}, Feng and {Gendron}, {\'E}ric and {Gillessen}, Stefan and {Gont{\'e}}, Fr{\'e}d{\'e}ric and {Gordo}, Paulo and {Gr{\"o}zinger}, Ulrich and {Guajardo}, Patricia and {Guieu}, Sylvain and {Haguenauer}, Pierre and {Hans}, Oliver and {Haubois}, Xavier and {Haug}, Marcus and {Hau{\ss}mann}, Frank and {Henning}, Thomas and {Hippler}, Stefan and {Horrobin}, Matthew and {Huber}, Armin and {Hubert}, Zoltan and {Hubin}, Norbert and {Jakob}, Gerd and {Jochum}, Lieselotte and {Jocou}, Laurent and {Kaufer}, Andreas and {Kellner}, Stefan and {Kendrew}, Sarah and {Kern}, Lothar and {Kervella}, Pierre and {Kiekebusch}, Mario and {Klein}, Ralf and {K{\"o}hler}, Rainer and {Kolb}, Johan and {Kulas}, Martin and {Lacour}, Sylvestre and {Lapeyr{\`e}re}, Vincent and {Lazareff}, Bernard and {Le Bouquin}, Jean-Baptiste and {L{\'e}na}, Pierre and {Lenzen}, Rainer and {L{\'e}v{\^e}que}, Samuel and {Lin}, Chien-Cheng and {Lippa}, Magdalena and {Magnard}, Yves and {Mehrgan}, Leander and {M{\'e}rand}, Antoine and {Moulin}, Thibaut and {M{\"u}ller}, Eric and {M{\"u}ller}, Friedrich and {Neumann}, Udo and {Oberti}, Sylvain and {Ott}, Thomas and {Pallanca}, Laurent and {Panduro}, Johana and {Pasquini}, Luca and {Paumard}, Thibaut and {Percheron}, Isabelle and {Perraut}, Karine and {Perrin}, Guy and {Pfl{\"u}ger}, Andreas and {Duc}, Thanh Phan and {Plewa}, Philipp M. and {Popovic}, Dan and {Rabien}, Sebastian and {Ram{\'\i}rez}, Andr{\'e}s and {Ramos}, Jose and {Rau}, Christian and {Riquelme}, Miguel and {Rodr{\'\i}guez-Coira}, Gustavo and {Rohloff}, Ralf-Rainer and {Rosales}, Alejandra and {Rousset}, G{\'e}rard and {Sanchez-Bermudez}, Joel and {Scheithauer}, Silvia and {Sch{\"o}ller}, Markus and {Schuhler}, Nicolas and {Spyromilio}, Jason and {Straub}, Odele and {Straubmeier}, Christian and {Sturm}, Eckhard and {Suarez}, Marcos and {Tristram}, Konrad R.~W. and {Ventura}, Noel and {Vincent}, Fr{\'e}d{\'e}ric and {Waisberg}, Idel and {Wank}, Imke and {Widmann}, Felix and {Wieprecht}, Ekkehard and {Wiest}, Michael and {Wiezorrek}, Erich and {Wittkowski}, Markus and {Woillez}, Julien and {Wolff}, Burkhard and {Yazici}, Senol and {Ziegler}, Denis and {Zins}, G{\'e}rard},
        title = "{Multiple star systems in the Orion nebula}",
      journal = {\aap},
         year = 2018,
        month = dec,
       volume = {620},
          eid = {A116},
        pages = {A116},
          doi = {10.1051/0004-6361/201833575},
archivePrefix = {arXiv},
       eprint = {1809.10376},
 primaryClass = {astro-ph.SR},
       adsurl = {https://ui.adsabs.harvard.edu/abs/2018A&A...620A.116G}
}

@ARTICLE{shenar2022,
       author = {{Shenar}, T. and {Sana}, H. and {Mahy}, L. and {Ma{\'\i}z Apell{\'a}niz}, J. and {Crowther}, Paul A. and {Gromadzki}, M. and {Herrero}, A. and {Langer}, N. and {Marchant}, P. and {Schneider}, F.~R.~N. and {Sen}, K. and {Soszy{\'n}ski}, I. and {Toonen}, S.},
        title = "{The Tarantula Massive Binary Monitoring. VI. Characterisation of hidden companions in 51 single-lined O-type binaries: A flat mass-ratio distribution and black-hole binary candidates}",
      journal = {\aap},
         year = 2022,
        month = sep,
       volume = {665},
          eid = {A148},
        pages = {A148},
          doi = {10.1051/0004-6361/202244245},
archivePrefix = {arXiv},
       eprint = {2207.07674},
 primaryClass = {astro-ph.SR},
       adsurl = {https://ui.adsabs.harvard.edu/abs/2022A&A...665A.148S}
}

@ARTICLE{almeida2017,
       author = {{Almeida}, L.~A. and {Sana}, H. and {Taylor}, W. and {Barb{\'a}}, R. and {Bonanos}, A.~Z. and {Crowther}, P. and {Damineli}, A. and {de Koter}, A. and {de Mink}, S.~E. and {Evans}, C.~J. and {Gieles}, M. and {Grin}, N.~J. and {H{\'e}nault-Brunet}, V. and {Langer}, N. and {Lennon}, D. and {Lockwood}, S. and {Ma{\'\i}z Apell{\'a}niz}, J. and {Moffat}, A.~F.~J. and {Neijssel}, C. and {Norman}, C. and {Ram{\'\i}rez-Agudelo}, O.~H. and {Richardson}, N.~D. and {Schootemeijer}, A. and {Shenar}, T. and {Soszy{\'n}ski}, I. and {Tramper}, F. and {Vink}, J.~S.},
        title = "{The Tarantula Massive Binary Monitoring. I. Observational campaign and OB-type spectroscopic binaries}",
      journal = {\aap},
         year = 2017,
        month = feb,
       volume = {598},
          eid = {A84},
        pages = {A84},
          doi = {10.1051/0004-6361/201629844},
archivePrefix = {arXiv},
       eprint = {1610.03500},
 primaryClass = {astro-ph.SR},
       adsurl = {https://ui.adsabs.harvard.edu/abs/2017A&A...598A..84A}
}

@ARTICLE{WhiteGhez2001,
       author = {{White}, R.~J. and {Ghez}, A.~M.},
        title = "{Observational Constraints on the Formation and Evolution of Binary Stars}",
      journal = {\apj},
         year = 2001,
        month = jul,
       volume = {556},
       number = {1},
        pages = {265-295},
          doi = {10.1086/321542},
archivePrefix = {arXiv},
       eprint = {astro-ph/0103098},
 primaryClass = {astro-ph},
       adsurl = {https://ui.adsabs.harvard.edu/abs/2001ApJ...556..265W}
}

@ARTICLE{ElBadryRix2019a,
       author = {{El-Badry}, Kareem and {Rix}, Hans-Walter and {Tian}, Haijun and {Duch{\^e}ne}, Gaspard and {Moe}, Maxwell},
        title = "{Discovery of an equal-mass `twin' binary population reaching 1000 + au separations}",
      journal = {\mnras},
         year = 2019,
        month = nov,
       volume = {489},
       number = {4},
        pages = {5822-5857},
          doi = {10.1093/mnras/stz2480},
archivePrefix = {arXiv},
       eprint = {1906.10128},
 primaryClass = {astro-ph.SR},
       adsurl = {https://ui.adsabs.harvard.edu/abs/2019MNRAS.489.5822E}
}

@ARTICLE{ElBadryRix2019b,
       author = {{El-Badry}, Kareem and {Rix}, Hans-Walter},
        title = "{The wide binary fraction of solar-type stars: emergence of metallicity dependence at a < 200 au}",
      journal = {\mnras},
         year = 2019,
        month = jan,
       volume = {482},
       number = {1},
        pages = {L139-L144},
          doi = {10.1093/mnrasl/sly206},
archivePrefix = {arXiv},
       eprint = {1809.06860},
 primaryClass = {astro-ph.SR},
       adsurl = {https://ui.adsabs.harvard.edu/abs/2019MNRAS.482L.139E}
}

@ARTICLE{oepik,
       author = {{{\"O}pik}, E.},
      journal = {Tartu Obs. Publ},
         year = 1924,
       volume = {26},
       number = {6}
}

@ARTICLE{perets2025,
       author = {{Perets}, Hagai B.},
        title = "{Evolution of Triple Stars}",
      journal = {arXiv e-prints},
         year = 2025,
        month = apr,
          eid = {arXiv:2504.02939},
        pages = {arXiv:2504.02939},
          doi = {10.48550/arXiv.2504.02939},
archivePrefix = {arXiv},
       eprint = {2504.02939},
 primaryClass = {astro-ph.SR},
       adsurl = {https://ui.adsabs.harvard.edu/abs/2025arXiv250402939P}
}

@ARTICLE{moe2017,
       author = {{Moe}, Maxwell and {Di Stefano}, Rosanne},
        title = "{Mind Your Ps and Qs: The Interrelation between Period (P) and Mass-ratio (Q) Distributions of Binary Stars}",
      journal = {\apjs},
         year = 2017,
        month = jun,
       volume = {230},
       number = {2},
          eid = {15},
        pages = {15},
          doi = {10.3847/1538-4365/aa6fb6},
archivePrefix = {arXiv},
       eprint = {1606.05347},
 primaryClass = {astro-ph.SR},
       adsurl = {https://ui.adsabs.harvard.edu/abs/2017ApJS..230...15M}
}

@ARTICLE{ekstroem2012,
       author = {{Ekstr{\"o}m}, S. and {Georgy}, C. and {Eggenberger}, P. and {Meynet}, G. and {Mowlavi}, N. and {Wyttenbach}, A. and {Granada}, A. and {Decressin}, T. and {Hirschi}, R. and {Frischknecht}, U. and {Charbonnel}, C. and {Maeder}, A.},
        title = "{Grids of stellar models with rotation. I. Models from 0.8 to 120 M$_{{\ensuremath{\odot}}}$ at solar metallicity (Z = 0.014)}",
      journal = {\aap},
         year = 2012,
        month = jan,
       volume = {537},
          eid = {A146},
        pages = {A146},
          doi = {10.1051/0004-6361/201117751},
archivePrefix = {arXiv},
       eprint = {1110.5049},
 primaryClass = {astro-ph.SR},
       adsurl = {https://ui.adsabs.harvard.edu/abs/2012A&A...537A.146E}
}

@ARTICLE{frost2024,
       author = {{Frost}, A.~J. and {Sana}, H. and {Mahy}, L. and {Wade}, G. and {Barron}, J. and {Le Bouquin}, J. -B. and {M{\'e}rand}, A. and {Schneider}, F.~R.~N. and {Shenar}, T. and {Barb{\'a}}, R.~H. and {Bowman}, D.~M. and {Fabry}, M. and {Farhang}, A. and {Marchant}, P. and {Morrell}, N.~I. and {Smoker}, J.~V.},
        title = "{A magnetic massive star has experienced a stellar merger}",
      journal = {Science},
         year = 2024,
        month = apr,
       volume = {384},
       number = {6692},
        pages = {214-217},
          doi = {10.1126/science.adg7700},
archivePrefix = {arXiv},
       eprint = {2404.10167},
 primaryClass = {astro-ph.SR},
       adsurl = {https://ui.adsabs.harvard.edu/abs/2024Sci...384..214F}
}

@ARTICLE{als3,
       author = {{Pantaleoni Gonz{\'a}lez}, M. and {Ma{\'\i}z Apell{\'a}niz}, J. and {Barb{\'a}}, R.~H. and {Reed}, B. Cameron and {Berlanas}, S.~R. and {Parras Rico}, A. and {Bodaghee}, A.},
        title = "{The Alma catalogue of OB stars ─ III. A cross-match with Gaia DR3 and an extension based on new spectral classifications}",
      journal = {\mnras},
         year = 2025,
        month = oct,
       volume = {543},
       number = {1},
        pages = {63-82},
          doi = {10.1093/mnras/staf1409},
archivePrefix = {arXiv},
       eprint = {2508.14875},
 primaryClass = {astro-ph.SR},
       adsurl = {https://ui.adsabs.harvard.edu/abs/2025MNRAS.543...63P}
}

@ARTICLE{pauwels2024,
       author = {{Pauwels}, Tinne and {Reggiani}, Maddalena and {Sana}, Hugues and {Mahy}, Laurent},
        title = "{Low-mass Stellar and Substellar Candidate Companions around Massive Stars in Sco OB1 and M17}",
      journal = {\aj},
         year = 2024,
        month = nov,
       volume = {168},
       number = {5},
          eid = {209},
        pages = {209},
          doi = {10.3847/1538-3881/ad6f06},
archivePrefix = {arXiv},
       eprint = {2409.16212},
 primaryClass = {astro-ph.SR},
       adsurl = {https://ui.adsabs.harvard.edu/abs/2024AJ....168..209P}
}

@ARTICLE{rainot2022,
       author = {{Rainot}, A. and {Reggiani}, M. and {Sana}, H. and {Bodensteiner}, J. and {Absil}, O.},
        title = "{Carina High-contrast Imaging Project for massive Stars (CHIPS). II. O stars in Trumpler 14}",
      journal = {\aap},
         year = 2022,
        month = feb,
       volume = {658},
          eid = {A198},
        pages = {A198},
          doi = {10.1051/0004-6361/202141562},
archivePrefix = {arXiv},
       eprint = {2111.12361},
 primaryClass = {astro-ph.SR},
       adsurl = {https://ui.adsabs.harvard.edu/abs/2022A&A...658A.198R}
}

@ARTICLE{MA2019,
       author = {{Ma{\'\i}z Apell{\'a}niz}, J. and {Trigueros P{\'a}ez}, E. and {Negueruela}, I. and {Barb{\'a}}, R.~H. and {Sim{\'o}n-D{\'\i}az}, S. and {Lorenzo}, J. and {Sota}, A. and {Gamen}, R.~C. and {Fari{\~n}a}, C. and {Salas}, J. and {Caballero}, J.~A. and {Morrell}, N.~I. and {Pellerin}, A. and {Alfaro}, E.~J. and {Herrero}, A. and {Arias}, J.~I. and {Marco}, A.},
        title = "{MONOS: Multiplicity Of Northern O-type Spectroscopic systems. I. Project description and spectral classifications and visual multiplicity of previously known objects}",
      journal = {\aap},
         year = 2019,
        month = jun,
       volume = {626},
          eid = {A20},
        pages = {A20},
          doi = {10.1051/0004-6361/201935359},
archivePrefix = {arXiv},
       eprint = {1904.11385},
 primaryClass = {astro-ph.SR},
       adsurl = {https://ui.adsabs.harvard.edu/abs/2019A&A...626A..20M}
}

@ARTICLE{MA2018,
       author = {{Ma{\'\i}z Apell{\'a}niz}, J. and {Barb{\'a}}, R.~H. and {Sim{\'o}n-D{\'\i}az}, S. and {Sota}, A. and {Trigueros P{\'a}ez}, E. and {Caballero}, J.~A. and {Alfaro}, E.~J.},
        title = "{Lucky Spectroscopy, an equivalent technique to Lucky Imaging. Spatially resolved spectroscopy of massive close visual binaries using the William Herschel Telescope}",
      journal = {\aap},
         year = 2018,
        month = jul,
       volume = {615},
          eid = {A161},
        pages = {A161},
          doi = {10.1051/0004-6361/201832885},
archivePrefix = {arXiv},
       eprint = {1804.03133},
 primaryClass = {astro-ph.SR},
       adsurl = {https://ui.adsabs.harvard.edu/abs/2018A&A...615A.161M}
}

@ARTICLE{kiminki2018,
       author = {{Kiminki}, Megan M. and {Smith}, Nathan},
        title = "{A radial velocity survey of the Carina Nebula's O-type stars}",
      journal = {\mnras},
         year = 2018,
        month = jun,
       volume = {477},
       number = {2},
        pages = {2068-2086},
          doi = {10.1093/mnras/sty748},
archivePrefix = {arXiv},
       eprint = {1803.07057},
 primaryClass = {astro-ph.SR},
       adsurl = {https://ui.adsabs.harvard.edu/abs/2018MNRAS.477.2068K}
}

@ARTICLE{kobulnicky2014,
       author = {{Kobulnicky}, Henry A. and {Kiminki}, Daniel C. and {Lundquist}, Michael J. and {Burke}, Jamison and {Chapman}, James and {Keller}, Erica and {Lester}, Kathryn and {Rolen}, Emily K. and {Topel}, Eric and {Bhattacharjee}, Anirban and {Smullen}, Rachel A. and {Vargas {\'A}lvarez}, Carlos A. and {Runnoe}, Jessie C. and {Dale}, Daniel A. and {Brotherton}, Michael M.},
        title = "{Toward Complete Statistics of Massive Binary Stars: Penultimate Results from the Cygnus OB2 Radial Velocity Survey}",
      journal = {\apjs},
         year = 2014,
        month = aug,
       volume = {213},
       number = {2},
          eid = {34},
        pages = {34},
          doi = {10.1088/0067-0049/213/2/34},
archivePrefix = {arXiv},
       eprint = {1406.6655},
 primaryClass = {astro-ph.SR},
       adsurl = {https://ui.adsabs.harvard.edu/abs/2014ApJS..213...34K}
}

@ARTICLE{villasenor2025,
       author = {{Villase{\~n}or}, J.~I. and {Sana}, H. and {Mahy}, L. and {Shenar}, T. and {Bodensteiner}, J. and {Britavskiy}, N. and {Lennon}, D.~J. and {Moe}, M. and {Patrick}, L.~R. and {Pawlak}, M. and {Bowman}, D.~M. and {Crowther}, P.~A. and {de Mink}, S.~E. and {Deshmukh}, K. and {Evans}, C.~J. and {Fabry}, M. and {Fouesneau}, M. and {Holgado}, G. and {Langer}, N. and {Ma{\'\i}z Apell{\'a}niz}, J. and {Mandel}, I. and {Oskinova}, L.~M. and {Pauli}, D. and {Ramachandran}, V. and {Renzo}, M. and {Rix}, H. -W. and {Rocha}, D.~F. and {Sander}, A.~A.~C. and {Schneider}, F.~R.~N. and {Sen}, K. and {Sim{\'o}n-D{\'\i}az}, S. and {van Loon}, J. Th. and {Toonen}, S. and {Vink}, J.~S.},
        title = "{Binarity at LOw Metallicity (BLOeM): Enhanced multiplicity of early B-type dwarfs and giants at Z = 0.2 Z$_{{\ensuremath{\odot}}}$}",
      journal = {\aap},
         year = 2025,
        month = jun,
       volume = {698},
          eid = {A41},
        pages = {A41},
          doi = {10.1051/0004-6361/202453166},
archivePrefix = {arXiv},
       eprint = {2503.21936},
 primaryClass = {astro-ph.SR},
       adsurl = {https://ui.adsabs.harvard.edu/abs/2025A&A...698A..41V}
}

@ARTICLE{banyard2022,
       author = {{Banyard}, G. and {Sana}, H. and {Mahy}, L. and {Bodensteiner}, J. and {Villase{\~n}or}, J.~I. and {Evans}, C.~J.},
        title = "{The observed multiplicity properties of B-type stars in the Galactic young open cluster NGC 6231}",
      journal = {\aap},
         year = 2022,
        month = feb,
       volume = {658},
          eid = {A69},
        pages = {A69},
          doi = {10.1051/0004-6361/202141037},
archivePrefix = {arXiv},
       eprint = {2108.07814},
 primaryClass = {astro-ph.SR},
       adsurl = {https://ui.adsabs.harvard.edu/abs/2022A&A...658A..69B}
}

@ARTICLE{villasenor2021,
       author = {{Villase{\~n}or}, J.~I. and {Taylor}, W.~D. and {Evans}, C.~J. and {Ram{\'\i}rez-Agudelo}, O.~H. and {Sana}, H. and {Almeida}, L.~A. and {de Mink}, S.~E. and {Dufton}, P.~L. and {Langer}, N.},
        title = "{The B-type binaries characterization programme I. Orbital solutions for the 30 Doradus population}",
      journal = {\mnras},
         year = 2021,
        month = nov,
       volume = {507},
       number = {4},
        pages = {5348-5375},
          doi = {10.1093/mnras/stab2197},
archivePrefix = {arXiv},
       eprint = {2107.10170},
 primaryClass = {astro-ph.SR},
       adsurl = {https://ui.adsabs.harvard.edu/abs/2021MNRAS.507.5348V}
}

@ARTICLE{dunstall2015,
       author = {{Dunstall}, P.~R. and {Dufton}, P.~L. and {Sana}, H. and {Evans}, C.~J. and {Howarth}, I.~D. and {Sim{\'o}n-D{\'\i}az}, S. and {de Mink}, S.~E. and {Langer}, N. and {Ma{\'\i}z Apell{\'a}niz}, J. and {Taylor}, W.~D.},
        title = "{The VLT-FLAMES Tarantula Survey. XXII. Multiplicity properties of the B-type stars}",
      journal = {\aap},
         year = 2015,
        month = aug,
       volume = {580},
          eid = {A93},
        pages = {A93},
          doi = {10.1051/0004-6361/201526192},
archivePrefix = {arXiv},
       eprint = {1505.07121},
 primaryClass = {astro-ph.SR},
       adsurl = {https://ui.adsabs.harvard.edu/abs/2015A&A...580A..93D}
}

@ARTICLE{frost2025,
       author = {{Frost}, A.~J. and {Sana}, H. and {Le Bouquin}, J. -B. and {Perets}, H.~B. and {Bodensteiner}, J. and {Igoshev}, A.~P. and {Banyard}, G. and {Mahy}, L. and {M{\'e}rand}, A. and {Ram{\'\i}rez-Agudelo}, O.~H.},
        title = "{An interferometric study of B star multiplicity}",
      journal = {\aap},
         year = 2025,
        month = sep,
       volume = {701},
          eid = {A171},
        pages = {A171},
          doi = {10.1051/0004-6361/202554344},
archivePrefix = {arXiv},
       eprint = {2505.02300},
 primaryClass = {astro-ph.SR},
       adsurl = {https://ui.adsabs.harvard.edu/abs/2025A&A...701A.171F}
}

@ARTICLE{langer2012,
       author = {{Langer}, N.},
        title = "{Presupernova Evolution of Massive Single and Binary Stars}",
      journal = {\araa},
         year = 2012,
        month = sep,
       volume = {50},
        pages = {107-164},
          doi = {10.1146/annurev-astro-081811-125534},
archivePrefix = {arXiv},
       eprint = {1206.5443},
 primaryClass = {astro-ph.SR},
       adsurl = {https://ui.adsabs.harvard.edu/abs/2012ARA&A..50..107L}
}

@ARTICLE{sana2014,
   author = {{Sana}, H. and {Le Bouquin}, J.-B. and {Lacour}, S. and {Berger}, J.-P. and 
	{Duvert}, G. and {Gauchet}, L. and {Norris}, B. and {Olofsson}, J. and 
	{Pickel}, D. and {Zins}, G. and {Absil}, O. and {de Koter}, A. and 
	{Kratter}, K. and {Schnurr}, O. and {Zinnecker}, H.},
    title = "{Southern Massive Stars at High Angular Resolution: Observational Campaign and Companion Detection}",
  journal = {\apjs},
archivePrefix = "arXiv",
   eprint = {1409.6304},
 primaryClass = "astro-ph.SR",
     year = 2014,
    month = nov,
   volume = 215,
      eid = {15},
    pages = {15},
      doi = {10.1088/0067-0049/215/1/15},
   adsurl = {http://adsabs.harvard.edu/abs/2014ApJS..215...15S}
}

@ARTICLE{fitzpatrick1999,
   author = {{Fitzpatrick}, E.~L.},
    title = "{Correcting for the Effects of Interstellar Extinction}",
  journal = {\pasp},
   eprint = {astro-ph/9809387},
     year = 1999,
    month = jan,
   volume = 111,
    pages = {63-75},
      doi = {10.1086/316293},
   adsurl = {http://adsabs.harvard.edu/abs/1999PASP..111...63F}
}

@ARTICLE{martins2006,
   author = {{Martins}, F. and {Plez}, B.},
    title = "{UBVJHK synthetic photometry of Galactic O stars}",
  journal = {\aap},
   eprint = {astro-ph/0606587},
     year = 2006,
    month = oct,
   volume = 457,
    pages = {637-644},
      doi = {10.1051/0004-6361:20065753},
   adsurl = {http://adsabs.harvard.edu/abs/2006A%26A...457..637M}
}

@ARTICLE{morrell2001,
   author = {{Morrell}, N.~I. and {Barb{\'a}}, R.~H. and {Niemela}, V.~S. and 
	{Corti}, M.~A. and {Albacete Colombo}, J.~F. and {Rauw}, G. and 
	{Corcoran}, M. and {Morel}, T. and {Bertrand}, J.-F. and {Moffat}, A.~F.~J. and 
	{St-Louis}, N.},
    title = "{Optical spectroscopy of X-Mega targets - II. The massive double-lined O-type binary HD 93205}",
  journal = {\mnras},
   eprint = {astro-ph/0105014},
     year = 2001,
    month = sep,
   volume = 326,
    pages = {85-94},
      doi = {10.1046/j.1365-8711.2001.04500.x},
   adsurl = {http://adsabs.harvard.edu/abs/2001MNRAS.326...85M}
}

@ARTICLE{sana2011,
   author = {{Sana}, H. and {James}, G. and {Gosset}, E.},
    title = "{The massive star binary fraction in young open clusters - III. IC 2944 and the Cen OB2 association}",
  journal = {\mnras},
archivePrefix = "arXiv",
   eprint = {1109.2899},
 primaryClass = "astro-ph.SR",
     year = 2011,
    month = sep,
   volume = 416,
    pages = {817-831},
      doi = {10.1111/j.1365-2966.2011.18698.x},
   adsurl = {http://adsabs.harvard.edu/abs/2011MNRAS.416..817S}
}

@ARTICLE{penny2002,
   author = {{Penny}, L.~R. and {Gies}, D.~R. and {Wise}, J.~H. and {Stickland}, D.~J. and 
	{Lloyd}, C.},
    title = "{Tomographic Separation of Composite Spectra. IX. The Massive Close Binary HD 115071}",
  journal = {\apj},
   eprint = {astro-ph/0201480},
     year = 2002,
    month = aug,
   volume = 575,
    pages = {1050-1056},
      doi = {10.1086/341432},
   adsurl = {http://adsabs.harvard.edu/abs/2002ApJ...575.1050P}
}

@ARTICLE{penny2001,
   author = {{Penny}, L.~R. and {Seyle}, D. and {Gies}, D.~R. and {Harvin}, J.~A. and 
	{Bagnuolo}, Jr., W.~G. and {Thaller}, M.~L. and {Fullerton}, A.~W. and 
	{Kaper}, L.},
    title = "{Tomographic Separation of Composite Spectra. VII. The Physical Properties of the Massive Triple System HD 135240 ({$\delta$} Circini)}",
  journal = {\apj},
     year = 2001,
    month = feb,
   volume = 548,
    pages = {889-899},
      doi = {10.1086/319031},
   adsurl = {http://adsabs.harvard.edu/abs/2001ApJ...548..889P}
}

@ARTICLE{sana2013,
   author = {{Sana}, H. and {Le Bouquin}, J.-B. and {Mahy}, L. and {Absil}, O. and 
	{De Becker}, M. and {Gosset}, E.},
    title = "{Three-dimensional orbits of the triple-O stellar system HD 150136}",
  journal = {\aap},
archivePrefix = "arXiv",
   eprint = {1304.3457},
 primaryClass = "astro-ph.SR",
     year = 2013,
    month = may,
   volume = 553,
      eid = {A131},
    pages = {A131},
      doi = {10.1051/0004-6361/201321189},
   adsurl = {http://adsabs.harvard.edu/abs/2013A%26A...553A.131S}
}

@ARTICLE{nasseri2014,
   author = {{Nasseri}, A. and {Chini}, R. and {Harmanec}, P. and {Mayer}, P. and 
	{Nemravov{\'a}}, J.~A. and {Dembsky}, T. and {Lehmann}, H. and 
	{Sana}, H. and {Le Bouquin}, J.-B.},
    title = "{HD 152246: a new high-mass triple system and its basic properties}",
  journal = {\aap},
archivePrefix = "arXiv",
   eprint = {1408.0956},
 primaryClass = "astro-ph.SR",
     year = 2014,
    month = aug,
   volume = 568,
      eid = {A94},
    pages = {A94},
      doi = {10.1051/0004-6361/201424382},
   adsurl = {http://adsabs.harvard.edu/abs/2014A%26A...568A..94N}
}

@ARTICLE{sana2001,
   author = {{Sana}, H. and {Rauw}, G. and {Gosset}, E.},
    title = "{HD 152248: Evidence for a colliding wind interaction}",
  journal = {\aap},
     year = 2001,
    month = apr,
   volume = 370,
    pages = {121-135},
      doi = {10.1051/0004-6361:20010221},
   adsurl = {http://adsabs.harvard.edu/abs/2001A%26A...370..121S}
}

@ARTICLE{mason1998,
   author = {{Mason}, B.~D. and {Gies}, D.~R. and {Hartkopf}, W.~I. and {Bagnuolo}, Jr., W.~G. and 
	{ten Brummelaar}, T. and {McAlister}, H.~A.},
    title = "{ICCD speckle observations of binary stars. XIX - an astrometric/spectroscopic survey of O stars}",
  journal = {\aj},
     year = 1998,
    month = feb,
   volume = 115,
    pages = {821},
      doi = {10.1086/300234},
   adsurl = {http://adsabs.harvard.edu/abs/1998AJ....115..821M}
}

@ARTICLE{ferrero2013,
   author = {{Ferrero}, G. and {Gamen}, R. and {Benvenuto}, O. and {Fern{\'a}ndez-Laj{\'u}s}, E.
	},
    title = "{Apsidal motion in massive close binary systems - I. HD 165052, an extreme case?}",
  journal = {\mnras},
archivePrefix = "arXiv",
   eprint = {1305.1911},
 primaryClass = "astro-ph.SR",
     year = 2013,
    month = aug,
   volume = 433,
    pages = {1300-1311},
      doi = {10.1093/mnras/stt812},
   adsurl = {http://adsabs.harvard.edu/abs/2013MNRAS.433.1300F}
}

@ARTICLE{mahy2010,
   author = {{Mahy}, L. and {Rauw}, G. and {Martins}, F. and {Naz{\'e}}, Y. and 
	{Gosset}, E. and {De Becker}, M. and {Sana}, H. and {Eenens}, P.
	},
    title = "{A New Investigation of the Binary HD 48099}",
  journal = {\apj},
archivePrefix = "arXiv",
   eprint = {0912.0605},
 primaryClass = "astro-ph.SR",
     year = 2010,
    month = jan,
   volume = 708,
    pages = {1537-1544},
      doi = {10.1088/0004-637X/708/2/1537},
   adsurl = {http://adsabs.harvard.edu/abs/2010ApJ...708.1537M}
}

@ARTICLE{sana2012,
   author = {{Sana}, H. and {de Mink}, S.~E. and {de Koter}, A. and {Langer}, N. and 
	{Evans}, C.~J. and {Gieles}, M. and {Gosset}, E. and {Izzard}, R.~G. and 
	{Le Bouquin}, J.-B. and {Schneider}, F.~R.~N.},
    title = "{Binary Interaction Dominates the Evolution of Massive Stars}",
  journal = {Science},
archivePrefix = "arXiv",
   eprint = {1207.6397},
 primaryClass = "astro-ph.SR",
     year = 2012,
    month = jul,
   volume = 337,
    pages = {444-},
      doi = {10.1126/science.1223344},
   adsurl = {http://adsabs.harvard.edu/abs/2012Sci...337..444S}
}

@ARTICLE{linder2008,
   author = {{Linder}, N. and {Rauw}, G. and {Martins}, F. and {Sana}, H. and 
	{De Becker}, M. and {Gosset}, E.},
    title = "{High-resolution optical spectroscopy of Plaskett's star}",
  journal = {\aap},
archivePrefix = "arXiv",
   eprint = {0807.4823},
     year = 2008,
    month = oct,
   volume = 489,
    pages = {713-723},
      doi = {10.1051/0004-6361:200810003},
   adsurl = {http://adsabs.harvard.edu/abs/2008A%26A...489..713L}
}

@ARTICLE{martins2005,
   author = {{Martins}, F. and {Schaerer}, D. and {Hillier}, D.~J.},
    title = "{A new calibration of stellar parameters of Galactic O stars}",
  journal = {\aap},
   eprint = {astro-ph/0503346},
     year = 2005,
    month = jun,
   volume = 436,
    pages = {1049-1065},
      doi = {10.1051/0004-6361:20042386},
   adsurl = {http://adsabs.harvard.edu/abs/2005A%26A...436.1049M}
}

@ARTICLE{brott2011,
   author = {{Brott}, I. and {de Mink}, S.~E. and {Cantiello}, M. and {Langer}, N. and 
	{de Koter}, A. and {Evans}, C.~J. and {Hunter}, I. and {Trundle}, C. and 
	{Vink}, J.~S.},
    title = "{Rotating massive main-sequence stars. I. Grids of evolutionary models and isochrones}",
  journal = {\aap},
archivePrefix = "arXiv",
   eprint = {1102.0530},
 primaryClass = "astro-ph.SR",
     year = 2011,
    month = jun,
   volume = 530,
      eid = {A115},
    pages = {A115},
      doi = {10.1051/0004-6361/201016113},
   adsurl = {http://adsabs.harvard.edu/abs/2011A%26A...530A.115B}
}

@ARTICLE{melnik2009,
   author = {{Mel'Nik}, A.~M. and {Dambis}, A.~K.},
    title = "{Kinematics of OB-associations and the new reduction of the Hipparcos data}",
  journal = {\mnras},
archivePrefix = "arXiv",
   eprint = {0909.0618},
 primaryClass = "astro-ph.GA",
     year = 2009,
    month = nov,
   volume = 400,
    pages = {518-523},
      doi = {10.1111/j.1365-2966.2009.15484.x},
   adsurl = {http://adsabs.harvard.edu/abs/2009MNRAS.400..518M}
}

@ARTICLE{kharchenko2005,
   author = {{Kharchenko}, N.~V. and {Piskunov}, A.~E. and {R{\"o}ser}, S. and 
	{Schilbach}, E. and {Scholz}, R.-D.},
    title = "{Astrophysical parameters of Galactic open clusters}",
  journal = {\aap},
   eprint = {astro-ph/0501674},
     year = 2005,
    month = aug,
   volume = 438,
    pages = {1163-1173},
      doi = {10.1051/0004-6361:20042523},
   adsurl = {http://adsabs.harvard.edu/abs/2005A%26A...438.1163K}
}

@ARTICLE{penny2008,
   author = {{Penny}, L.~R. and {Ouzts}, C. and {Gies}, D.~R.},
    title = "{Tomographic Separation of Composite Spectra. XI. The Physical Properties of the Massive Close Binary HD 100213 (TU Muscae)}",
  journal = {\apj},
archivePrefix = "arXiv",
   eprint = {0905.3687},
 primaryClass = "astro-ph.SR",
     year = 2008,
    month = jul,
   volume = 681,
      eid = {554-561},
    pages = {554-561},
      doi = {10.1086/587509},
   adsurl = {http://adsabs.harvard.edu/abs/2008ApJ...681..554P}
}

@ARTICLE{rauw2016,
   author = {{Rauw}, G. and {Rosu}, S. and {Noels}, A. and {Mahy}, L. and 
	{Schmitt}, J.~H.~M.~M. and {Godart}, M. and {Dupret}, M.-A. and 
	{Gosset}, E.},
    title = "{Apsidal motion in the massive binary HD 152218}",
  journal = {\aap},
archivePrefix = "arXiv",
   eprint = {1609.02735},
 primaryClass = "astro-ph.SR",
     year = 2016,
    month = oct,
   volume = 594,
      eid = {A33},
    pages = {A33},
      doi = {10.1051/0004-6361/201628766},
   adsurl = {http://adsabs.harvard.edu/abs/2016A%26A...594A..33R}
}

@ARTICLE{penny2016,
   author = {{Penny}, L.~R. and {Epps}, J.~G. and {Snyder}, J.~D.},
    title = "{Tomographic Separation of Composite Spectra. XII. The Physical Properties and Spectral Phase Variability of the Massive Close Binary HD 159176}",
  journal = {\apj},
     year = 2016,
    month = dec,
   volume = 832,
      eid = {211},
    pages = {211},
      doi = {10.3847/0004-637X/832/2/211},
   adsurl = {http://adsabs.harvard.edu/abs/2016ApJ...832..211P}
}

@ARTICLE{johnston2017,
   author = {{Johnston}, C. and {Buysschaert}, B. and {Tkachenko}, A. and 
	{Aerts}, C. and {Neiner}, C.},
    title = "{Detection of intrinsic variability in the eclipsing massive main sequence O+B binary HD 165246}",
  journal = {ArXiv e-prints},
archivePrefix = "arXiv",
   eprint = {1704.07592},
 primaryClass = "astro-ph.SR",
     year = 2017,
    month = apr,
   adsurl = {http://adsabs.harvard.edu/abs/2017arXiv170407592J}
}

@ARTICLE{mason2009,
   author = {{Mason}, B.~D. and {Hartkopf}, W.~I. and {Gies}, D.~R. and {Henry}, T.~J. and 
	{Helsel}, J.~W.},
    title = "{The High Angular Resolution Multiplicity of Massive Stars}",
  journal = {\aj},
archivePrefix = "arXiv",
   eprint = {0811.0492},
     year = 2009,
    month = feb,
   volume = 137,
    pages = {3358-3377},
      doi = {10.1088/0004-6256/137/2/3358},
   adsurl = {http://adsabs.harvard.edu/abs/2009AJ....137.3358M}
}

@INPROCEEDINGS{sanaevans2011,
   author = {{Sana}, H. and {Evans}, C.~J.},
    title = "{The multiplicity of massive stars}",
booktitle = {Active OB Stars: Structure, Evolution, Mass Loss, and Critical Limits},
     year = 2011,
   series = {IAU Symposium},
   volume = 272,
archivePrefix = "arXiv",
   eprint = {1009.4197},
 primaryClass = "astro-ph.SR",
   editor = {{Neiner}, C. and {Wade}, G. and {Meynet}, G. and {Peters}, G.
	},
    month = jul,
    pages = {474-485},
      doi = {10.1017/S1743921311011124},
   adsurl = {http://adsabs.harvard.edu/abs/2011IAUS..272..474S}
}

@ARTICLE{lacour2011,
   author = {{Lacour}, S. and {Tuthill}, P. and {Ireland}, M. and {Amico}, P. and 
	{Girard}, J.},
    title = "{Sparse Aperture Masking on Paranal}",
  journal = {The Messenger},
     year = 2011,
    month = dec,
   volume = 146,
    pages = {18-23},
   adsurl = {http://adsabs.harvard.edu/abs/2011Msngr.146...18L}
}

@ARTICLE{lebouquin2011,
   author = {{Le Bouquin}, J.-B. and {Berger}, J.-P. and {Lazareff}, B. and 
	{Zins}, G. and {Haguenauer}, P. and {Jocou}, L. and {Kern}, P. and 
	{Millan-Gabet}, R. and {Traub}, W. and {Absil}, O. and {Augereau}, J.-C. and 
	{Benisty}, M. and {Blind}, N. and {Bonfils}, X. and {Bourget}, P. and 
	{Delboulbe}, A. and {Feautrier}, P. and {Germain}, M. and {Gitton}, P. and 
	{Gillier}, D. and {Kiekebusch}, M. and {Kluska}, J. and {Knudstrup}, J. and 
	{Labeye}, P. and {Lizon}, J.-L. and {Monin}, J.-L. and {Magnard}, Y. and 
	{Malbet}, F. and {Maurel}, D. and {M{\'e}nard}, F. and {Micallef}, M. and 
	{Michaud}, L. and {Montagnier}, G. and {Morel}, S. and {Moulin}, T. and 
	{Perraut}, K. and {Popovic}, D. and {Rabou}, P. and {Rochat}, S. and 
	{Rojas}, C. and {Roussel}, F. and {Roux}, A. and {Stadler}, E. and 
	{Stefl}, S. and {Tatulli}, E. and {Ventura}, N.},
    title = "{PIONIER: a 4-telescope visitor instrument at VLTI}",
  journal = {\aap},
archivePrefix = "arXiv",
   eprint = {1109.1918},
 primaryClass = "astro-ph.IM",
     year = 2011,
    month = nov,
   volume = 535,
      eid = {A67},
    pages = {A67},
      doi = {10.1051/0004-6361/201117586},
   adsurl = {http://adsabs.harvard.edu/abs/2011A%26A...535A..67L}
}

@ARTICLE{mahy2017,
   author = {{Mahy}, L. and {Damerdji}, Y. and {Gosset}, E. and {Nitschelm}, C. and 
	{Eenens}, P. and {Sana}, H. and {Klotz}, A.},
    title = "{A modern study of HD166734: a massive supergiant system}",
  journal = {ArXiv e-prints},
archivePrefix = "arXiv",
   eprint = {1707.02060},
 primaryClass = "astro-ph.SR",
     year = 2017,
    month = jul,
   adsurl = {http://adsabs.harvard.edu/abs/2017arXiv170702060M}
}

@article{astropy:2013,
Adsurl = {http://adsabs.harvard.edu/abs/2013A%26A...558A..33A},
Archiveprefix = {arXiv},
Author = {{Astropy Collaboration} and {Robitaille}, T.~P. and {Tollerud}, E.~J. and {Greenfield}, P. and {Droettboom}, M. and {Bray}, E. and {Aldcroft}, T. and {Davis}, M. and {Ginsburg}, A. and {Price-Whelan}, A.~M. and {Kerzendorf}, W.~E. and {Conley}, A. and {Crighton}, N. and {Barbary}, K. and {Muna}, D. and {Ferguson}, H. and {Grollier}, F. and {Parikh}, M.~M. and {Nair}, P.~H. and {Unther}, H.~M. and {Deil}, C. and {Woillez}, J. and {Conseil}, S. and {Kramer}, R. and {Turner}, J.~E.~H. and {Singer}, L. and {Fox}, R. and {Weaver}, B.~A. and {Zabalza}, V. and {Edwards}, Z.~I. and {Azalee Bostroem}, K. and {Burke}, D.~J. and {Casey}, A.~R. and {Crawford}, S.~M. and {Dencheva}, N. and {Ely}, J. and {Jenness}, T. and {Labrie}, K. and {Lim}, P.~L. and {Pierfederici}, F. and {Pontzen}, A. and {Ptak}, A. and {Refsdal}, B. and {Servillat}, M. and {Streicher}, O.},
Doi = {10.1051/0004-6361/201322068},
Eid = {A33},
Eprint = {1307.6212},
Journal = {\aap},
Month = oct,
Pages = {A33},
Primaryclass = {astro-ph.IM},
Title = {{Astropy: A community Python package for astronomy}},
Volume = 558,
Year = 2013}

@ARTICLE{astropy:2018,
       author = {{Astropy Collaboration} and {Price-Whelan}, A.~M. and
         {Sip{\H{o}}cz}, B.~M. and {G{\"u}nther}, H.~M. and {Lim}, P.~L. and
         {Crawford}, S.~M. and {Conseil}, S. and {Shupe}, D.~L. and
         {Craig}, M.~W. and {Dencheva}, N. and {Ginsburg}, A. and {Vand
        erPlas}, J.~T. and {Bradley}, L.~D. and {P{\'e}rez-Su{\'a}rez}, D. and
         {de Val-Borro}, M. and {Aldcroft}, T.~L. and {Cruz}, K.~L. and
         {Robitaille}, T.~P. and {Tollerud}, E.~J. and {Ardelean}, C. and
         {Babej}, T. and {Bach}, Y.~P. and {Bachetti}, M. and {Bakanov}, A.~V. and
         {Bamford}, S.~P. and {Barentsen}, G. and {Barmby}, P. and
         {Baumbach}, A. and {Berry}, K.~L. and {Biscani}, F. and {Boquien}, M. and
         {Bostroem}, K.~A. and {Bouma}, L.~G. and {Brammer}, G.~B. and
         {Bray}, E.~M. and {Breytenbach}, H. and {Buddelmeijer}, H. and
         {Burke}, D.~J. and {Calderone}, G. and {Cano Rodr{\'\i}guez}, J.~L. and
         {Cara}, M. and {Cardoso}, J.~V.~M. and {Cheedella}, S. and {Copin}, Y. and
         {Corrales}, L. and {Crichton}, D. and {D'Avella}, D. and {Deil}, C. and
         {Depagne}, {\'E}. and {Dietrich}, J.~P. and {Donath}, A. and
         {Droettboom}, M. and {Earl}, N. and {Erben}, T. and {Fabbro}, S. and
         {Ferreira}, L.~A. and {Finethy}, T. and {Fox}, R.~T. and
         {Garrison}, L.~H. and {Gibbons}, S.~L.~J. and {Goldstein}, D.~A. and
         {Gommers}, R. and {Greco}, J.~P. and {Greenfield}, P. and
         {Groener}, A.~M. and {Grollier}, F. and {Hagen}, A. and {Hirst}, P. and
         {Homeier}, D. and {Horton}, A.~J. and {Hosseinzadeh}, G. and {Hu}, L. and
         {Hunkeler}, J.~S. and {Ivezi{\'c}}, {\v{Z}}. and {Jain}, A. and
         {Jenness}, T. and {Kanarek}, G. and {Kendrew}, S. and {Kern}, N.~S. and
         {Kerzendorf}, W.~E. and {Khvalko}, A. and {King}, J. and {Kirkby}, D. and
         {Kulkarni}, A.~M. and {Kumar}, A. and {Lee}, A. and {Lenz}, D. and
         {Littlefair}, S.~P. and {Ma}, Z. and {Macleod}, D.~M. and
         {Mastropietro}, M. and {McCully}, C. and {Montagnac}, S. and
         {Morris}, B.~M. and {Mueller}, M. and {Mumford}, S.~J. and {Muna}, D. and
         {Murphy}, N.~A. and {Nelson}, S. and {Nguyen}, G.~H. and
         {Ninan}, J.~P. and {N{\"o}the}, M. and {Ogaz}, S. and {Oh}, S. and
         {Parejko}, J.~K. and {Parley}, N. and {Pascual}, S. and {Patil}, R. and
         {Patil}, A.~A. and {Plunkett}, A.~L. and {Prochaska}, J.~X. and
         {Rastogi}, T. and {Reddy Janga}, V. and {Sabater}, J. and
         {Sakurikar}, P. and {Seifert}, M. and {Sherbert}, L.~E. and
         {Sherwood-Taylor}, H. and {Shih}, A.~Y. and {Sick}, J. and
         {Silbiger}, M.~T. and {Singanamalla}, S. and {Singer}, L.~P. and
         {Sladen}, P.~H. and {Sooley}, K.~A. and {Sornarajah}, S. and
         {Streicher}, O. and {Teuben}, P. and {Thomas}, S.~W. and
         {Tremblay}, G.~R. and {Turner}, J.~E.~H. and {Terr{\'o}n}, V. and
         {van Kerkwijk}, M.~H. and {de la Vega}, A. and {Watkins}, L.~L. and
         {Weaver}, B.~A. and {Whitmore}, J.~B. and {Woillez}, J. and
         {Zabalza}, V. and {Astropy Contributors}},
        title = "{The Astropy Project: Building an Open-science Project and Status of the v2.0 Core Package}",
      journal = {\aj},
         year = 2018,
        month = sep,
       volume = {156},
       number = {3},
          eid = {123},
        pages = {123},
          doi = {10.3847/1538-3881/aabc4f},
archivePrefix = {arXiv},
       eprint = {1801.02634},
 primaryClass = {astro-ph.IM},
       adsurl = {https://ui.adsabs.harvard.edu/abs/2018AJ....156..123A}
}

@ARTICLE{astropy:2022,
       author = {{Astropy Collaboration} and {Price-Whelan}, Adrian M. and {Lim}, Pey Lian and {Earl}, Nicholas and {Starkman}, Nathaniel and {Bradley}, Larry and {Shupe}, David L. and {Patil}, Aarya A. and {Corrales}, Lia and {Brasseur}, C.~E. and {N{"o}the}, Maximilian and {Donath}, Axel and {Tollerud}, Erik and {Morris}, Brett M. and {Ginsburg}, Adam and {Vaher}, Eero and {Weaver}, Benjamin A. and {Tocknell}, James and {Jamieson}, William and {van Kerkwijk}, Marten H. and {Robitaille}, Thomas P. and {Merry}, Bruce and {Bachetti}, Matteo and {G{"u}nther}, H. Moritz and {Aldcroft}, Thomas L. and {Alvarado-Montes}, Jaime A. and {Archibald}, Anne M. and {B{'o}di}, Attila and {Bapat}, Shreyas and {Barentsen}, Geert and {Baz{'a}n}, Juanjo and {Biswas}, Manish and {Boquien}, M{'e}d{'e}ric and {Burke}, D.~J. and {Cara}, Daria and {Cara}, Mihai and {Conroy}, Kyle E. and {Conseil}, Simon and {Craig}, Matthew W. and {Cross}, Robert M. and {Cruz}, Kelle L. and {D'Eugenio}, Francesco and {Dencheva}, Nadia and {Devillepoix}, Hadrien A.~R. and {Dietrich}, J{"o}rg P. and {Eigenbrot}, Arthur Davis and {Erben}, Thomas and {Ferreira}, Leonardo and {Foreman-Mackey}, Daniel and {Fox}, Ryan and {Freij}, Nabil and {Garg}, Suyog and {Geda}, Robel and {Glattly}, Lauren and {Gondhalekar}, Yash and {Gordon}, Karl D. and {Grant}, David and {Greenfield}, Perry and {Groener}, Austen M. and {Guest}, Steve and {Gurovich}, Sebastian and {Handberg}, Rasmus and {Hart}, Akeem and {Hatfield-Dodds}, Zac and {Homeier}, Derek and {Hosseinzadeh}, Griffin and {Jenness}, Tim and {Jones}, Craig K. and {Joseph}, Prajwel and {Kalmbach}, J. Bryce and {Karamehmetoglu}, Emir and {Ka{l}uszy{'n}ski}, Miko{l}aj and {Kelley}, Michael S.~P. and {Kern}, Nicholas and {Kerzendorf}, Wolfgang E. and {Koch}, Eric W. and {Kulumani}, Shankar and {Lee}, Antony and {Ly}, Chun and {Ma}, Zhiyuan and {MacBride}, Conor and {Maljaars}, Jakob M. and {Muna}, Demitri and {Murphy}, N.~A. and {Norman}, Henrik and {O'Steen}, Richard and {Oman}, Kyle A. and {Pacifici}, Camilla and {Pascual}, Sergio and {Pascual-Granado}, J. and {Patil}, Rohit R. and {Perren}, Gabriel I. and {Pickering}, Timothy E. and {Rastogi}, Tanuj and {Roulston}, Benjamin R. and {Ryan}, Daniel F. and {Rykoff}, Eli S. and {Sabater}, Jose and {Sakurikar}, Parikshit and {Salgado}, Jes{'u}s and {Sanghi}, Aniket and {Saunders}, Nicholas and {Savchenko}, Volodymyr and {Schwardt}, Ludwig and {Seifert-Eckert}, Michael and {Shih}, Albert Y. and {Jain}, Anany Shrey and {Shukla}, Gyanendra and {Sick}, Jonathan and {Simpson}, Chris and {Singanamalla}, Sudheesh and {Singer}, Leo P. and {Singhal}, Jaladh and {Sinha}, Manodeep and {Sip{H{o}}cz}, Brigitta M. and {Spitler}, Lee R. and {Stansby}, David and {Streicher}, Ole and {{{S}}umak}, Jani and {Swinbank}, John D. and {Taranu}, Dan S. and {Tewary}, Nikita and {Tremblay}, Grant R. and {Val-Borro}, Miguel de and {Van Kooten}, Samuel J. and {Vasovi{'c}}, Zlatan and {Verma}, Shresth and {de Miranda Cardoso}, Jos{'e} Vin{'i}cius and {Williams}, Peter K.~G. and {Wilson}, Tom J. and {Winkel}, Benjamin and {Wood-Vasey}, W.~M. and {Xue}, Rui and {Yoachim}, Peter and {Zhang}, Chen and {Zonca}, Andrea and {Astropy Project Contributors}},
        title = "{The Astropy Project: Sustaining and Growing a Community-oriented Open-source Project and the Latest Major Release (v5.0) of the Core Package}",
      journal = {\apj},
         year = 2022,
        month = aug,
       volume = {935},
       number = {2},
          eid = {167},
        pages = {167},
          doi = {10.3847/1538-4357/ac7c74},
archivePrefix = {arXiv},
       eprint = {2206.14220},
 primaryClass = {astro-ph.IM},
       adsurl = {https://ui.adsabs.harvard.edu/abs/2022ApJ...935..167A}
}

\begin{appendix}
\nolinenumbers

\section{Absolute magnitude calibration}\label{sec:absmag}

To obtain the absolute $K$- and $H$-band magnitudes that are needed for the distance calculations, we used the \cite{martins2006} calibration. As \cite{martins2006} only gives values for luminosity classes V, III, and I, the first step was to obtain values for luminosity classes II and IV. To do this, we fitted a second degree polynomial of the form
\begin{equation}
M = a\times \mathrm{LC}^2 + b \times \mathrm{LC} + c \label{eq:LCvsMk}
\end{equation}
\noindent to the magnitudes of each of the spectral subtypes listed in \cite{martins2006}. Here, LC is an integer in the range 1 to 5, representing the luminosity classes I to V.

Table~\ref{tab:LCvsMk} lists the obtained values of $a$, $b$, and $c$ for each of the spectral subtypes listed in \cite{martins2006}. Figure~\ref{fig:LCvsMk} shows the corresponding LC versus $M_{\mathrm{K}}$ relations.

\begin{table}[!b]
\centering
\caption{Obtained coefficients of Eq.~\ref{eq:LCvsMk} for the K band.}\label{tab:LCvsMk}
\begin{tabular}{l c c c}
\hline\hline \\[-8pt]
SpT                     &       $a$     &       $b$             & $c$\\
\hline \\[-8pt]
3       & 0.0100                & 0.080 & $-$5.630              \\
4       & 0.0175                & 0.105 & $-$5.653              \\
5       & 0.0213                & 0.155 & $-$5.696              \\
5.5     & 0.0188                & 0.200 & $-$5.739              \\
6       & 0.0213                & 0.220 & $-$5.761              \\
6.5     & 0.0163                & 0.290 & $-$5.836              \\
7       & 0.0188                & 0.375 & $-$5.836              \\
7.5     & 0.0200                & 0.340 & $-$5.900              \\
8       & 0.0188                & 0.375 & $-$5.914              \\
8.5     & 0.0200                & 0.410 & $-$5.960              \\
9       & 0.0200                & 0.440 & $-$5.980              \\
9.5     & 0.0238                & 0.450 & $-$5.994              \\
\hline
\end{tabular}
\end{table}

\begin{figure}[b!]
\centering
        \resizebox{0.85\hsize}{!}{\includegraphics{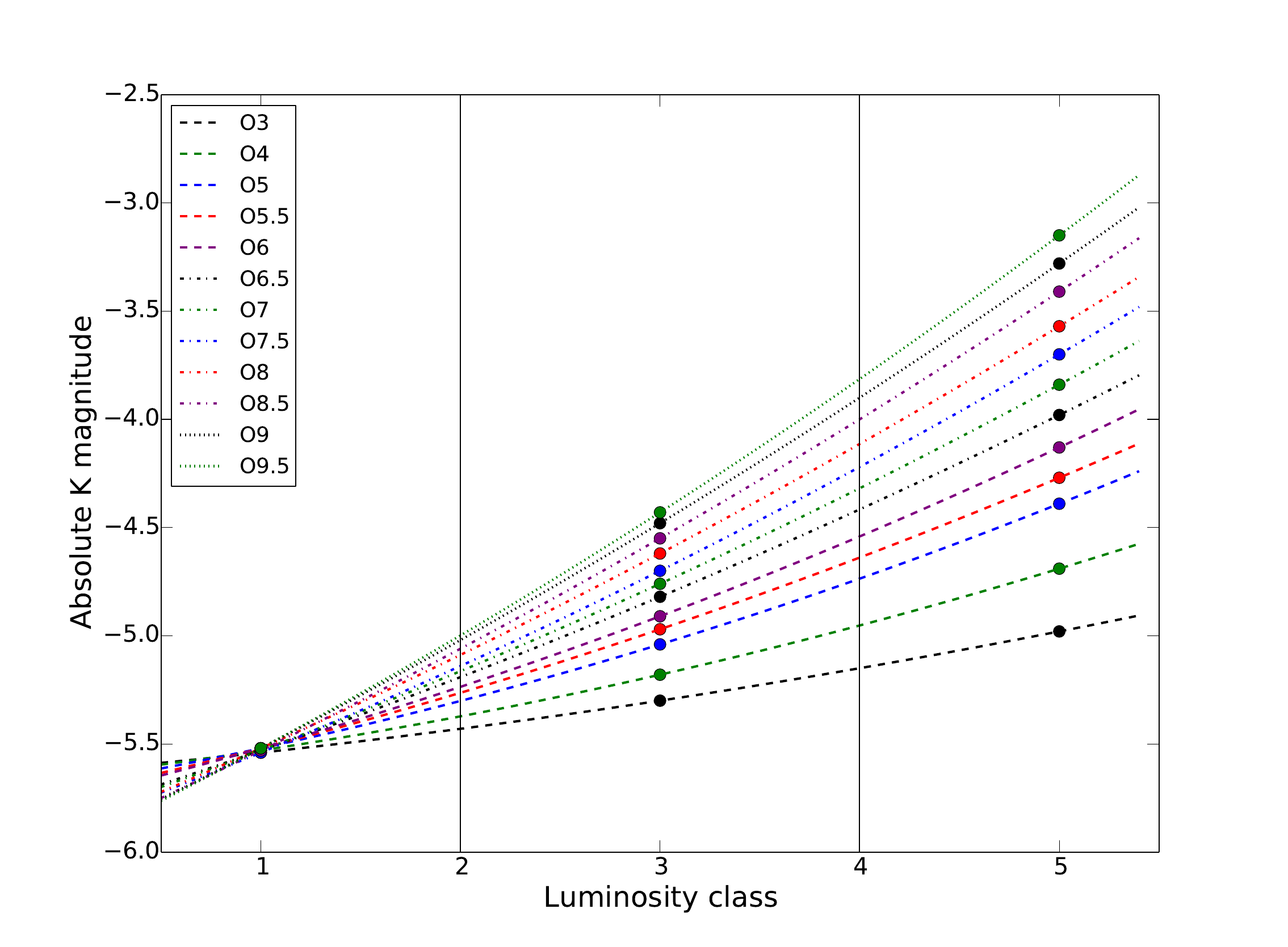}}
        \caption{LC versus $M_{\mathrm{K}}$ calibration for each of the spectral subtypes listed in \cite{martins2006}.}
        \label{fig:LCvsMk}
\end{figure}

\begin{figure}[h!]
\centering
        \resizebox{0.85\hsize}{!}{\includegraphics{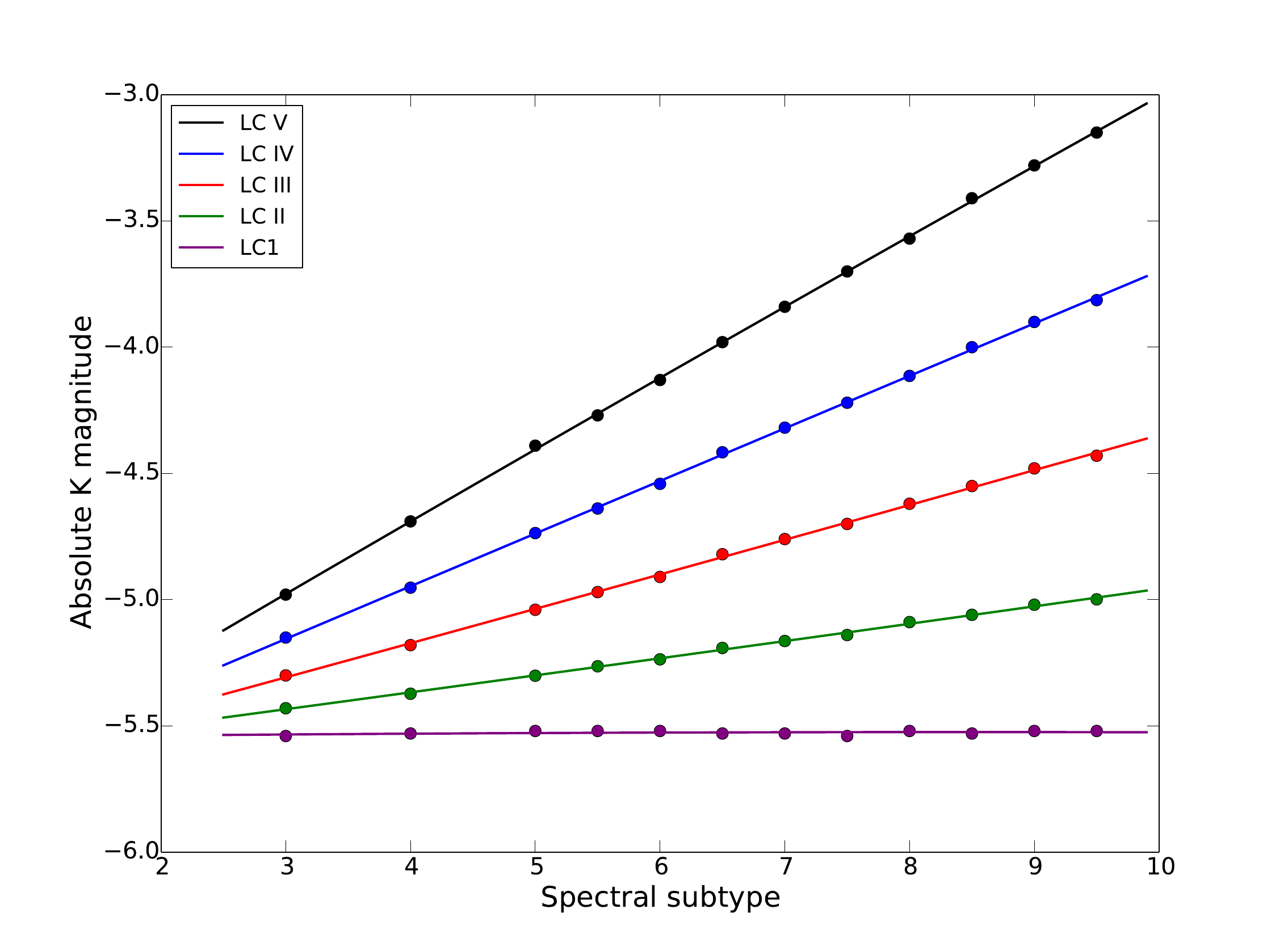}}
        \caption{SpT versus $M_{\mathrm{K}}$ calibration for each of the luminosity classes.}
        \label{fig:SpTvsMk}
\end{figure}

We used an identical approach to obtain relations for the absolute magnitude as a function of the spectral subtype (SpT). This was done to obtain absolute magnitudes for the spectral subtypes not listed in \cite{martins2006}, that is, O2, O2.5, O3.5, O9.2, and O9.7. We fitted a second-degree polynomial of the form

\begin{equation}
M = a\times \mathrm{SpT}^2 + b \times \mathrm{SpT} + c \label{eq:SpTvsMk}
\end{equation}

\noindent to the absolute magnitudes of each of the luminosity classes. For LC V, III, and I we used the \cite{martins2006} values, and for LC IV and II the values obtained from the LC versus $M_{\mathrm{K}}$ calibration presented above. This resulted in five SpT versus $M_{\mathrm{K}}$ relations, which are presented in Fig.~\ref{fig:SpTvsMk}. The corresponding coefficients of the polynomials are given in Table~\ref{tab:SpTvsMk}.

\begin{table}
\centering
\caption{Obtained coefficients of Eq.~\ref{eq:SpTvsMk} for the K band.}\label{tab:SpTvsMk}
\begin{tabular}{l c c c}
\hline\hline \\[-8pt]
LC                      &       $a$     &       $b$             & $c$\\
\hline \\[-8pt]
V       & $-$0.00096    & 0.29395       & $-$5.85071    \\
IV      & $-$0.00099    & 0.20948       & $-$5.78312    \\
III     & 0.00030               & 0.13323       & $-$5.71012    \\
II      & 0.00022       & 0.06522       & $-$5.63150    \\
I       & $-$0.00032    & 0.00542       & $-$5.54732    \\
\hline
\end{tabular}
\end{table}

The obtained relations also provide the absolute $H$-band magnitudes using $(H-K)_0 = -0.10$, which is valid for all spectral subtypes and luminosity classes in the O-star regime \citep{martins2006}.

\section{Magnitude correction}\label{sec:magcorrect}

Before the distance can be derived from the H-band magnitude, this magnitude first needs to be corrected for the contribution to the 2MASS magnitude from companions within the 4\arcsec\ aperture. The next subsection briefly describes this process for companions that are interferometrically resolved, and thus have a measured magnitude difference. Section~\ref{sec:magcorrectSB} describes the correction for the unresolved companions (i.e., the SBs).

\subsection{Correction for resolved companions}\label{sec:magcorrectresolved}

We corrected for all observed companions that are within 4\arcsec\ of the primary star, and for which either $\Delta m_{\mathrm{K_s}}$ or $\Delta m_{\mathrm{H}}$ is available. As we derived the distance from the H-band magnitude, we assume that $\Delta m_{\mathrm{K_s}} = \Delta m_{\mathrm{H}}$, which is valid when the companion is in the Rayleigh-Jeans domain in the H and K bands. While this assumption may not hold for the faintest companions, these stars have such a large magnitude difference from the primary that their contribution to the observed magnitude is negligible.

For a single companion, the corrected magnitude ($m_1$) of the central star is given by

\begin{equation}
m_1 = m_{\mathrm{obs}} + 2.5\log{(1 + 10^{-0.4 \Delta m})},          \label{eq:magcorr}
\end{equation}

\noindent where $m_{\mathrm{obs}}$ is the observed magnitude, and $\Delta m$ the magnitude difference between the central star and the companion. If multiple companions are present within the aperture, the correction was iterated for each of the components.

\subsection{Correction for unresolved companions (SBs)}\label{sec:magcorrectSB}

Apart from the resolved companions, the observed magnitude also contains a contribution from the unresolved companions, i.e., the spectroscopic binaries. To correct the magnitude for this, we used the calibration given in Appendix~\ref{sec:absmag} to estimate the magnitude difference. Therefore, we could only do this for systems for which the spectral types of both components are known. Thus, we could not correct for SB1 systems, apart from four SB1 companions that have been resolved by \pionier. However, that the companion is not seen in the spectrum implies that it is likely faint compared to the primary star, and thus will only have a small contribution to the combined magnitude, i.e., the derived distance. These systems have been marked in Table~\ref{tab:distance}. 

\section{Extinction correction}\label{sec:extinction}

To estimate the extinction toward each of the stars, we used the available $VJHK$-band photometry from 2MASS. Using Eqs. A3, A4, and A5 from \cite{fitzpatrick1999} we derived the following relations for $E(B-V)$:
\begin{subequations}
\begin{equation}
E(B-V) = 1.39 \times E(V-J) / (R + 0.02)\label{eq:EVminJ}
\end{equation}
\begin{equation}
E(B-V) = 1.19 \times E(V-H) / (R - 0.04)\label{eq:EVminH}
\end{equation}
\begin{equation}
E(B-V) = 1.12 \times E(V-K) / (R - 0.02)\label{eq:EVminK}
\end{equation}
\end{subequations}
\begin{figure}[b!]
        \resizebox{0.85\hsize}{!}{\includegraphics{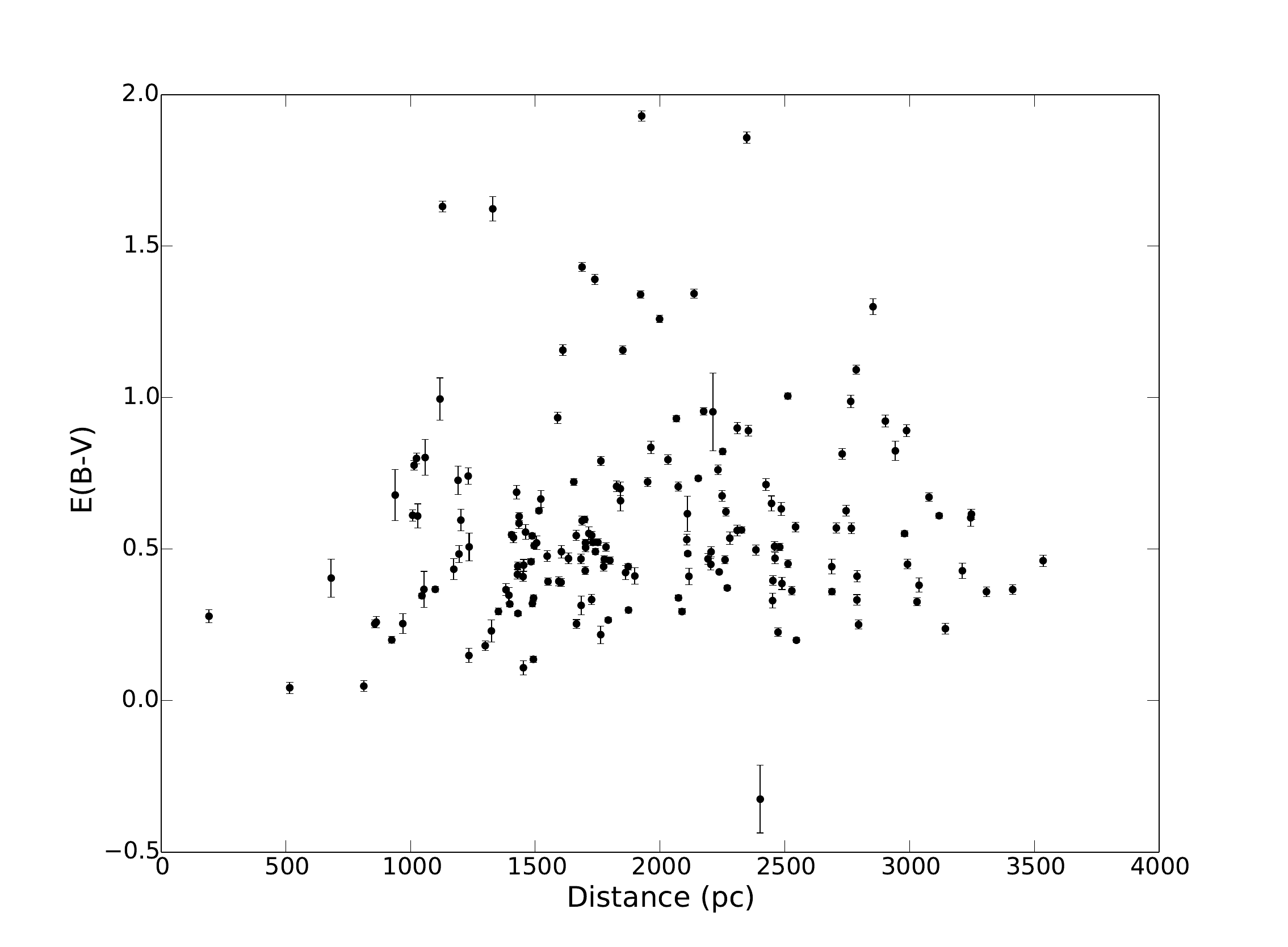}}
        \caption{Derived $E(B-V)$ for all stars as a function of the derived distance.}
        \label{fig:EBminV}
\end{figure}
To use Eq.~\ref{eq:EVminK}, we applied the average offset of 0.04 mag between the 2MASS $K_s$ band and the $K$ band, i.e., $K_s = K + 0.04$. We assumed the average value of the total-to-selective extinction of $R=3.1$ for all systems. Figure~\ref{fig:EBminV} shows the derived values of $E(B-V)$ for each of the systems. Here, the given value is the mean $E(B-V)$ derived from the three colors, and the error bars correspond to the standard deviation.

The resulting values of $E(B-V)$ were then used to derive $A_{\mathrm{H}}$ and $A_{\mathrm{K}}$ using the model curve values of \cite[][their Table 2]{fitzpatrick1999}, which results in

\begin{subequations}
\begin{equation}
A_\mathrm{K} = 0.36 \times E(B-V)\label{eq:Ak}
\end{equation}
\begin{equation}
A_\mathrm{H} = 0.53 \times E(B-V)\label{eq:Ah}
\end{equation}
\end{subequations}

\section{Bolometric corrections}\label{sec:bolcor}

To convert the absolute magnitudes to bolometric magnitudes and hence the luminosity, we again used the \cite{martins2006} calibration to derive relations for the bolometric correction in the $H$ band ($BC_H$). As for the absolute magnitude calibration, we first fitted $BC_H$ as a function of the luminosity class using

\begin{equation}
BC_H = a\times \mathrm{LC}^2 + b \times \mathrm{LC} + c. \label{eq:LCvsBC}
\end{equation}

\noindent The results are shown in Fig.~\ref{fig:BC_LC} and the derived coefficients given in Table~\ref{tab:LCvsBC}. Using these results, we derived $BC_H$ for luminosity classes IV and II, which are not given in \cite{martins2006}. For each luminosity class we then fitted $BC_H$ as a function of spectral subtype using

\begin{equation}
BC_H = a\times \mathrm{SpT}^2 + b \times \mathrm{SpT} + c. \label{eq:SpTvsBC}
\end{equation}

\noindent The resulting relations are shown in Fig.~\ref{fig:BC_SpT} and the coefficients given in Table~\ref{tab:SpTvsBC}. 

Finally, to also be able to derive $BC_H$ for the companion stars, we fitted a second-degree polynomial for $BC_H$ as a function of the absolute H-band magnitude $M_H$ for each of the luminosity classes:

\begin{equation}
BC_H = a\times M_H^2 + b \times M_H + c. \label{eq:MHvsBC}
\end{equation}

\noindent The resulting coefficients are given in Table~\ref{tab:MHvsBC} and the relations are plotted in Fig.~\ref{fig:BC_H}.

\begin{table}[h!]
\centering
\caption{Obtained coefficients of Eq.~\ref{eq:LCvsBC} for the H band.}\label{tab:LCvsBC}
\begin{tabular}{l c c c}
\hline\hline \\[-8pt]
SpT                     &       $a$     &       $b$             & $c$\\
\hline \\[-8pt]
3       & 0.0163                & $-$0.145      & $-$4.461              \\
4       & 0.0150                & $-$0.135      & $-$4.330              \\
5       & 0.0125                & $-$0.120      & $-$4.193              \\
5.5     & 0.0112                & $-$0.110      & $-$4.131              \\
6       & 0.0075                & $-$0.090      & $-$4.068              \\
6.5     & 0.0050                & $-$0.070      & $-$4.015              \\
7       & 0.0025                & $-$0.055      & $-$3.948              \\
7.5     & 0.0025                & $-$0.055      & $-$3.858              \\
8       & $-$0.0013             & $-$0.030      & $-$3.799              \\
8.5     & $-$0.0038             & $-$0.015      & $-$3.721              \\
9       & $-$0.0063             & 0.000 & $-$3.644              \\
9.5     & $-$0.0100             & 0.025 & $-$3.575              \\
\hline
\end{tabular}
\end{table}

\begin{table}[h!]
\centering
\caption{Obtained coefficients of Eq.~\ref{eq:SpTvsBC} for the H band.}\label{tab:SpTvsBC}
\begin{tabular}{l c c c}
\hline\hline \\[-8pt]
LC                      &       $a$     &       $b$             & $c$\\
\hline \\[-8pt]
V       & 0.0041        & 0.1142        & $-$5.1557     \\
IV      & 0.0050        & 0.1125        & $-$5.1627     \\
III     & 0.0053        & 0.1112        & $-$5.1309     \\
II      &  0.0049       & 0.1103        & $-$5.0602     \\
I       & 0.0038        & 0.1097        & $-$4.9509     \\
\hline
\end{tabular}
\end{table}

\begin{table}[h!]
\centering
\caption{Obtained coefficients of Eq.~\ref{eq:MHvsBC} for the H band.}\label{tab:MHvsBC}
\begin{tabular}{l c c c}
\hline\hline \\[-8pt]
LC                      &       $a$     &       $b$             & $c$\\
\hline \\[-8pt]
V       &       0.0584 &        1.0719 &        $-$0.8398\\
IV      &       0.1184 &        1.9274&         2.0806\\
III     &       0.2639 &        3.9188 &        8.7255\\
II      &       0.9456 &        12.578 &        35.962\\
I       &       0 &             94.984 &        530.40\\
\hline
\end{tabular}
\end{table}

\begin{figure}[h!]
\centering
        \resizebox{0.85\hsize}{!}{\includegraphics{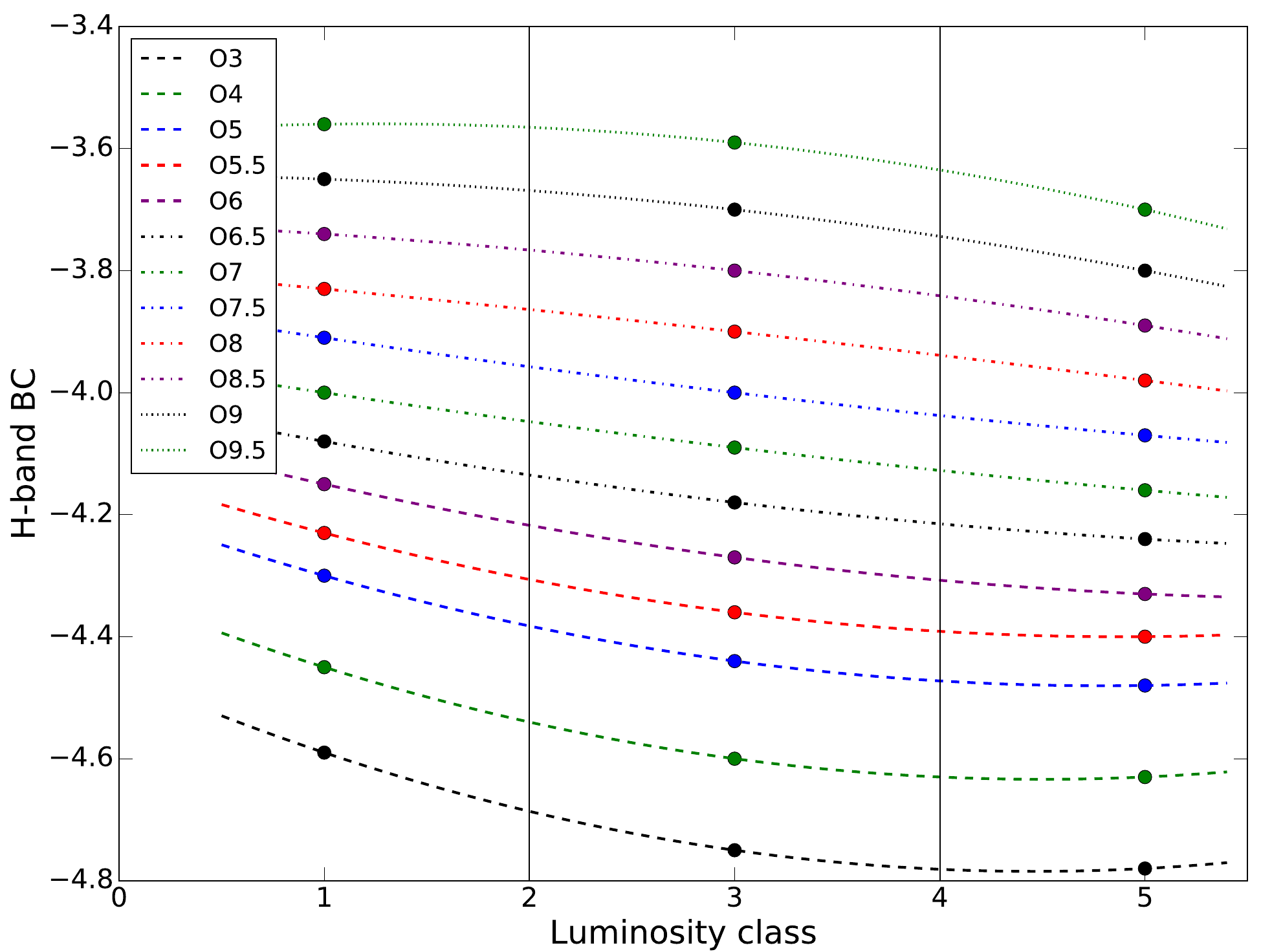}}
        \caption{LC versus $BC_H$ calibration for each of the spectral subtypes listed in \cite{martins2006}.}
        \label{fig:BC_LC}
\end{figure}

\begin{figure}[h!]
\centering
        \resizebox{0.85\hsize}{!}{\includegraphics{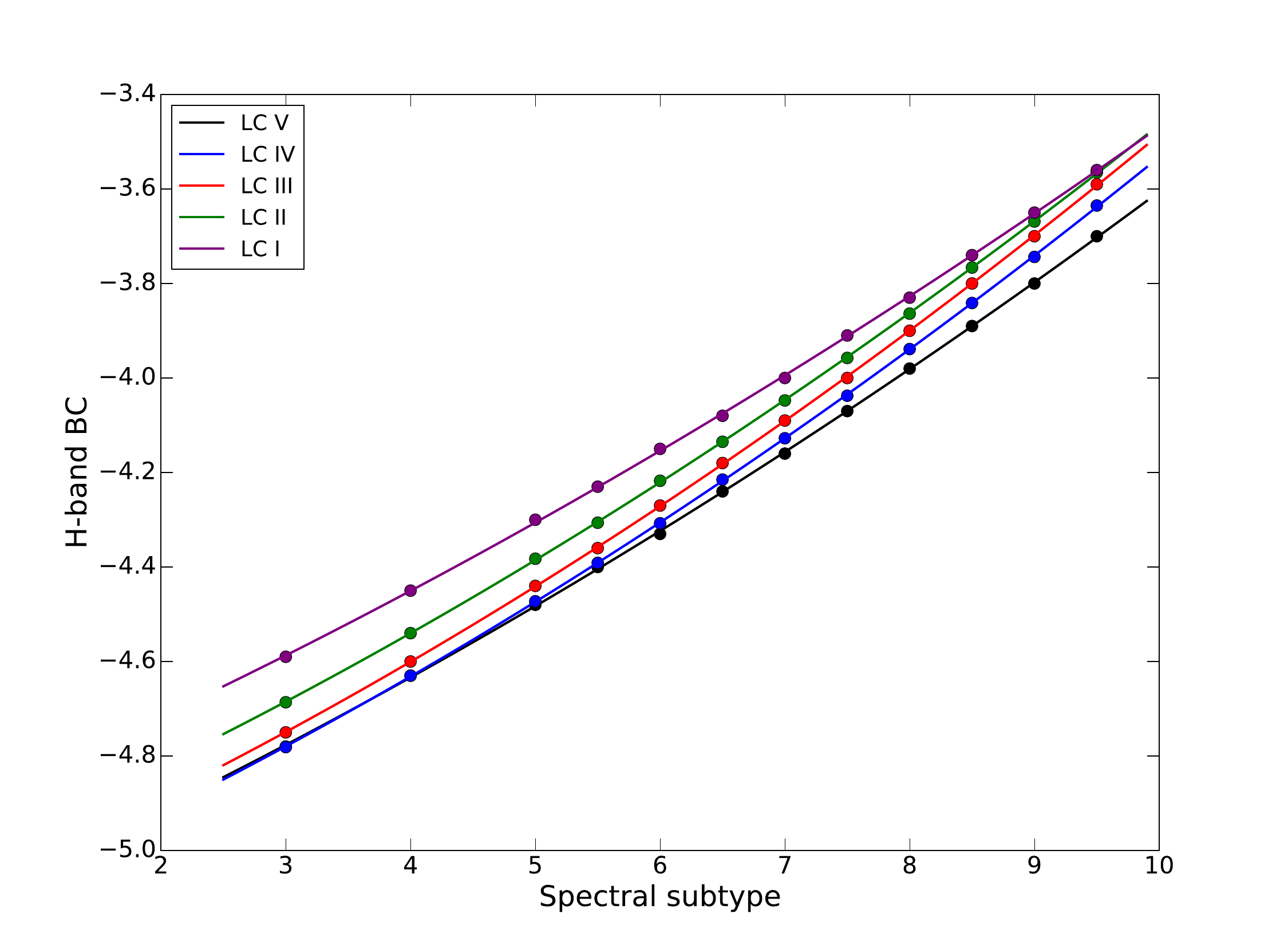}}
        \caption{SpT versus $BC_H$ for each of the luminosity classes.}
        \label{fig:BC_SpT}
\end{figure}

\begin{figure}[h!]
\centering
        \resizebox{0.85\hsize}{!}{\includegraphics{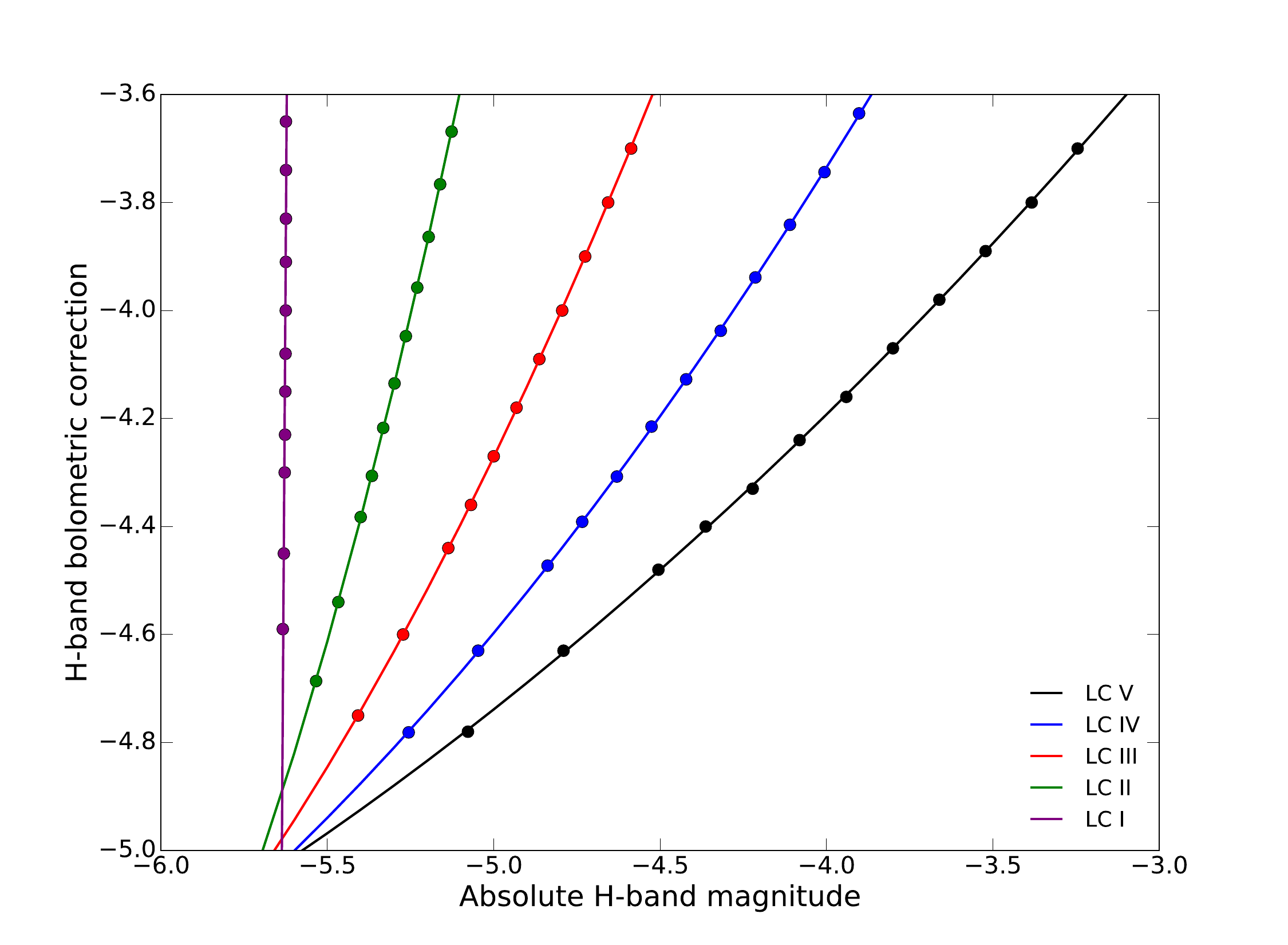}}
        \caption{$M_H$ versus $BC_H$ for each luminosity class.}
        \label{fig:BC_H}
\end{figure}

\section{Mass-luminosity relation}\label{sec:masslum}

To estimate the masses of the primaries and companions we used the mass-luminosity relation,

\begin{equation}
\frac{L}{L_\odot} = \left( \frac{M}{M_{\odot}} \right)^x. \label{eq:MLeq}
\end{equation}

\noindent The value of the exponent $x$ varies with the luminosity and evolutionary state, which is shown in Fig.~\ref{fig:ML} using the \cite{brott2011} evolutionary models and the \cite{martins2005} parameters. The exponent could be fitted well by a fourth-degree polynomial for the \cite{brott2011} models, i.e.,

\begin{equation}
x = a\times \left(\frac{L}{L_{\odot}}\right)^4 + b \times \left(\frac{L}{L_{\odot}}\right)^3 + c \times \left(\frac{L}{L_{\odot}}\right)^2 + d \times \left(\frac{L}{L_{\odot}}\right) + e. \label{eq:ML}
\end{equation}

The \cite{martins2005} relations are nearly linear, and were fitted using a second-degree polynomial (i.e., $a=b=0$). The coefficients of Eq.~\ref{eq:ML} for both the \cite{brott2011} models and the \cite{martins2005} parameters are given in Table~\ref{tab:ML}. 

\begin{figure}[h!]
\centering
        \resizebox{0.85\hsize}{!}{\includegraphics{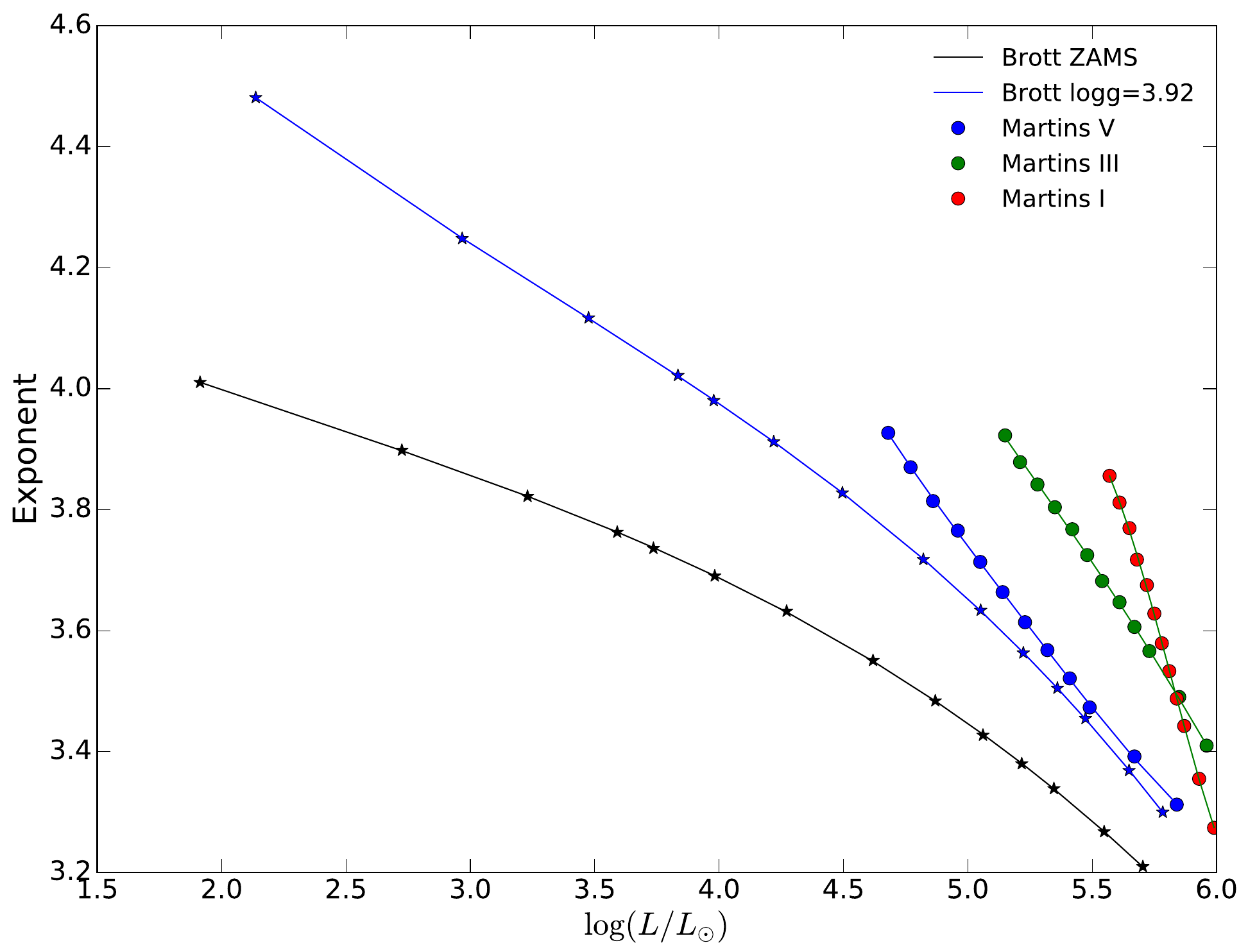}}
        \caption{Exponents of the mass-luminosity relations derived from \cite{brott2011} and \cite{martins2005}.}
        \label{fig:ML}
\end{figure}
\begin{figure}[h!]
\centering
        \resizebox{0.85\hsize}{!}{\includegraphics{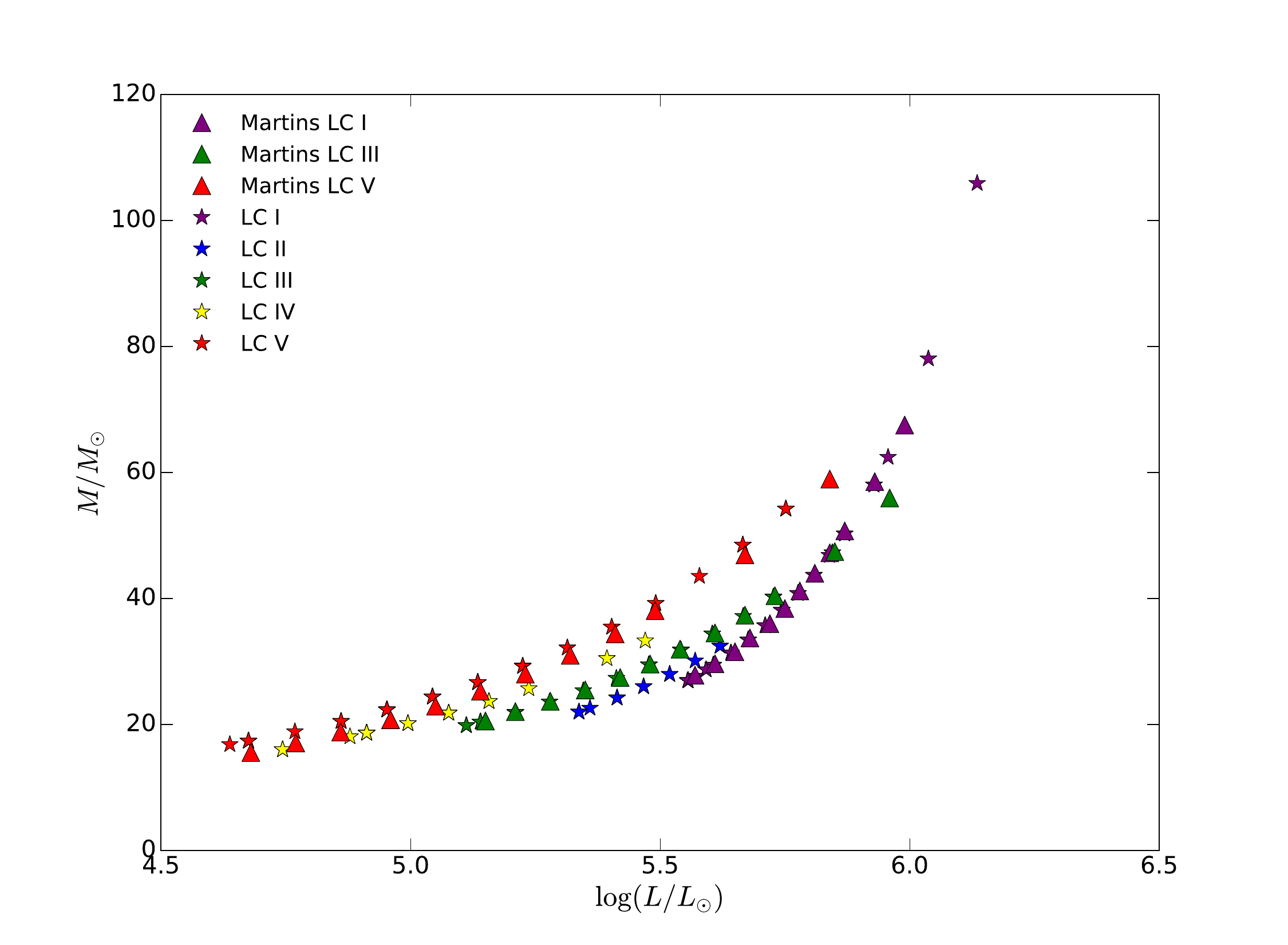}}
        \caption{Derived masses from the adopted mass-luminosity relations for the primary stars (star symbols) compared to \cite{martins2005} values (triangles) for each luminosity class.}
        \label{fig:LvsM}
\end{figure}

\onecolumn

\begin{table*}
\centering
\caption{Coefficients of Eq.~\ref{eq:ML}}\label{tab:ML}
\begin{tabular}{l c c c c c}
\hline\hline \\[-8pt]
                        &       $a$     &       $b$             & $c$   & $d$     & $e$\\
\hline \\[-8pt]
\cite{brott2011} ZAMS                   & $-$2.449$\times 10^{-4}$      & $-$3.502$\times 10^{-3}$        & 3.050$\times 10^{-2}$         & $-$0.211      & 4.331   \\
\cite{brott2011} $\log{g} = 3.92$               & $-$6.932$\times 10^{-4}$      & $-$2.968$\times 10^{-3}$        & 7.397$\times 10^{-2}$         & $-$0.552      & 5.366   \\
\cite{martins2005} LC V                 & 0                                             & 0                                               & 6.055$\times 10^{-2}$         & $-$1.166        & 8.057 \\
\cite{martins2005} LC III                       & 0                                             & 0                                               & $-$8.515$\times 10^{-2}$      & 0.321           & 4.522 \\
\cite{martins2005} LC I                 & 0                                             & 0                                               & $-$0.354              & 2.668           & $-$1.182$\times 10^{-2}$      \\
\hline
\end{tabular}
\end{table*}

\section{Derived distances, separations, and masses}\label{sec:distanceAppendix}

The distances, separations, and masses derived for each of the program stars and detected companions are given in Table~\ref{tab:distance}. The full table is available on CDS.

\begin{table*}[h!]
\centering
\caption{(Excerpt) Derived absolute H-band magnitude ($M_{H,p})$, distance ($d_H$), luminosity ($L_p$), and mass ($M_p$) for the primary stars. The indented lines show the adopted luminosity class ($LC_c$), absolute H-band magnitude ($M_{H,c}$), separation ($Sep$), luminosity ($L_c$), mass ($M_c$), and mass ratio ($Mc/Mp$) for each detected companion.}\label{tab:distance} 
\begin{tabular}{l l l l l l l}
\hline\hline \\[-8pt]

ID & SpT & $M_{H,p}$ & $d_H$ (kpc) & $\log(L_p/L_{\odot})$ & $M_p/M_{\odot}$ & \\ 
\ \ \ \it{Pair} &\ \ \ $LC_c$  &\ \ \ $M_{H,c}$ & \ \ \ \it{Sep (AU)} & \ \ \ $\log(L_c/L_{\odot})$ & \ \ \ $M_c/M_{\odot}$ & $M_c/M_p$\\ 
\hline 
HD~52266$^{c}$ &O9.5 III&-4.52 & 2088$\pm$291&5.1 & $20.4$\\ 
HD~53975$^{}$ &O7.5 V + B2-3 V&-3.8 & 1299$\pm$331&5.0 & $24.4$\\ 
HD~54662$^{}$ &O6.5 V + O7?9.5 V&-4.08 & 1397$\pm$309&5.2 & $29.3$\\ 
\ \ \ {\it  A-B } & \ \ \ \it{V} &\ \ \ \it{-3.85} &\ \ \ \it{3.6} &\ \ \ \it{5.1} &\ \ \ \it{25.2} & \it{0.86}\\ 
HD~55879$^{}$ &O9.7 III&-4.49 & 1492$\pm$212&5.1 & $19.8$\\ 
HD~57060$^{}$ &O7.5-8 Iab + O9.7 Ib&-5.62 & 1762$\pm$27&5.7 & $35.7$\\ 
HD~57061$^{}$ &O9 II + (B0.5 V + B0.5 V)&-5.13 & 1234$\pm$146&5.4 & $24.2$\\ 
\ \ \ {\it  Aa-Ab } & \ \ \ \it{IV} &\ \ \ \it{-4.06} &\ \ \ \it{140.7} &\ \ \ \it{5.0} &\ \ \ \it{21.0} & \it{0.86}\\ 
\ \ \ {\it  Aa-E } & \ \ \ \it{V} &\ \ \ \it{-0.66} &\ \ \ \it{1172.6} &\ \ \ \it{2.8} &\ \ \ \it{4.4} & \it{0.18}\\ 
\ \ \ {\it  Aa-B } & \ \ \ \it{V} &\ \ \ \it{0.17} &\ \ \ \it{10490.1} &\ \ \ \it{2.1} &\ \ \ \it{2.9} & \it{0.12}\\ 
\ \ \ {\it  Aa-C } & \ \ \ \it{V} &\ \ \ \it{1.17} &\ \ \ \it{18265.1} &\ \ \ \it{1.2} &\ \ \ \it{1.8} & \it{0.07}\\ 
\ \ \ {\it  Aa-D } & \ \ \ \it{V} &\ \ \ \it{-1.83} &\ \ \ \it{104901.0} &\ \ \ \it{3.7} &\ \ \ \it{8.0} & \it{0.33}\\ 
...\\
\hline 
\end{tabular} 

\tablefoot{
   \tablefoottext{a}{Magnitude not corrected for an interferometric companion.}
   \tablefoottext{b}{V-band magnitude taken from Simbad.} 
   \tablefoottext{c}{Magnitude not corrected for a SB companion.}
   }
\end{table*}
\end{appendix}
\end{document}